\documentclass[useAMS,usenatbib]{mn2e}
\usepackage[dvips]{graphicx}

\title[New flux-calibrated Lick fitting functions]{Empirical calibrations of optical absorption line indices based on the stellar library MILES}
\author[J. Johansson, D. Thomas and C. Maraston]{Jonas Johansson, Daniel Thomas and Claudia Maraston\\
Institute of Cosmology and Gravitation, Dennis Sciama Building, Burnaby Road, Portsmouth PO1 3FX}

\begin{document}

\pagerange{\pageref{firstpage}--\pageref{lastpage}} \pubyear{2010}
\maketitle
\label{firstpage}
\begin{abstract} 
Stellar population models of absorption line indices are an important  
tool for the analysis of stellar population spectra. They  
are most accurately modelled through empirical calibrations of  
absorption line indices with the stellar parameters effective  
temperature, metallicity, and surface gravity, the so-called fitting  
functions. Here we present new empirical fitting functions for the 25  
optical Lick absorption line indices based on the new stellar library MILES.  
The major improvements with respect to the Lick/IDS library are the  
better sampling of stellar parameter space, a generally higher signal-
to-noise, and a careful flux calibration. In fact we find that errors  
on individual index measurements in MILES are considerably smaller  
than in Lick/IDS. Instead we find the rms of the residuals between the final fitting
functions and the data to be dominated by errors in the stellar parameters. 
We provide fitting functions for both Lick/IDS and  
MILES spectral resolutions, and compare our results with other  
fitting functions in the literature. A Fortran 90 code is available  
online in order to simplify the  
implementation in stellar population models. We further calculate the  
offsets in index measurements between the Lick/IDS system to a flux  
calibrated system. For this purpose we use the three libraries MILES,  
ELODIE, and STELIB. We find that offsets are negligible in some  
cases, most notably for the widely used indices H$\beta$, Mg$b$,  
Fe5270, and Fe5335. In a number of cases, however, the difference  
between flux calibrated library and Lick/IDS is significant with the  
offsets depending on index strengths. Interestingly, there is no  
general agreement between the three libraries for a large number of  
indices, which hampers the derivation of a universal offset  
between the Lick/IDS and flux calibrated systems.
\end{abstract}


\section{introduction}
\label{intro}

Stellar population models of absorption line indices are a key tool
for the analysis of star cluster and galaxy absorption spectra. They
are used to derive the fundamental stellar population properties such
as age, metallicity and element abundance ratios. In particular,
optical absorption line diagnostics in the spectra of evolved stellar
populations have successfully been adopted in the past in studies on
galaxy evolution
\citep*[e.g.][]{worthey92,davies93,vazdekis97,kuntschner98,worthey98,trager98,henry99,
kuntschner00,trager00,thomas05} and globular cluster formation
\citep[e.g.][]{kpatig98,forbes01,kuntschner02,brodie05,puzia05}.  The
Lick/IDS system
\citep{burstein84,faber85,gorgas93,worthey94,worthey97,trager98} is
the standard set of absorption line indices that has been used
extensively during the last two decades for studying absorption
features of stellar populations. This system consists of index
definitions for 25 prominent absorption features between 4000 and
$6500\;$\AA\ present in the spectra of evolved stellar populations.

For studies of galaxy and star cluster evolution, absorption lines
need to be modelled for stellar populations \citep*[e.g.][]{maraston98,maraston05,bc03,vazdekis99,leitherer99,worthey94,worthey97,rose94,trager00,TMB03,thomas04}. A convenient way goes through the use of
empirical calibrations. This is motivated by the fact that theoretical
model atmospheres are known to suffer from incomplete line lists and
continuum uncertainties. \citep*[e.g.][]{korn05,coelho07,merino05,lee09,walcher09}.
Empirical calibrations on the other hand have the disadvantage to be
hardwired to the chemical abundance pattern of the Milky Way, which
can be overcome in a semi-empirical approach as in the models by
\citet{trager00}, \citet{TMB03,thomas04} and \citet{schiavon07}

Empirical calibrations can be inserted in the models in two ways.
In the first and most widely used approach, absorption line indices enter
stellar population modelling through calibrations of the
empirical relationship between the indices and the stellar atmospheric
parameters T$_{\rm eff}$, $\log g$ and [Fe/H] as provided by stellar
libraries. As these calibrations are usually obtained through
polynomial fitting procedures they are commonly referred to as
'fitting functions'. The quality of the final stellar population model
critically depends on the accuracy with which these relationships can
be inferred from stellar libraries, i.e.\ the coverage of stellar
parameter space and the reliability of the index measurements. The
computational procedure with which the fitting functions are
determined is a further crucial step in producing accurate models. A
number of studies in the literature are devoted to such empirical
calibrations for various stellar libraries, either for the Lick
indices, parts of the Lick indices or other prominent absorption
features
\citep*{buzzoni92,buzzoni94,worthey94,borges95,gorgas99,cenarro02,schiavon07,maraston09}.

Alternatively to the use of fitting functions, absorption line indices
can be measured directly on the synthetic spectral energy distribution (SED) from stellar 
population models that are based on empirical stellar libraries. The benefit
of this method is that the full SED can be compared pixel-by-pixel to observations 
\citep[e.g.][]{panter07,tojeiro07}. 

The major strength of fitting
functions, instead, lies in the fact that they allow for interpolation
between well populated regions of stellar parameter space which
increases the accuracy of the model in stellar parameter space that is
only sparsely sampled by empirical stellar libraries. Moreover, each
absorption index or spectral feature is represented by an individual
fitting function, which is optimised to best reproduce its behaviour
in stellar parameter space. Fitting functions are also easier to
implement in a stellar population synthesis code, and models based on fitting
functions are better comparable.

The widely used fitting functions of \citet{worthey94} and
\citet{worthey97} are based on the Lick/IDS stellar library
\citep{burstein84,faber85}. They are adopted in most stellar
population models
\citep{worthey94A,vazdekis96,trager00,TMB03,thomas04,thomas05,annibali07}
in the literature. Other fitting functions based on the same stellar
library exist \citep{buzzoni92,buzzoni94,borges95} and lead to overall
consistent results in the final stellar population model
\citep{maraston03}. Major progress has been made with the advent of a
new generation of stellar libraries
\citep{jones99,prugniel01,leBorgne03,miles} that have led to
considerable improvements regarding coverage of stellar parameter
space, spectral resolution, signal-to-noise ratio, and flux
calibration.

In particular the latter is a critical step forward. As the Lick/IDS system is not flux calibrated, observations have to be re-calibrated onto the Lick/IDS system through comparison with Lick standard stars. This requirement hampers the analysis of data samples for which spectra of such calibration stars are either not available at sufficient quality or do not cover the appropriate rest-frame wavelength range. This problem is most imminent in high redshift observations and in galaxy redshift surveys such as the Sloan Digital Sky Survey \citep{york00}. The new flux calibrated libraries allow the analysis of flux calibrated spectra at any redshift without spectroscopic standard stars.

Flux-calibrated stellar libraries in the literature that are suitable for stellar population modelling include the \emph{Jones} \citep{jones99}, 
\emph{ELODIE} \citep{prugniel01}, \emph{STELIB} \citep{leBorgne03} and \emph{MILES} \citep{miles} libraries. 
The \emph{MILES} library is particularly well suited for stellar population modelling of absorption line indices owing to its favourable combination of spectral resolution, wavelength range, stellar parameter coverage, and quality of flux calibration. In this paper we present new Lick index fitting functions based on the \emph{MILES} stellar library. To take advantage of the full spectral resolution of the \emph{MILES} library we have produced fitting functions for both the lower Lick/IDS resolution ($8-11\;$\AA\ FWHM) and the higher resolution of the \emph{MILES} library (2.3 \AA\ FWHM). A new version of the TMB stellar population model of absorption line indices based on these new fitting functions will be presented in a subsequent paper.

This paper is organized as follows. In Section~\ref{stlib} we present the Lick indices measured
on the \emph{MILES} library and a quality evaluation of the index measurements. We discuss offsets between the flux calibrated MILES and the Lick/IDS systems.
The empirical fitting method is presented in Section~\ref{fitf} along with the resulting fitting functions.
In Section~\ref{comps} we compare the fitting functions of this work with 
fitting functions from the literature. We summarise in Section~\ref{concs}.

\begin{table*}
\centering
\caption{Typical Lick index errors and offsets to the Lick/IDS library. M-$\sigma$ and L-$\sigma$ corresponds to index errors at the resolution of the \emph{MILES} and Lick/IDS libraries, respectively. T98-$\sigma$ are the index errors presented in \citet{trager98} for the Lick/IDS library. $I_{lib}$ are indices measured on the libraries (\emph{MILES}, \emph{ELODIE} and \emph{STELIB}) for which offsets to the Lick/IDS library are presented. $I_{L}$ are indices measured on the Lick/IDS library.}
\label{offtable}
\begin{tabular}{rlllllllr@{0}llr@{0}llr@{0}l}
\hline
\multicolumn{2}{|c|}{\bf INDEX} & & \multicolumn{3}{|c|}{\bf Error} & & \multicolumn{9}{|c|}{\bf Offset $I_{lib}=a\cdot I_{Lick}+b$}\\
  &      & &            &            &              & & \multicolumn{3}{|c|}{MILES} & \multicolumn{3}{|c|}{ELODIE} & \multicolumn{3}{|c|}{STELIB} \\
i & NAME & & \multicolumn{1}{|c|}{M-$\sigma$} & \multicolumn{1}{|c|}{L-$\sigma$} & \multicolumn{1}{|c|}{T98-$\sigma$} & & \multicolumn{1}{|c|}{$a$} & \multicolumn{2}{|c|}{$b$} & \multicolumn{1}{|c|}{$a$} & \multicolumn{2}{|c|}{$b$} & \multicolumn{1}{|c|}{$a$} & \multicolumn{2}{|c|}{$b$} \\
\hline
1 & H$\delta_{A}$  &  & 0.164  & 0.125  & 0.64  & & 0.960 & -&.054   & 0.955 &  &.721   & 0.940 &  &.823 \\
2 & H$\delta_{F}$  &  & 0.093  & 0.075  & 0.40  & & 0.965 &  &.049   & 0.936 &  &.397   & 0.956 &  &.242 \\
3 & CN$_{1}$       &  & 0.0042 & 0.0038 & 0.018 & & 0.912 &  &.008   & 0.897 & -&.012   & 0.986 & -&.010 \\
4 & CN$_{2}$       &  & 0.0050 & 0.0042 & 0.019 & & 0.907 &  &.006   & 0.900 & -&.008   & 0.985 & -&.013 \\
5 & Ca4227         &  & 0.063  & 0.047  & 0.25  & & 0.904 &  &.074   & 0.771 &  &.163   & 0.918 & -&.057 \\
6 & G4300          &  & 0.112  & 0.093  & 0.33  & & 0.858 &  &.625   & 0.870 &  &.646   & 0.924 &  &.565 \\
7 & H$\gamma_{A}$  &  & 0.142  & 0.107  & 0.48  & & 0.976 & -&.148   & 0.967 & -&.057   & 1.022 & -&.735 \\
8 & H$\gamma_{F}$  &  & 0.069  & 0.059  & 0.33  & & 0.963 & -&.038   & 0.962 &  &.016   & 0.999 & -&.238 \\
9 & Fe4383         &  & 0.155  & 0.127  & 0.46  & & 0.932 & -&.220   & 0.929 & -&.184   & 0.915 &  &.796 \\
10 & Ca4455        &  & 0.073  & 0.056  & 0.22  & & 0.747 & -&.067   & 0.785 & -&.105   & 0.891 & -&.228 \\
11 & Fe4531        &  & 0.122  & 0.096  & 0.37  & & 0.857 &  &.290   & 0.838 &  &.390   & 0.877 & -&.002 \\
12 & C$_{2}$4668   &  & 0.179  & 0.156  & 0.57  & & 0.903 &  &.484   & 0.913 &  &.295   & 0.992 &  &.512 \\
13 & H$\beta$      &  & 0.063  & 0.051  & 0.19  & & 0.981 &  &.126   & 0.996 &  &.015   & 1.004 &  &.032 \\
14 & Fe5015        &  & 0.139  & 0.115  & 0.41  & & 0.902 &  &.084   & 0.926 &  &.178   & 0.989 &  &.168 \\
15 & Mg$_{1}$      &  & 0.0017 & 0.0013 & 0.006 & & 0.911 &  &.0004  & 0.923 &  &.005   & 0.903 & -&.009 \\
16 & Mg$_{2}$      &  & 0.0023 & 0.0014 & 0.007 & & 0.918 & -&.003   & 0.940 &  &.0006  & 0.960 & -&.013 \\
17 & Mg$b$         &  & 0.053  & 0.045  & 0.20  & & 0.964 &  &.108   & 0.935 &  &.247   & 1.003 & -&.026 \\
18 & Fe5270        &  & 0.058  & 0.047  & 0.24  & & 0.923 &  &.101   & 0.919 &  &.180   & 0.932 &  &.173 \\
19 & Fe5335        &  & 0.063  & 0.044  & 0.22  & & 0.960 &  &.135   & 0.963 &  &.032   & 0.946 &  &.110 \\
20 & Fe5406        &  & 0.044  & 0.031  & 0.18  & & 0.874 &  &.269   & 0.913 &  &.165   & 0.853 &  &.264 \\
21 & Fe5709        &  & 0.060  & 0.050  & 0.16  & & 0.979 & -&.026   & 0.907 &  &.015   & 1.019 & -&.046 \\
22 & Fe5782        &  & 0.057  & 0.043  & 0.19  & & 0.920 &  &.037   & 0.879 & -&.004   & 0.906 &  &.088 \\
23 & Na D          &  & 0.082  & 0.064  & 0.21  & & 0.990 & -&.162   & 0.979 & -&.069   & 0.993 & -&.071 \\
24 & TiO$_{1}$     &  & 0.0021 & 0.0017 & 0.006 & & 0.918 & -&.005   & 0.895 & -&.006   & 0.918 &  &.0003 \\
25 & TiO$_{2}$     &  & 0.0022 & 0.0016 & 0.006 & & 0.904 &  &.0007  & 0.912 &  &.005   & 0.940 &  &.009 \\
\hline
\end{tabular} 
\end{table*}

\begin{figure*}
\begin{minipage}{17cm}
\centering
\includegraphics[scale=0.2]{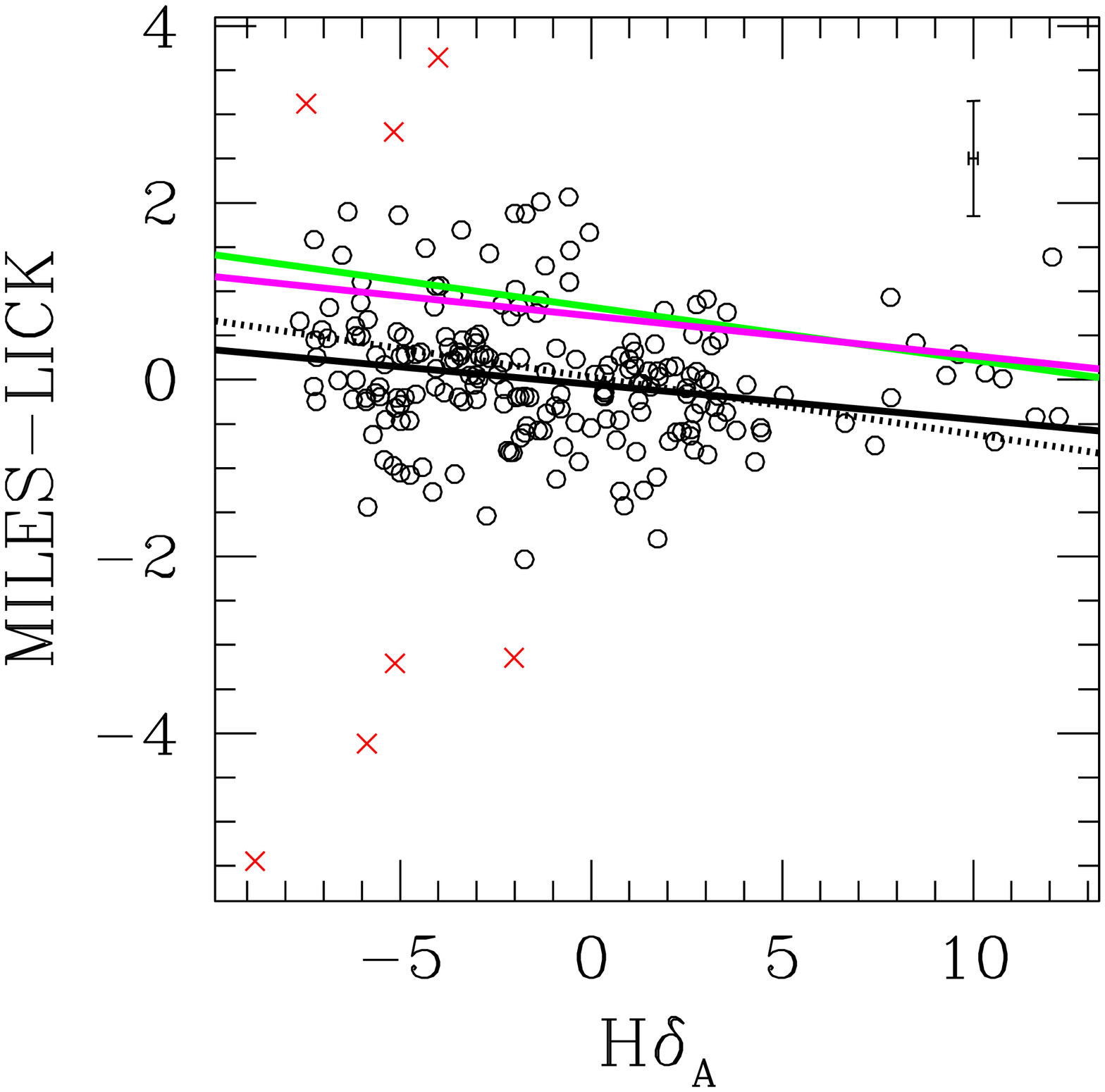}\includegraphics[scale=0.2]{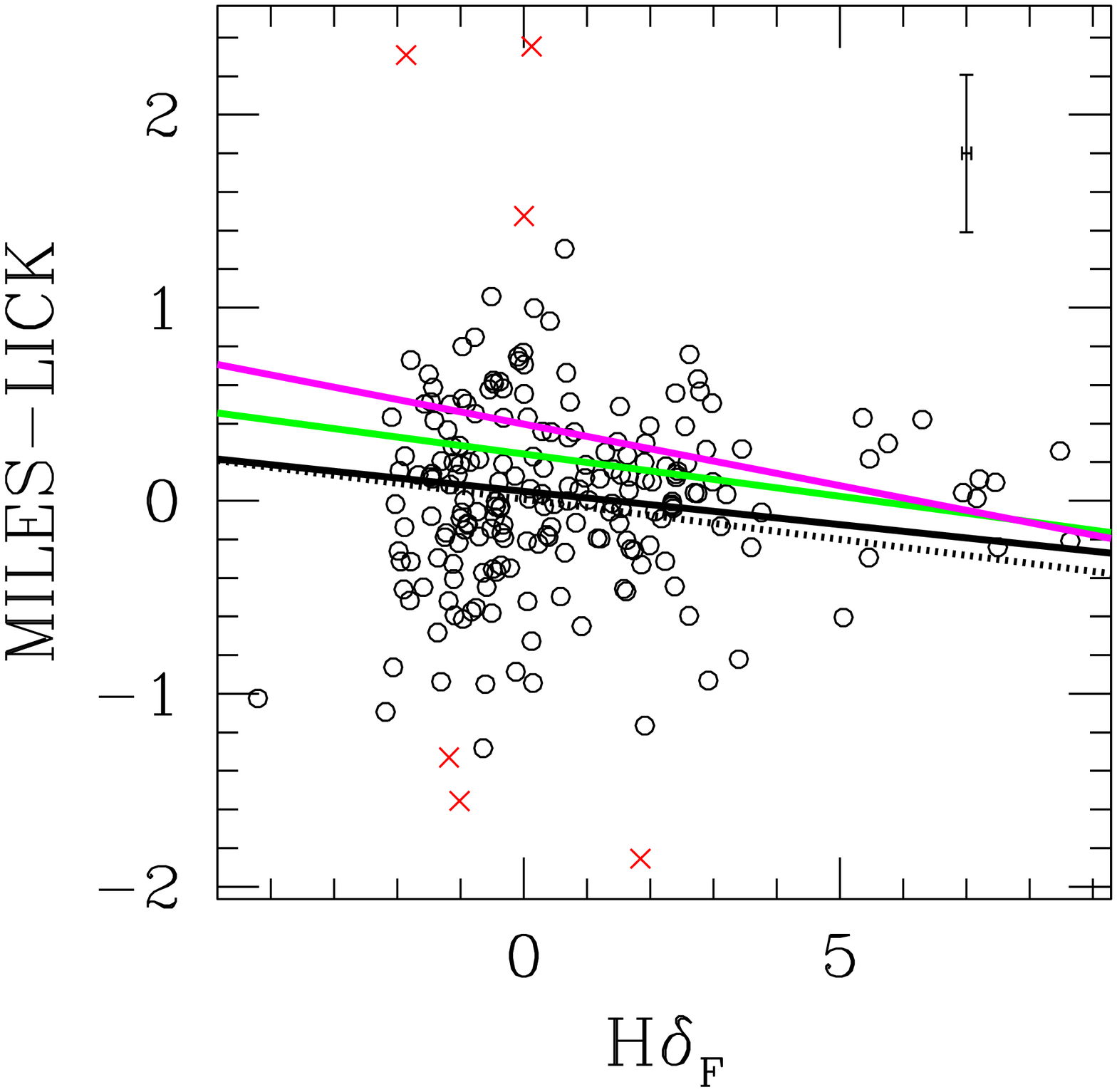}\includegraphics[scale=0.2]{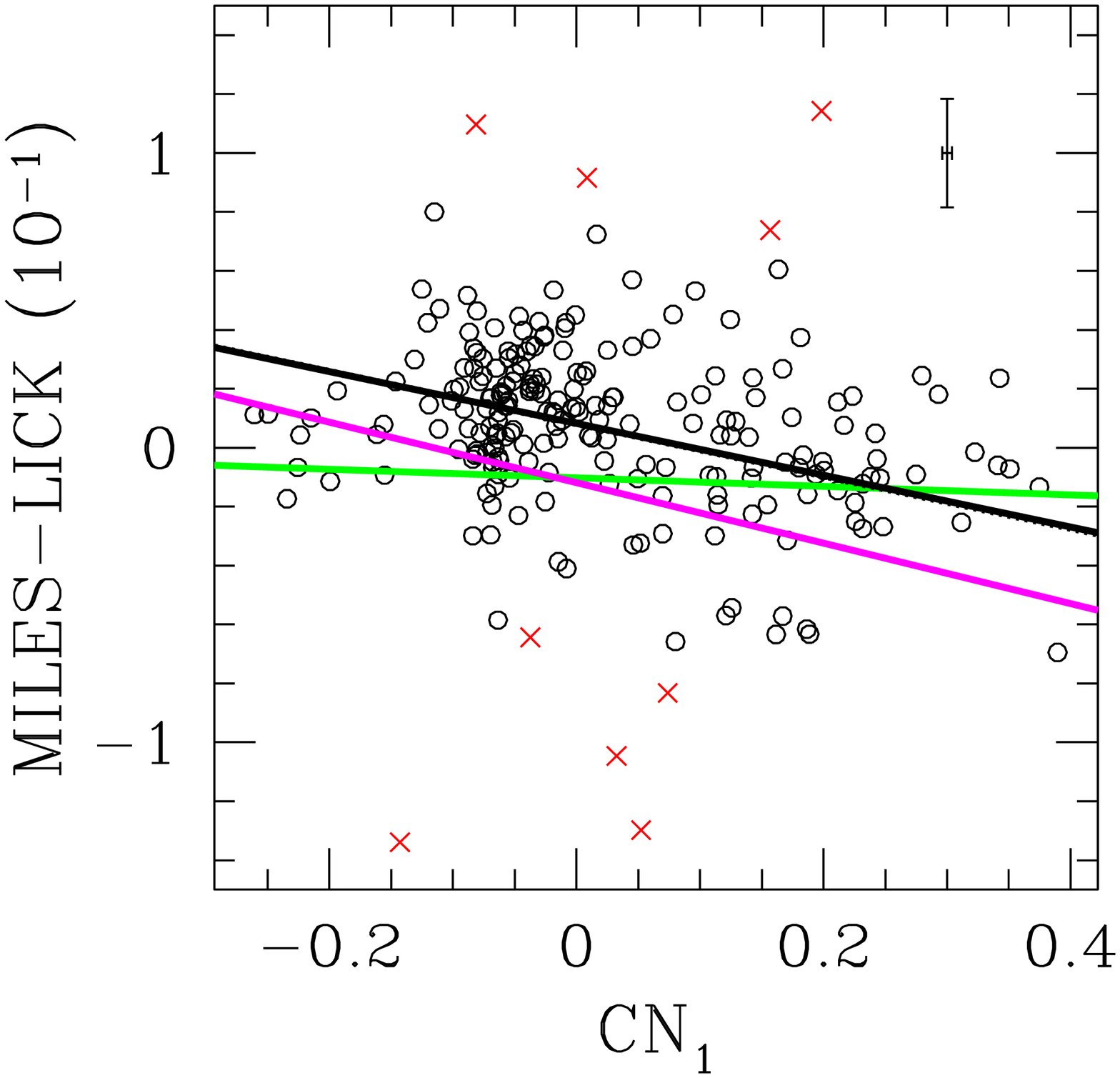}\includegraphics[scale=0.2]{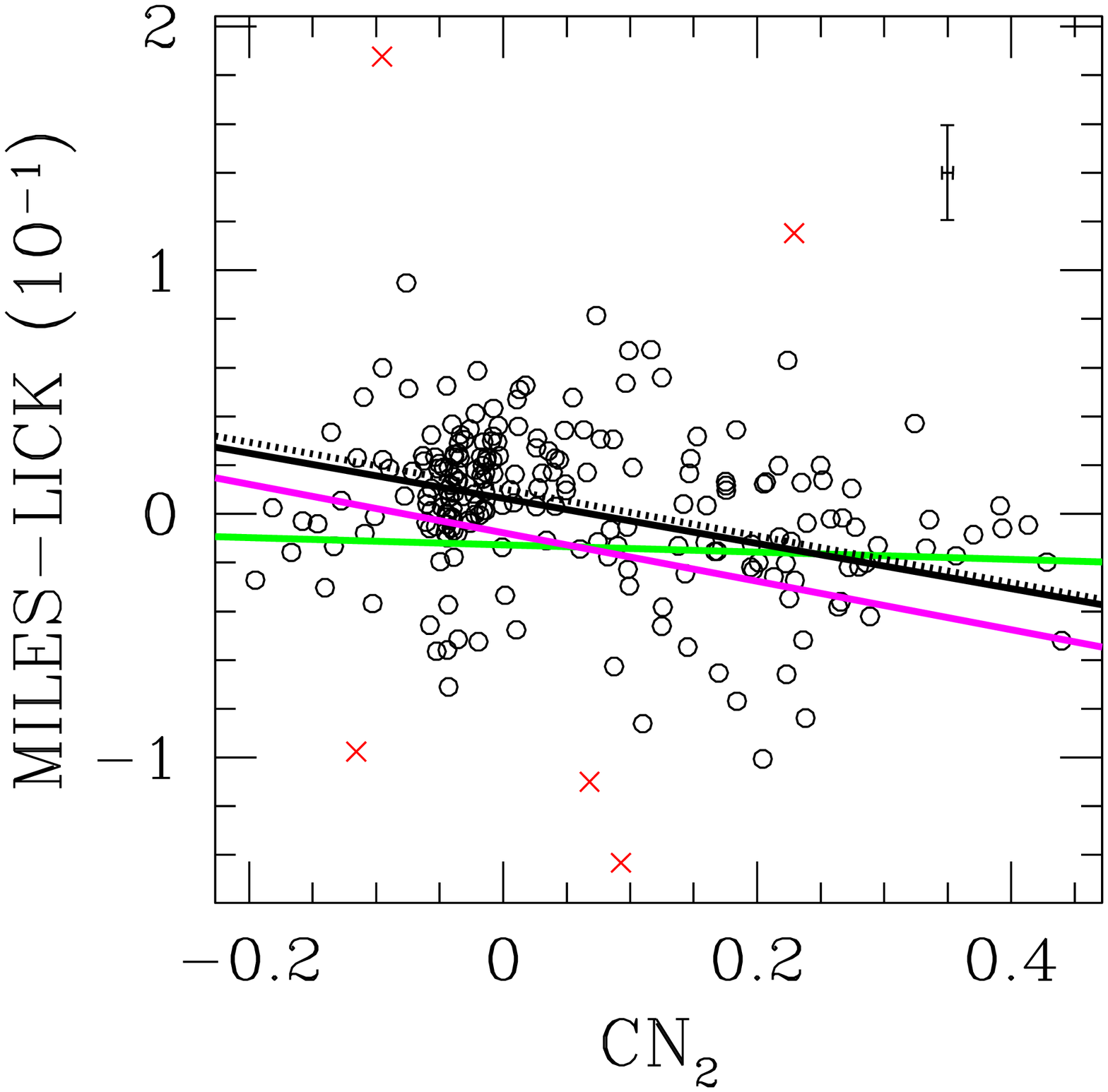}\\
\includegraphics[scale=0.2]{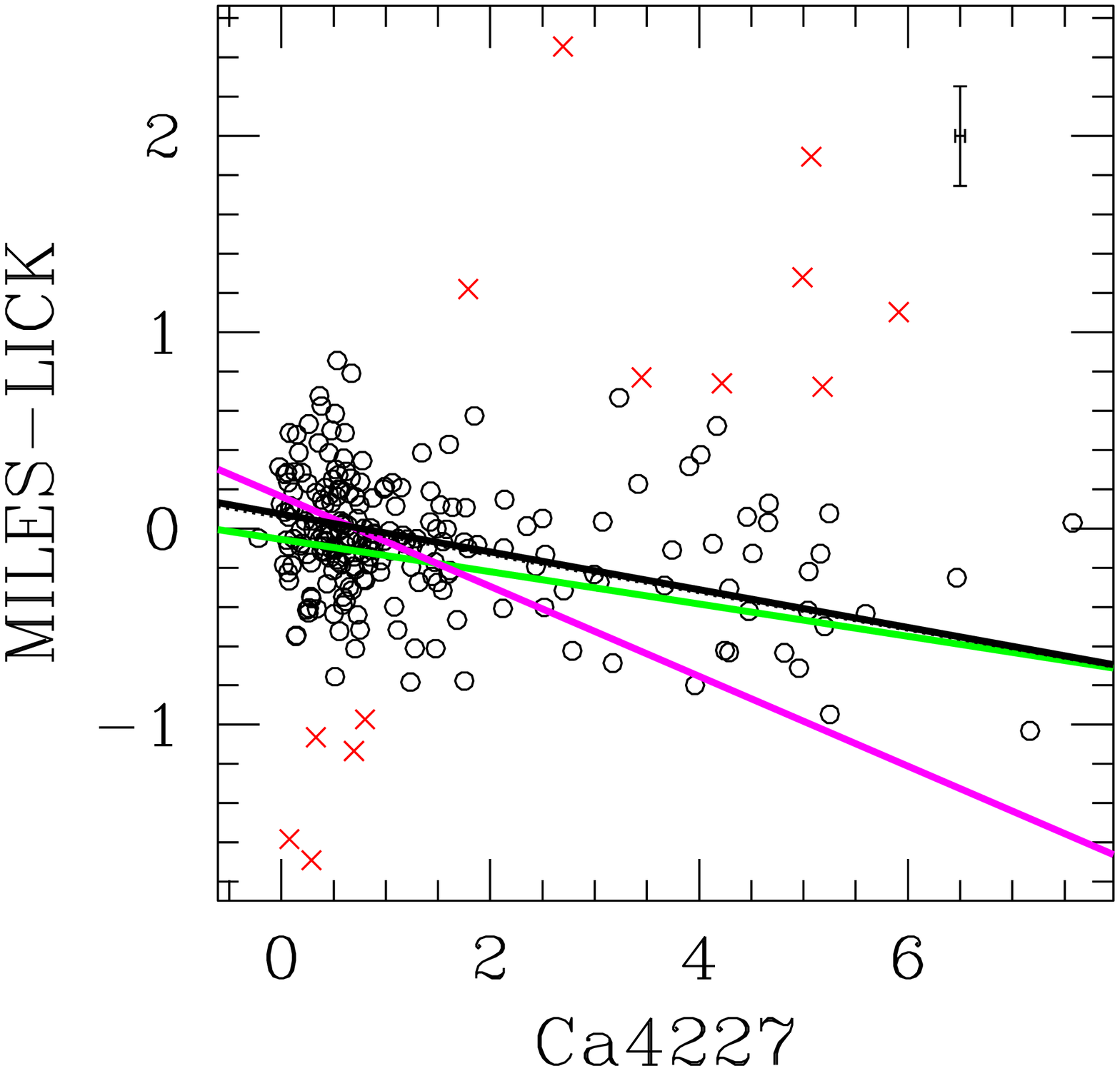}\includegraphics[scale=0.2]{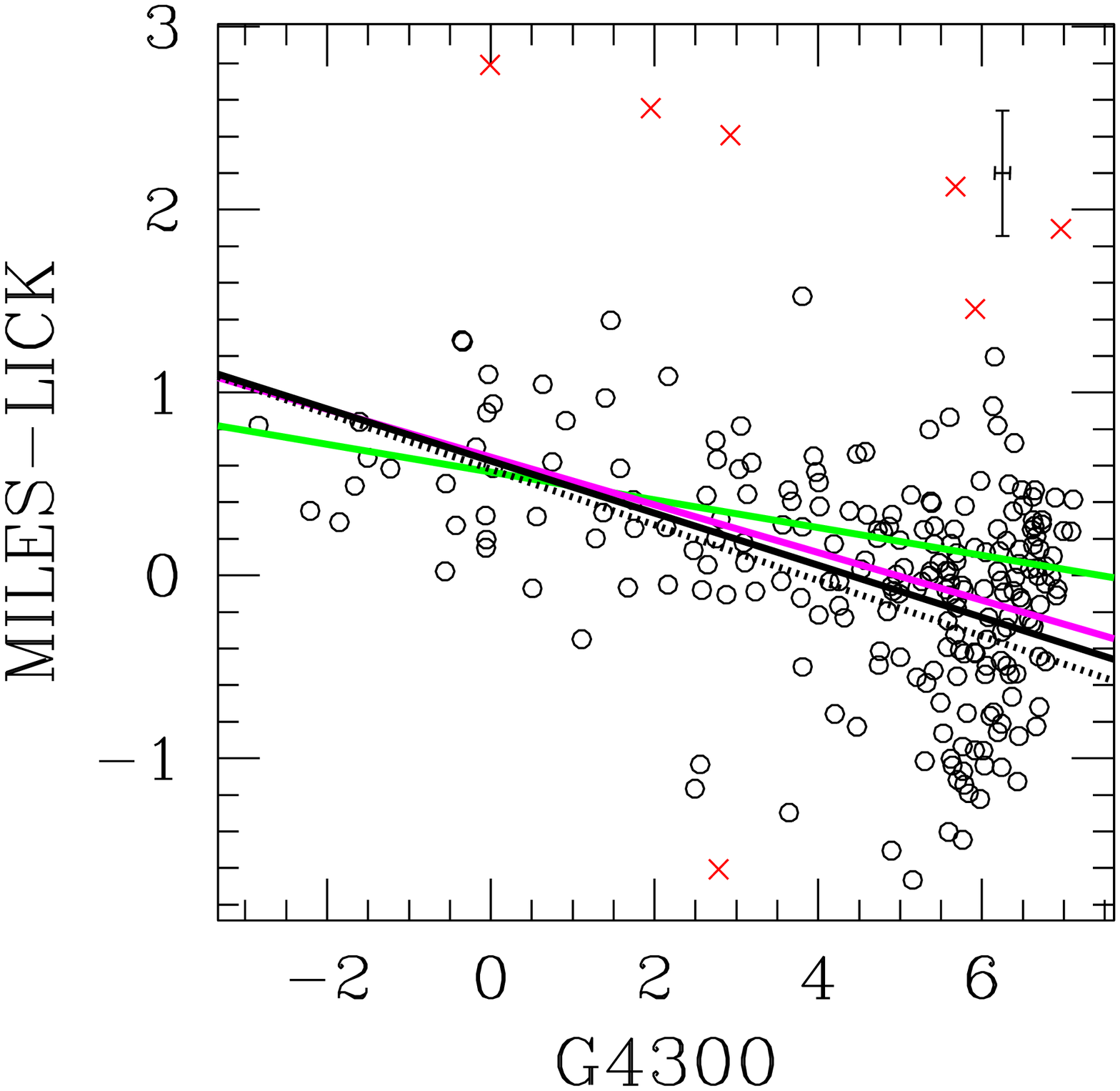}\includegraphics[scale=0.2]{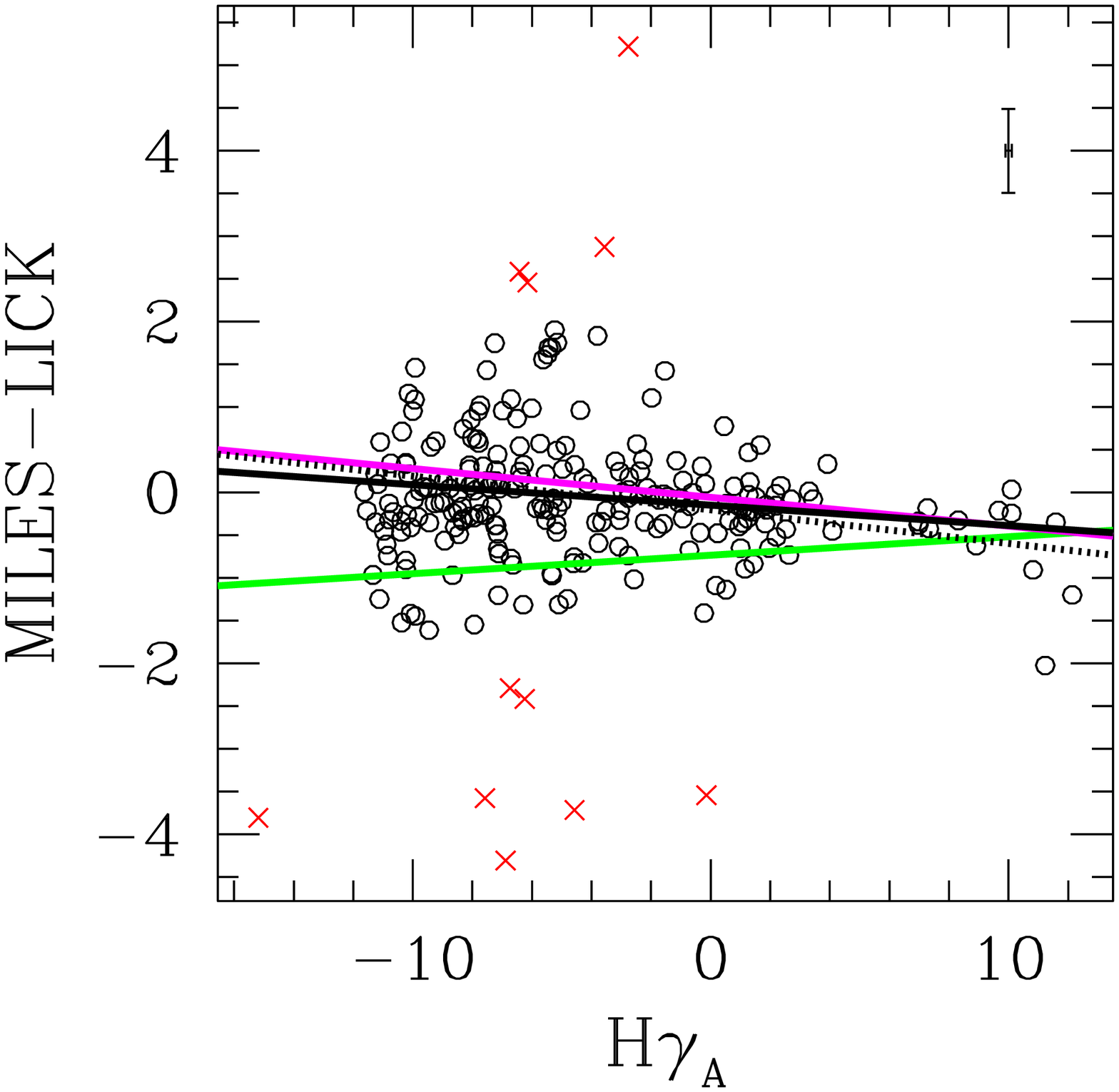}\includegraphics[scale=0.2]{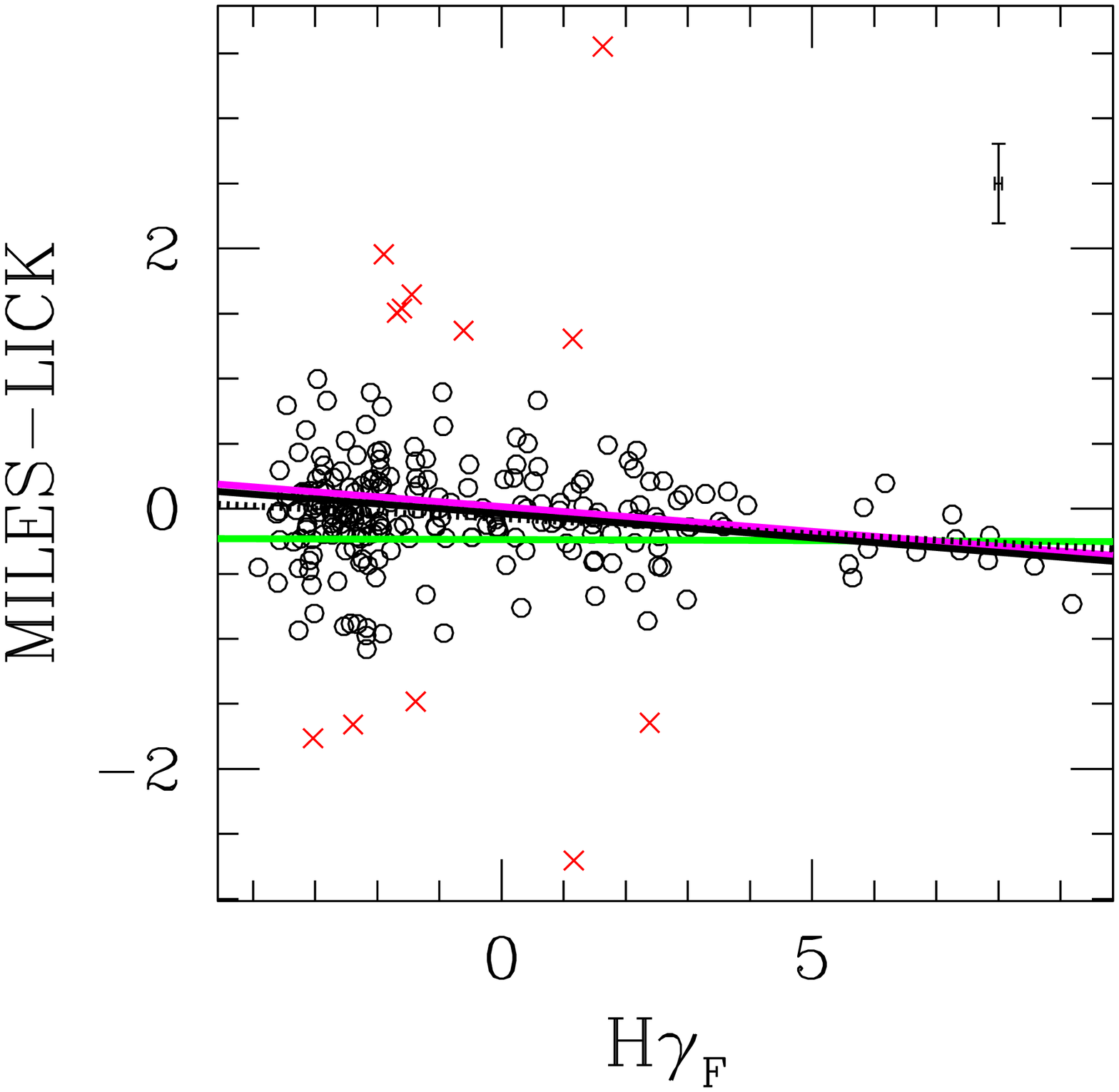}\\
\includegraphics[scale=0.2]{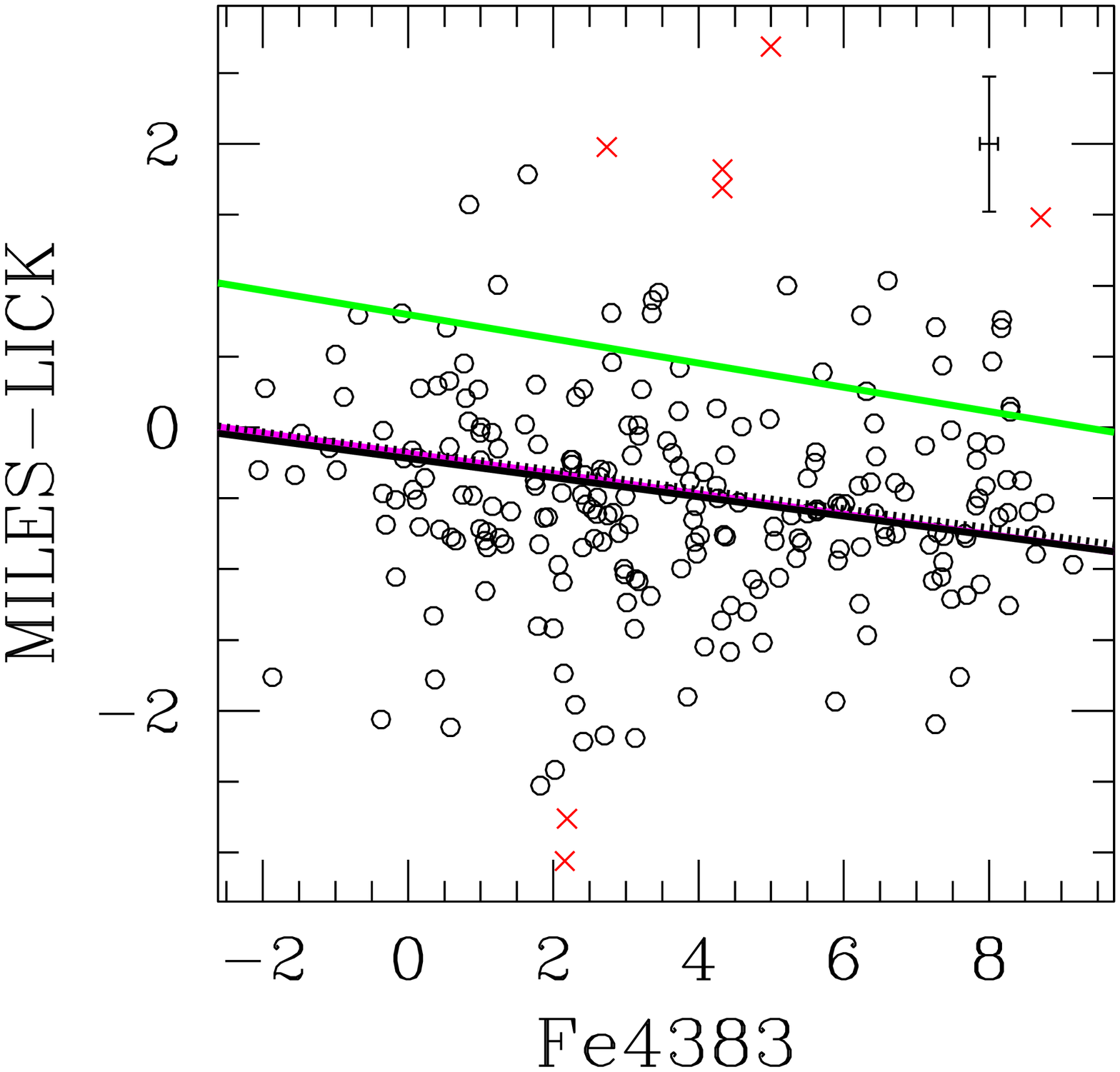}\includegraphics[scale=0.2]{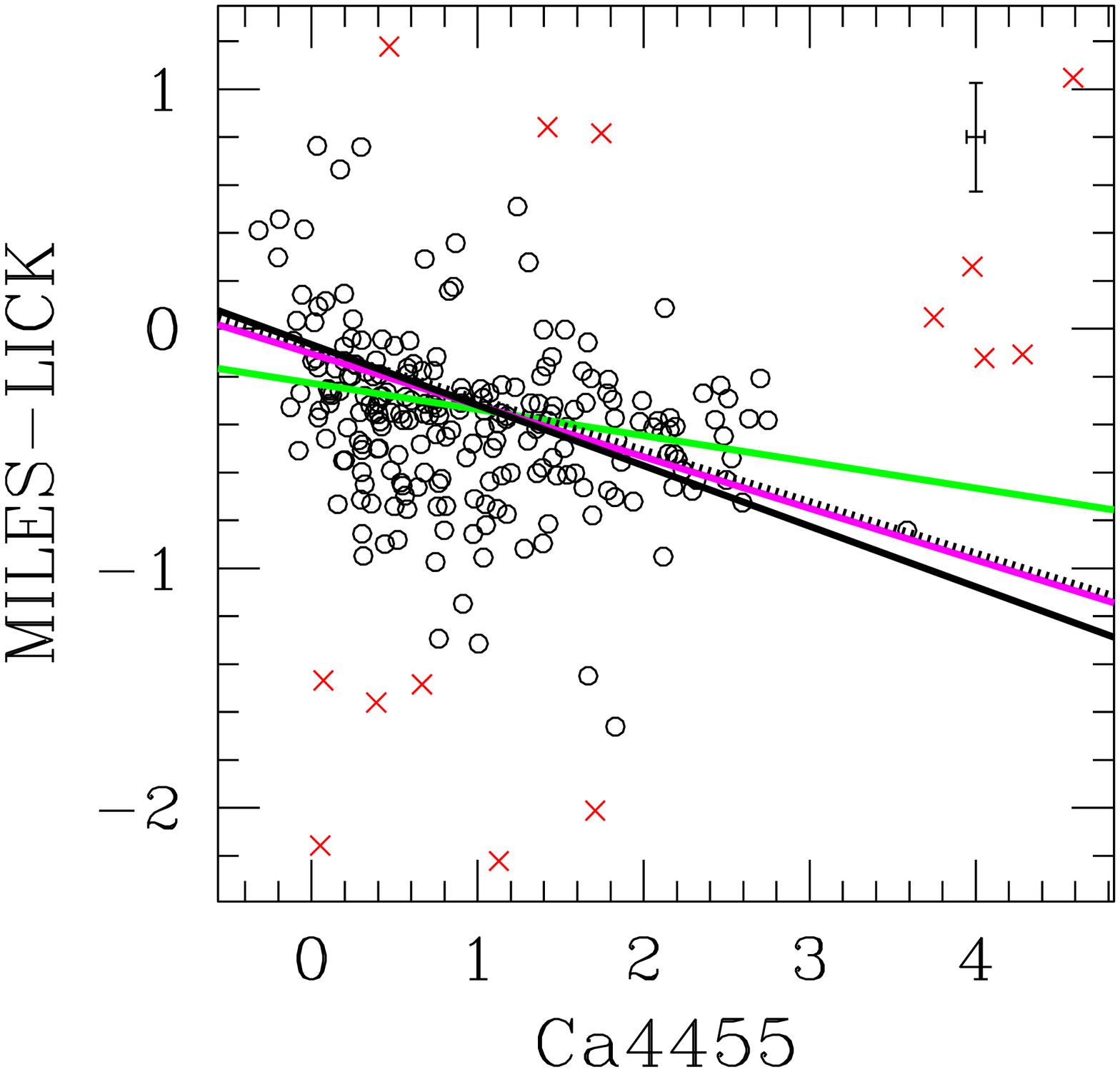}\includegraphics[scale=0.2]{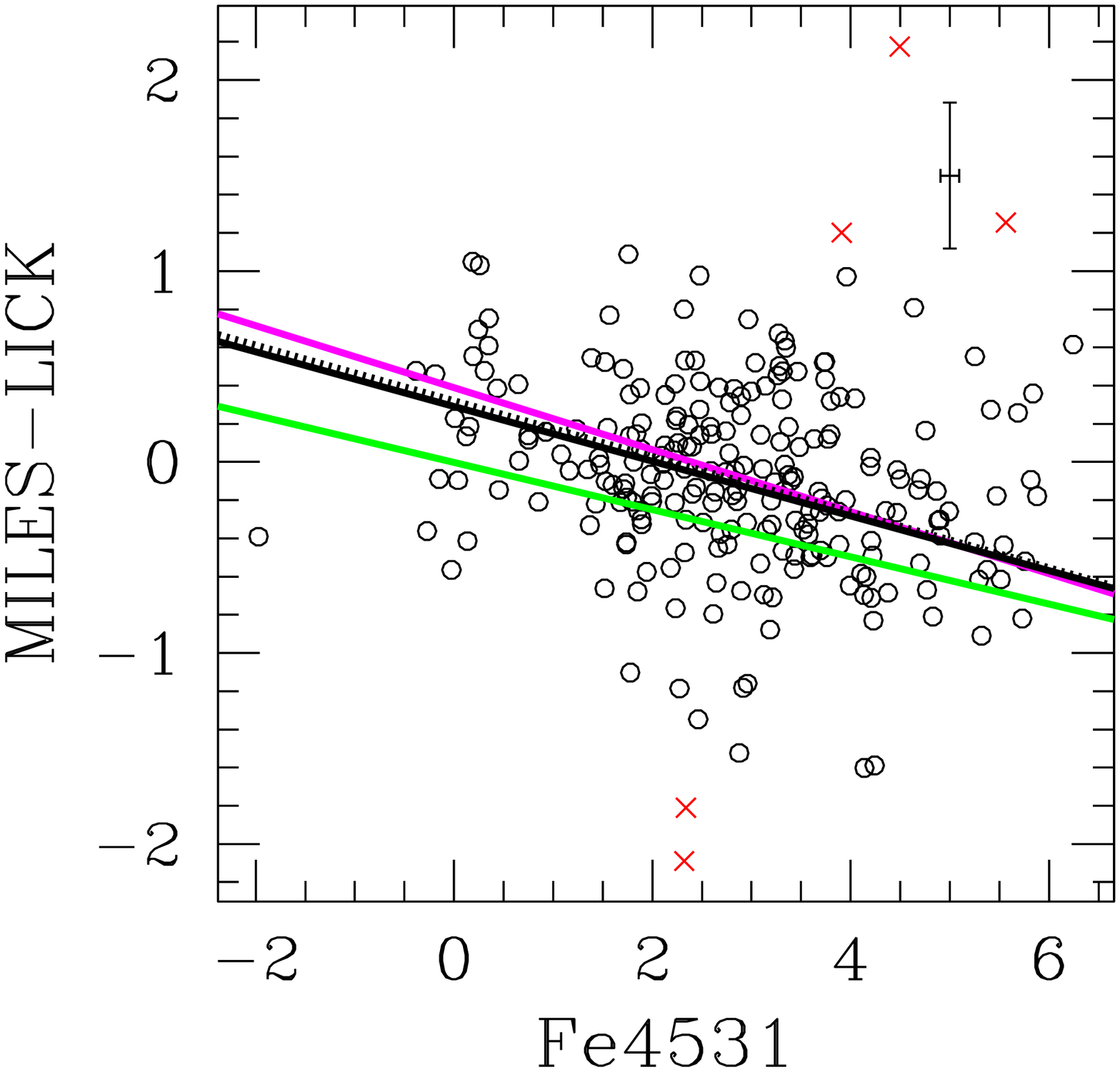}\includegraphics[scale=0.2]{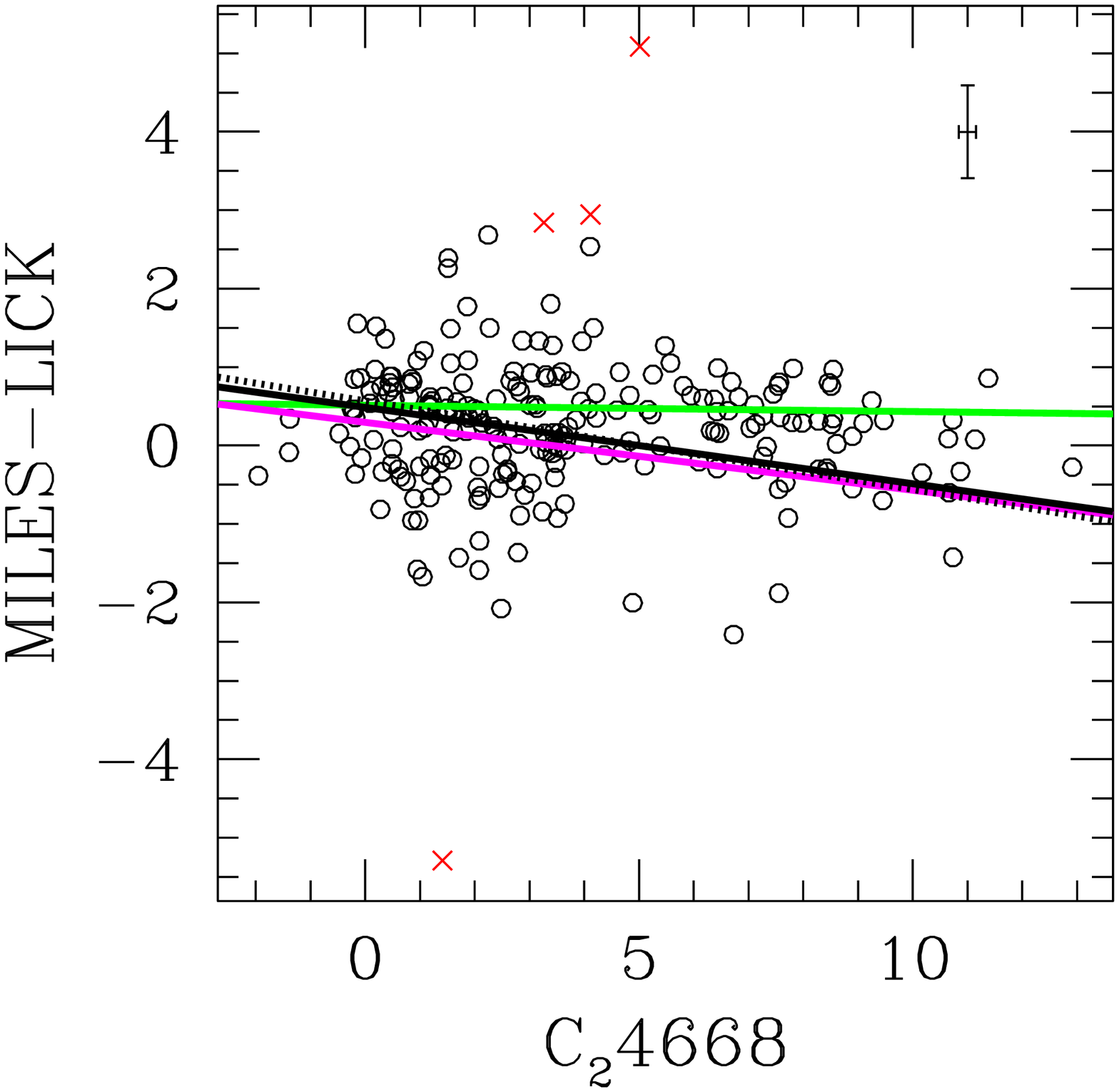}\\
\caption{Index by index comparison between index strengths measured on the MILES library and the Lick/IDS library. Each panel shows the residual as a function of index strength. Included are also least-square fits of the residuals (black lines, coefficients presented in Table~\ref{offtable}) that show clear index strength dependent offsets between the two libraries. Red crosses are sigma clipped data points in the least-square fitting routine. Typical index errors are indicated in the right top corners (see text for more details). Included are also offsets to the Lick/IDS library derived in this work for the \emph{ELODIE} (magenta lines) and \emph{STELIB} (green lines) libraries, as well as the offsets derived in \citet{sanchez09} for the \emph{MILES} library (dotted lines).}
\label{offig}
\end{minipage}
\end{figure*}

\begin{figure*}
\begin{minipage}{17cm}
\centering
\includegraphics[scale=0.2]{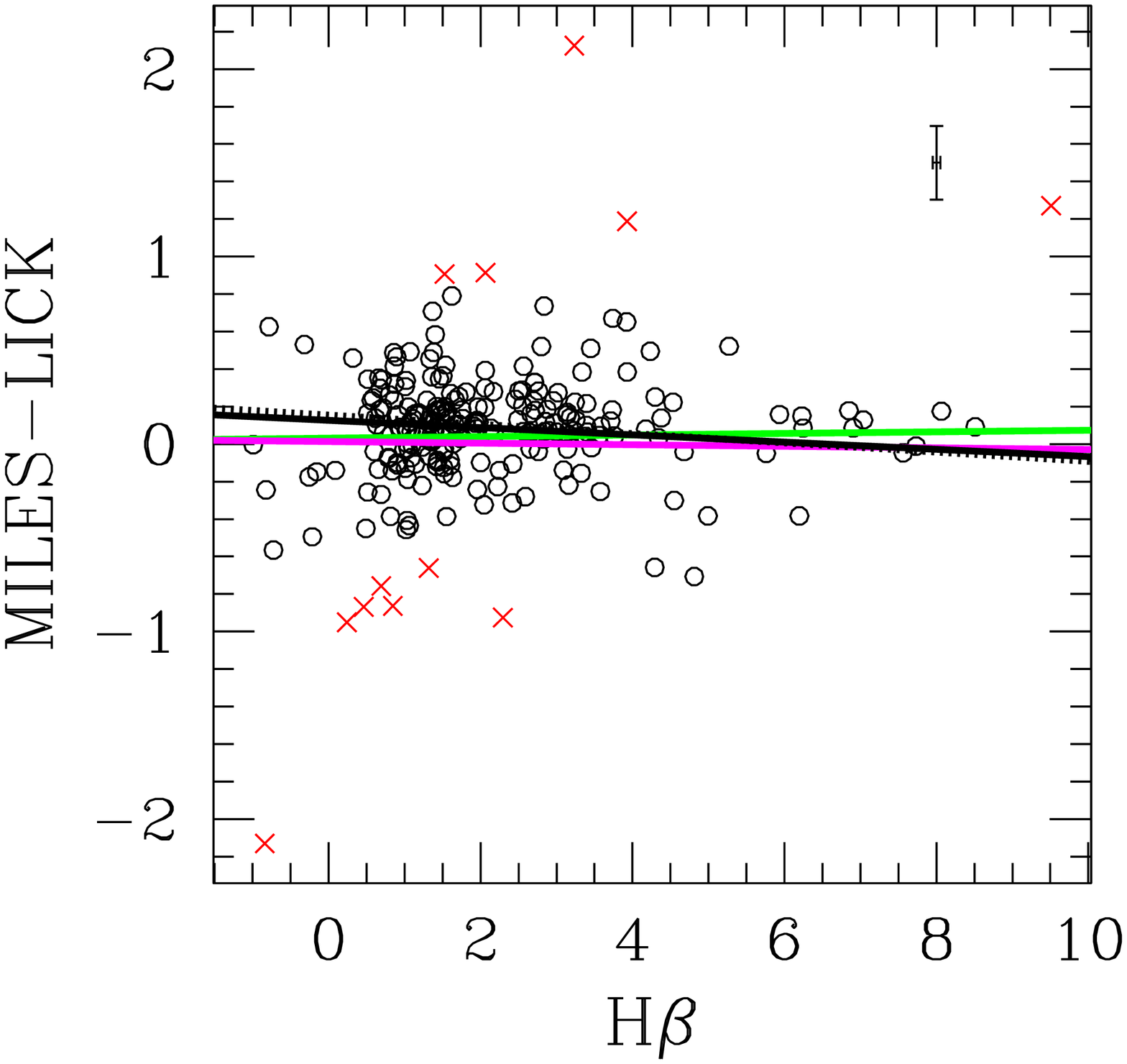}\includegraphics[scale=0.2]{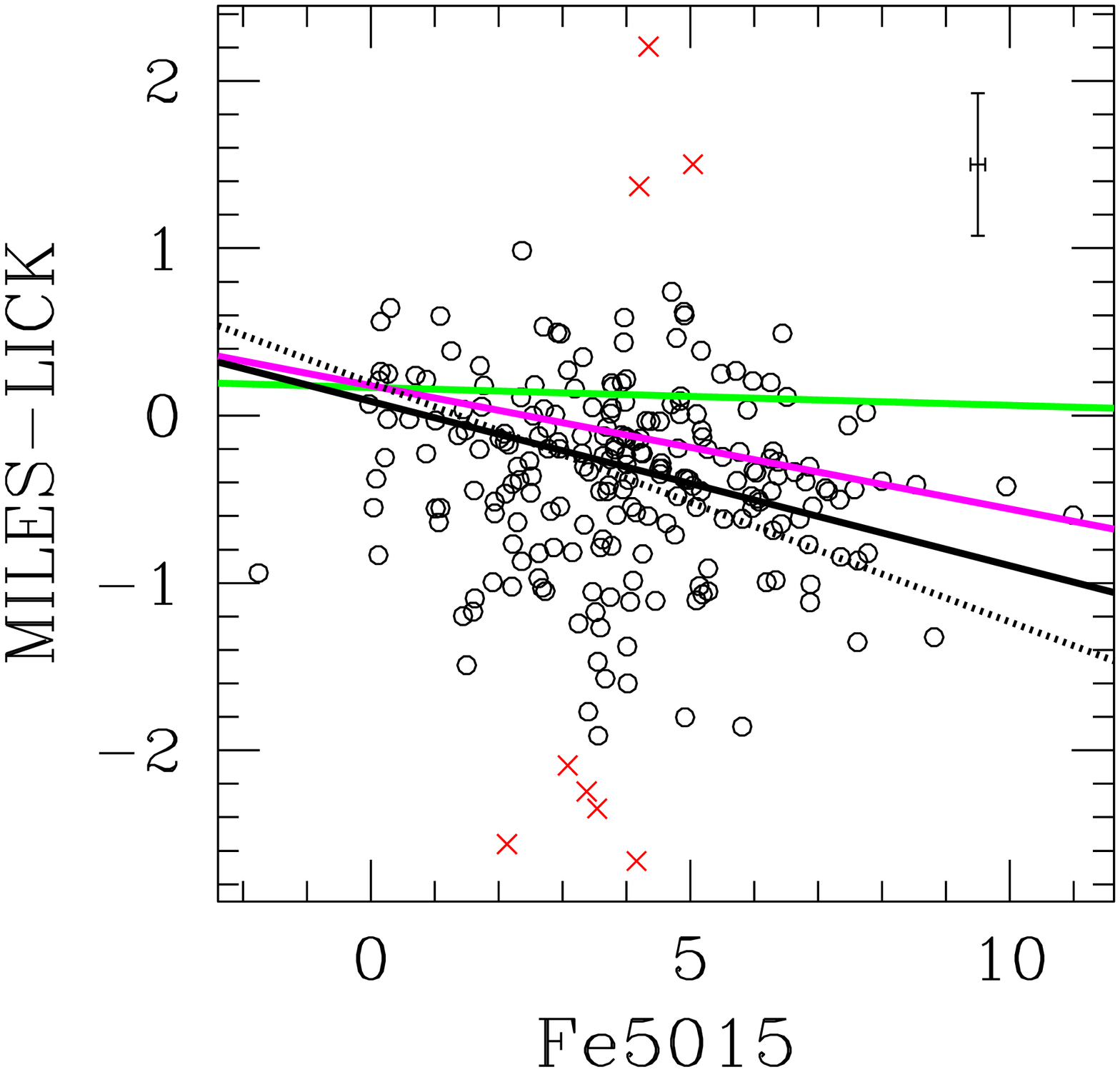}\includegraphics[scale=0.2]{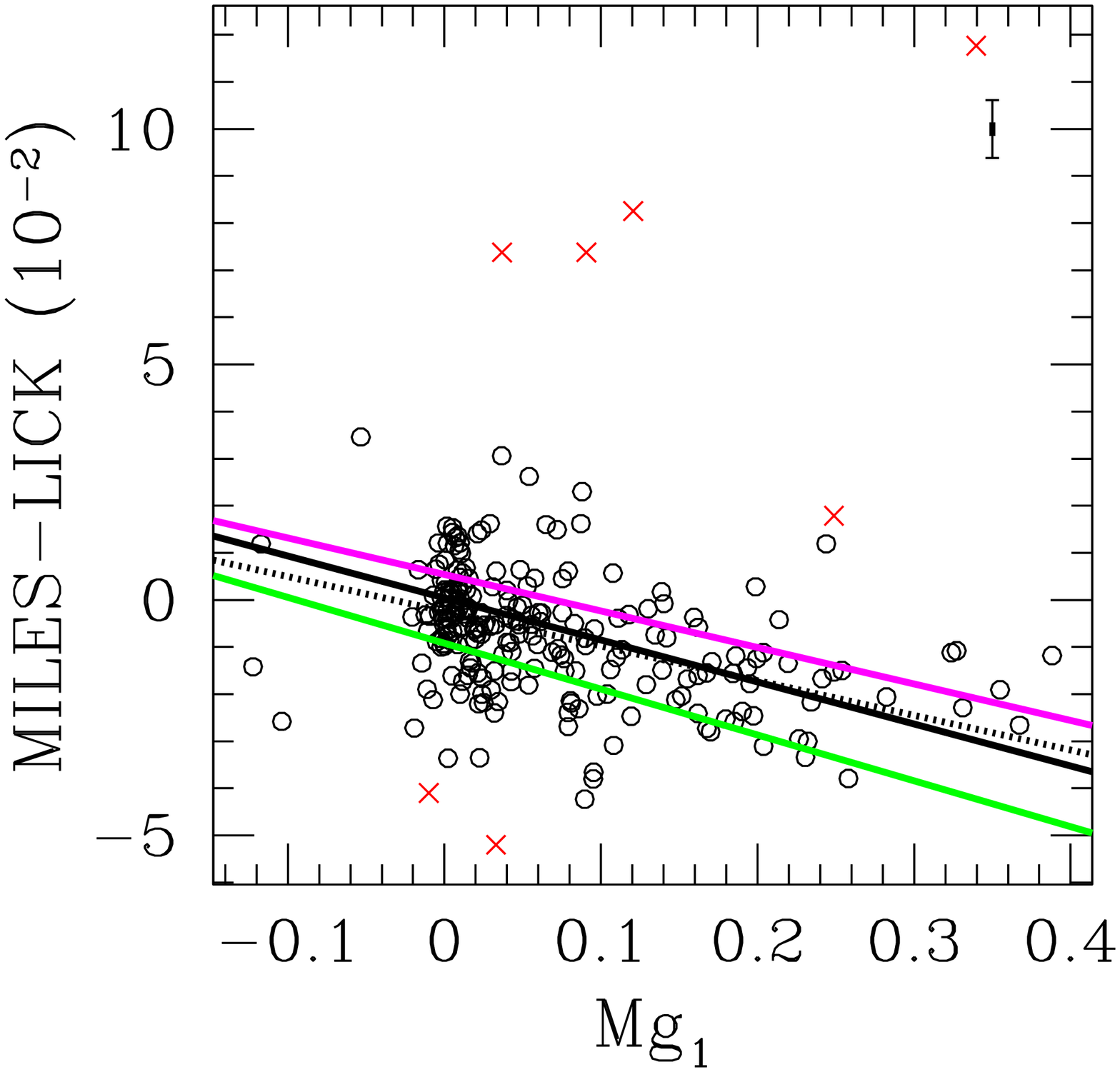}\includegraphics[scale=0.2]{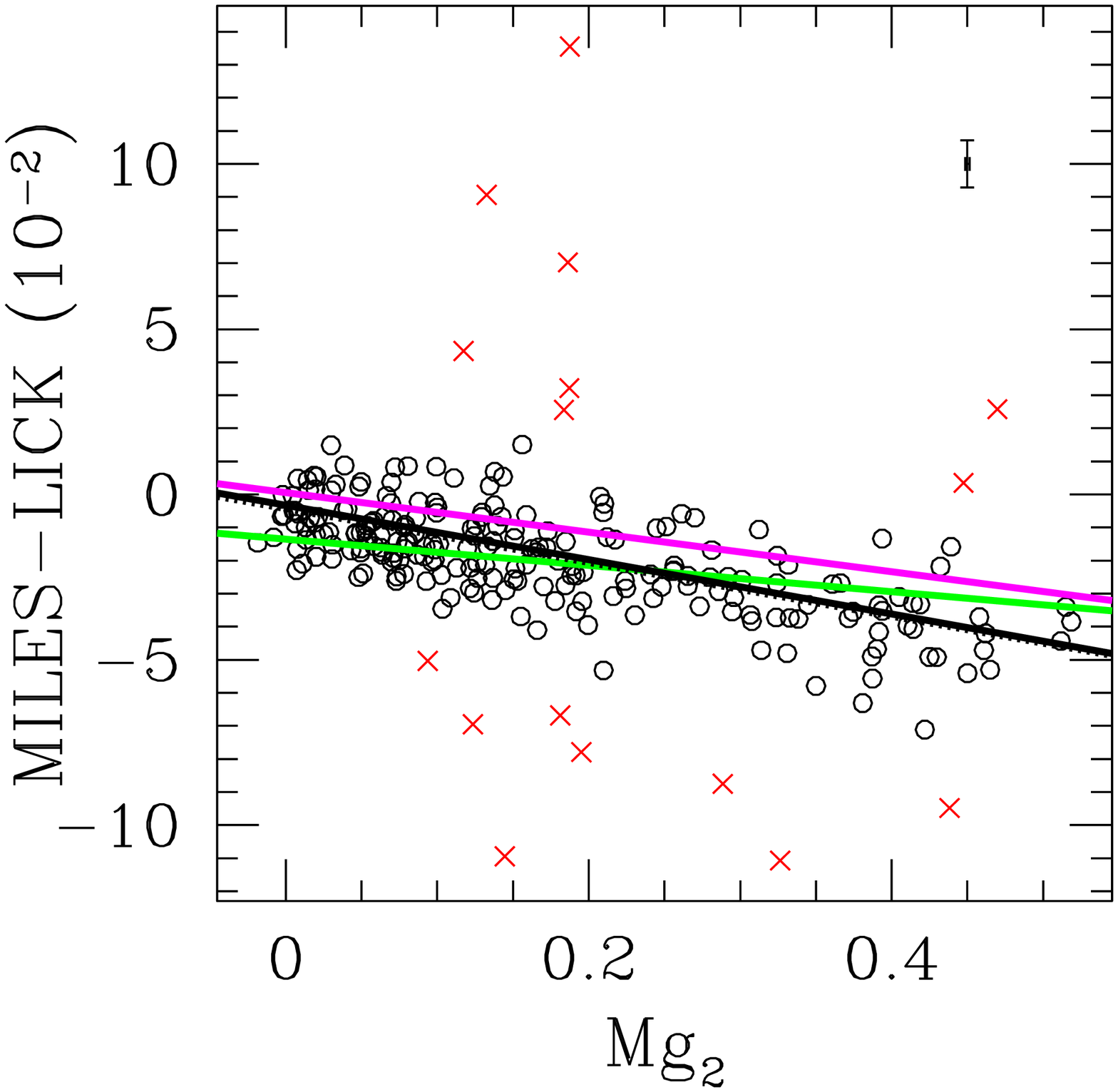}\\
\includegraphics[scale=0.2]{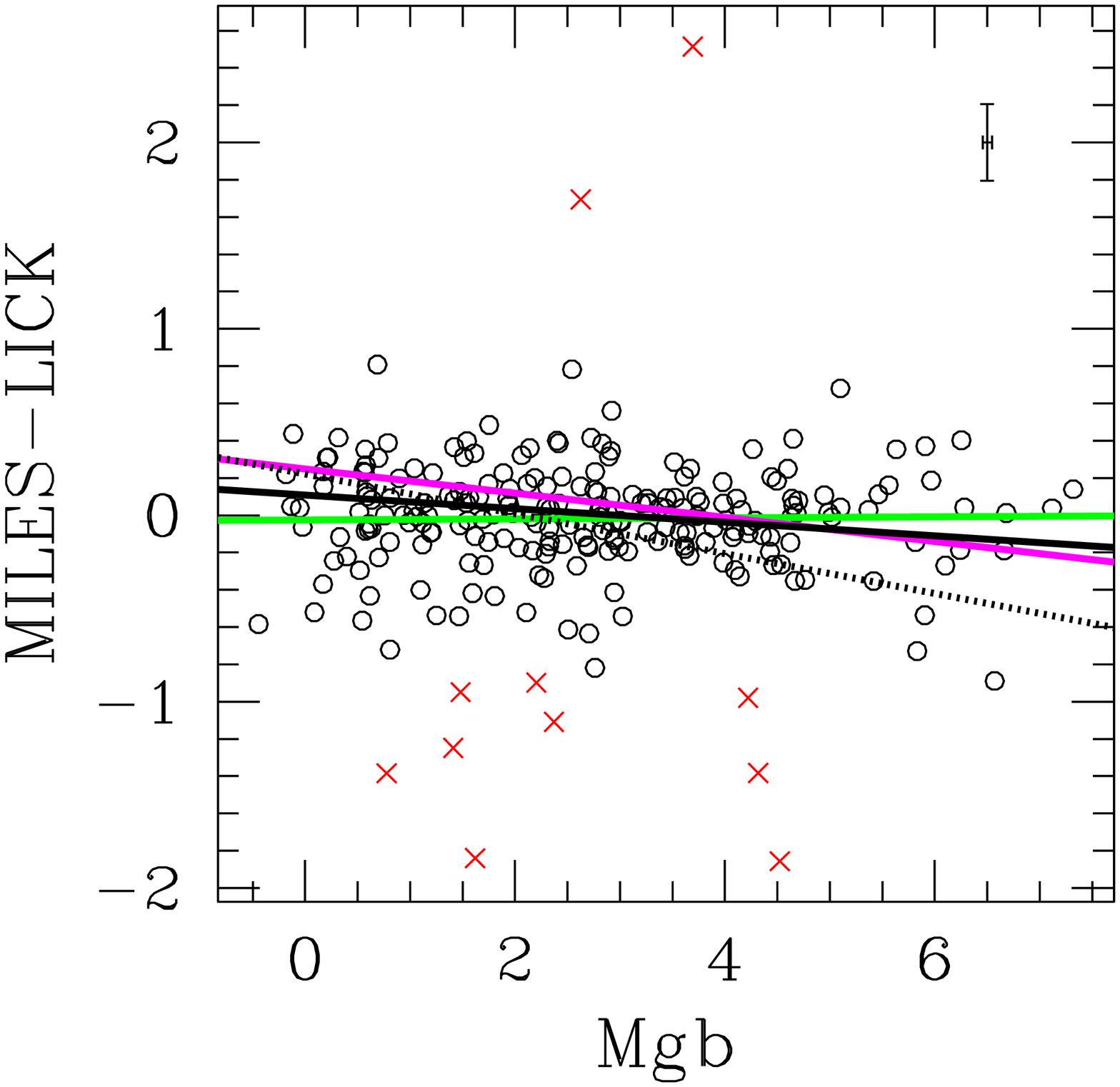}\includegraphics[scale=0.2]{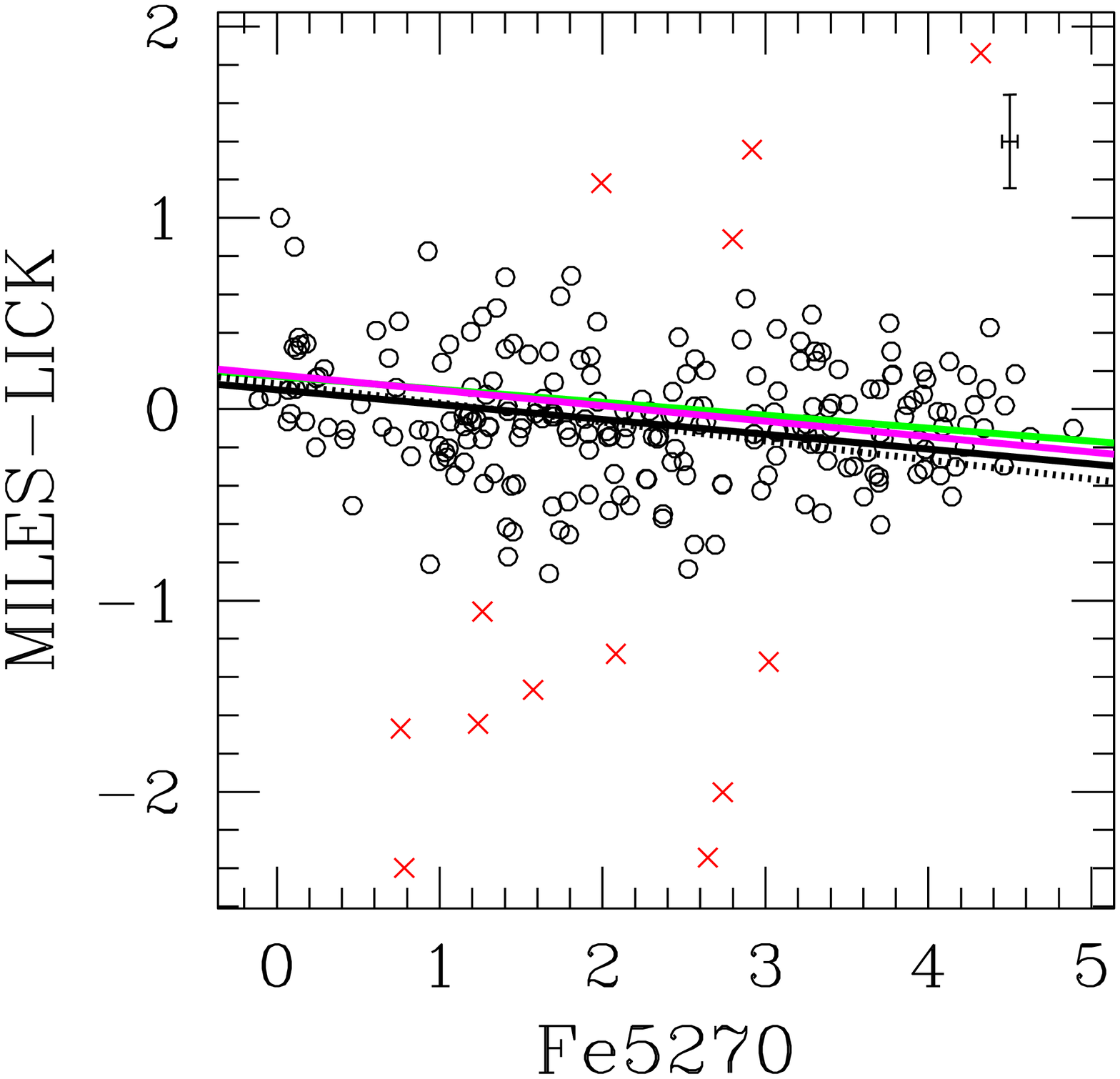}\includegraphics[scale=0.2]{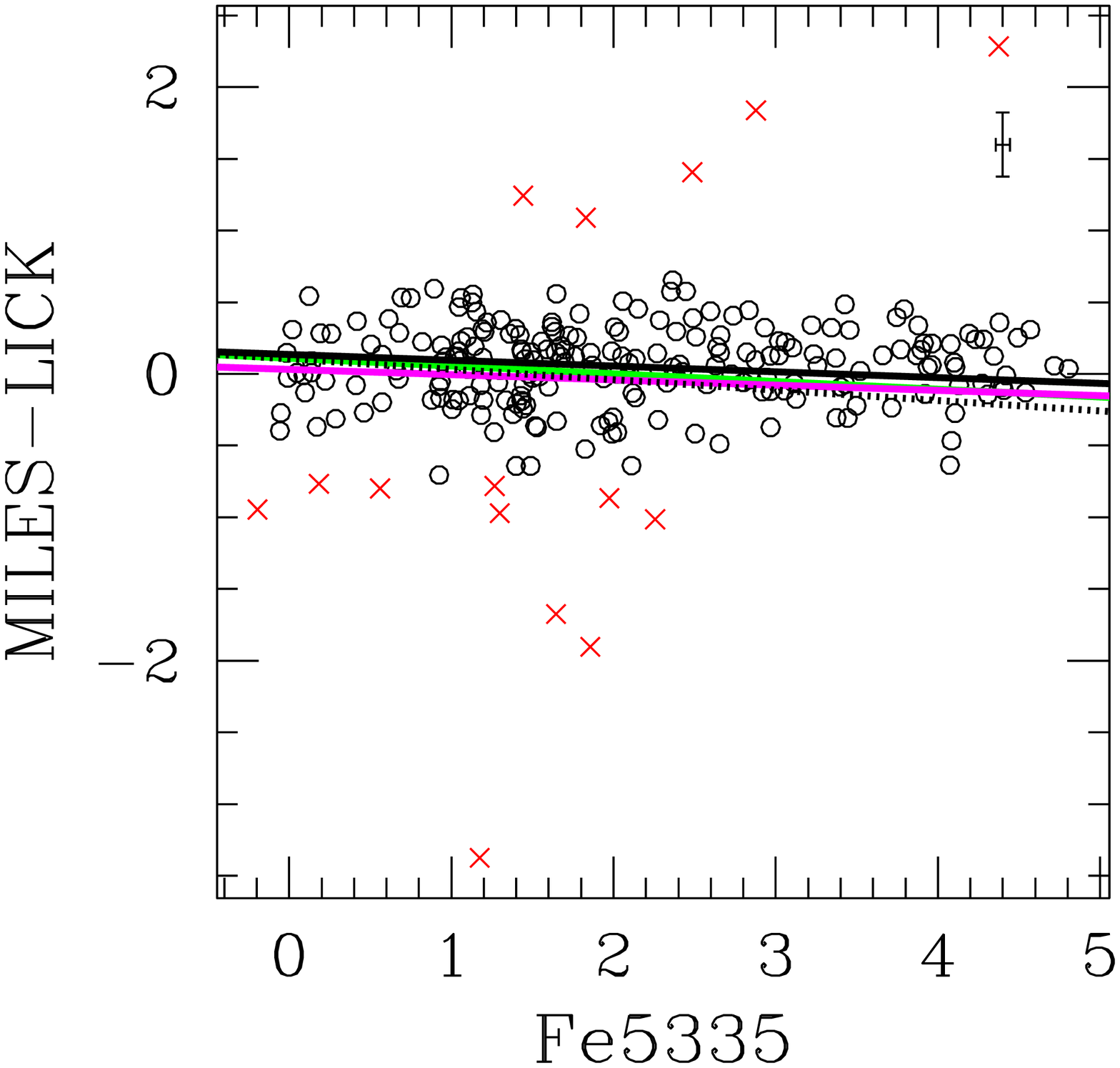}\includegraphics[scale=0.2]{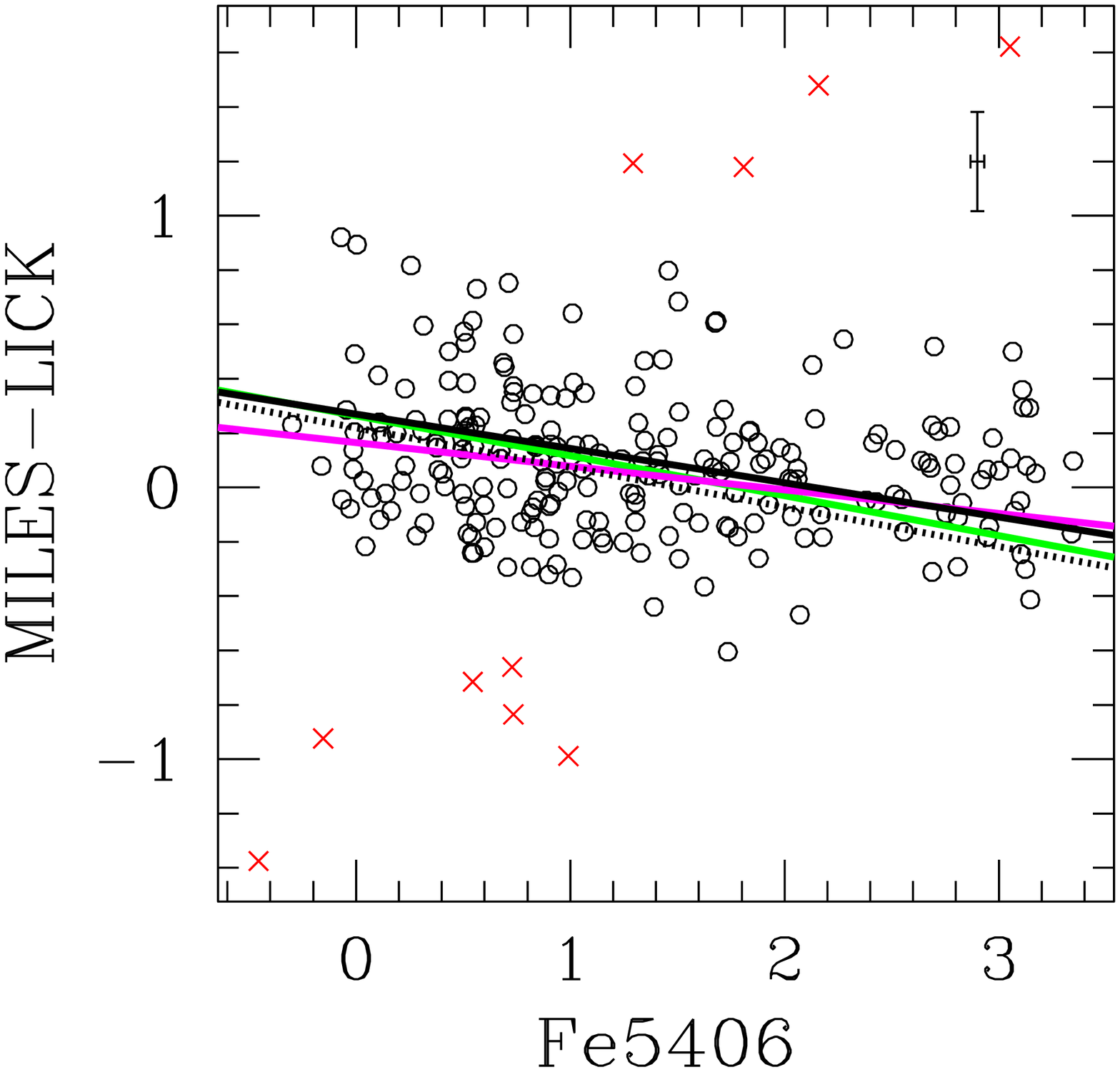}\\
\includegraphics[scale=0.2]{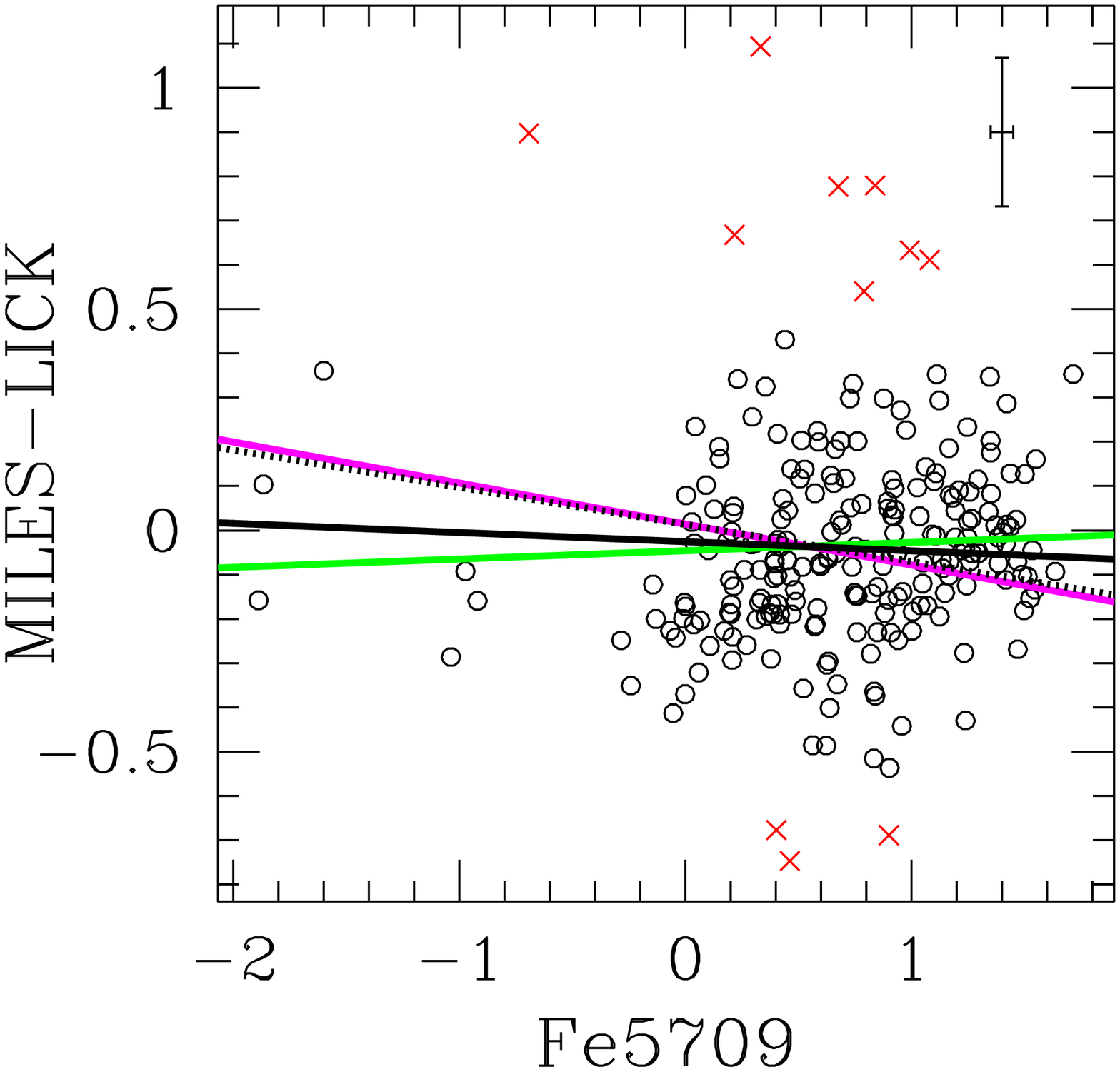}\includegraphics[scale=0.2]{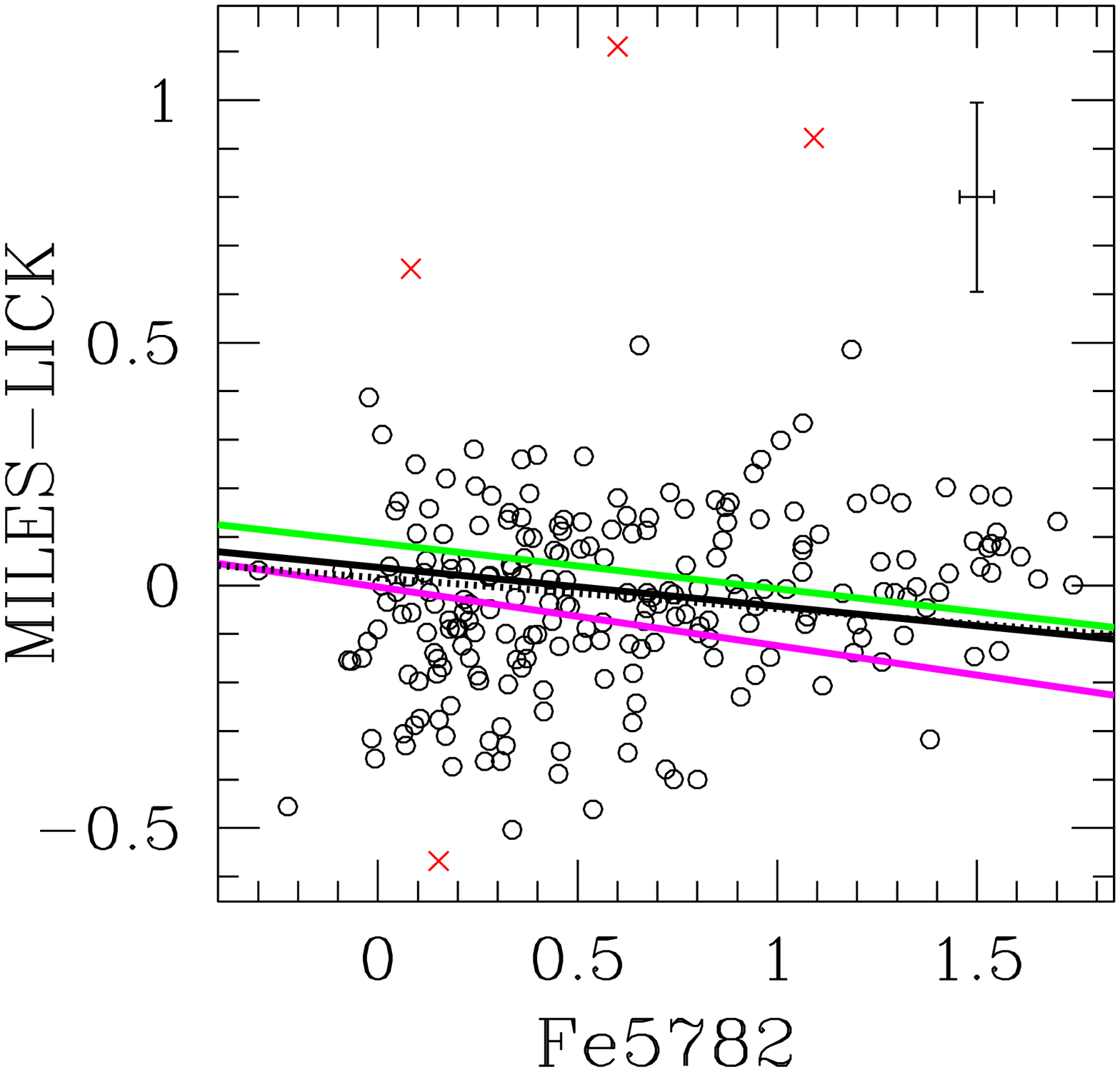}\includegraphics[scale=0.2]{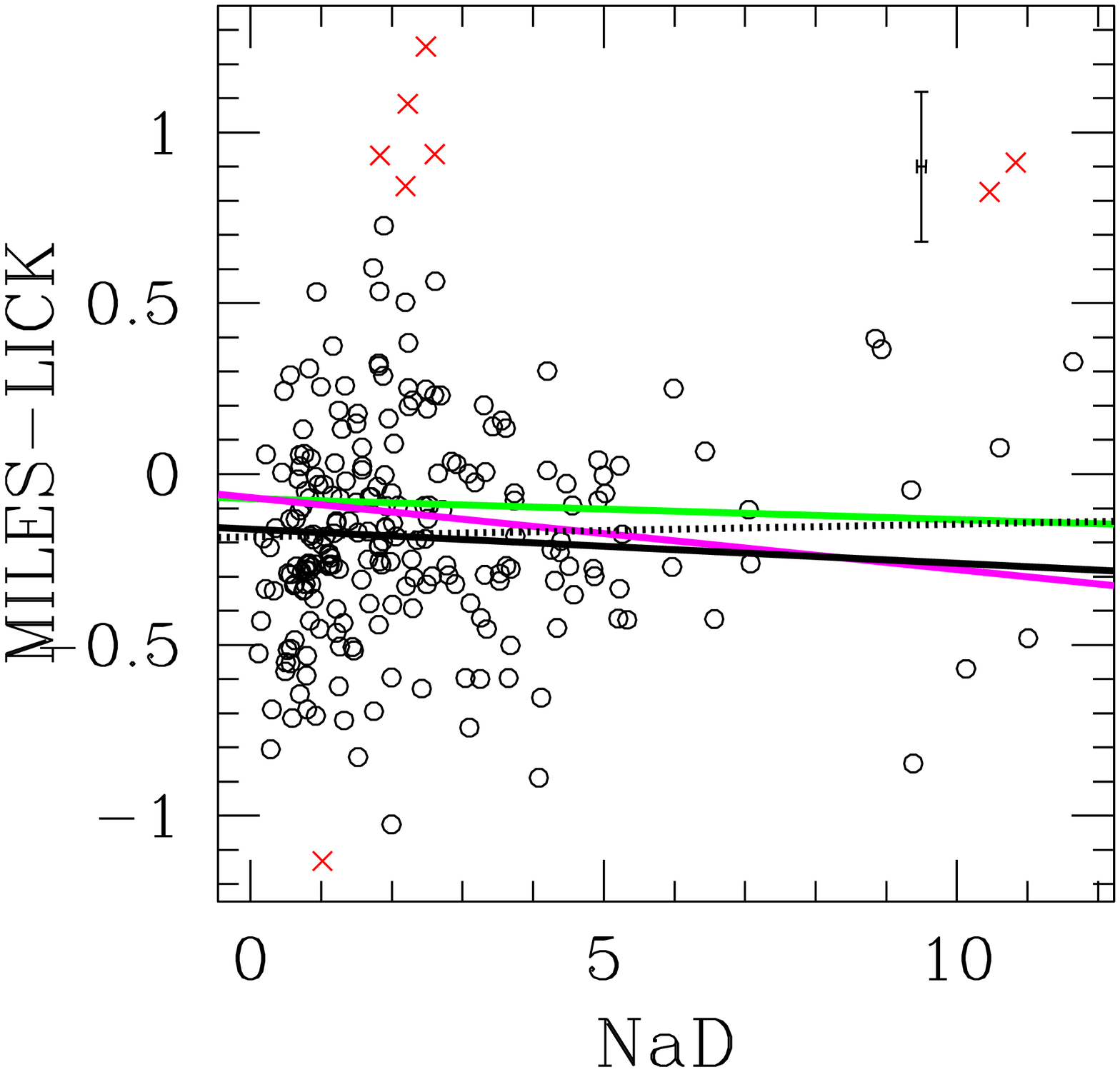}\includegraphics[scale=0.2]{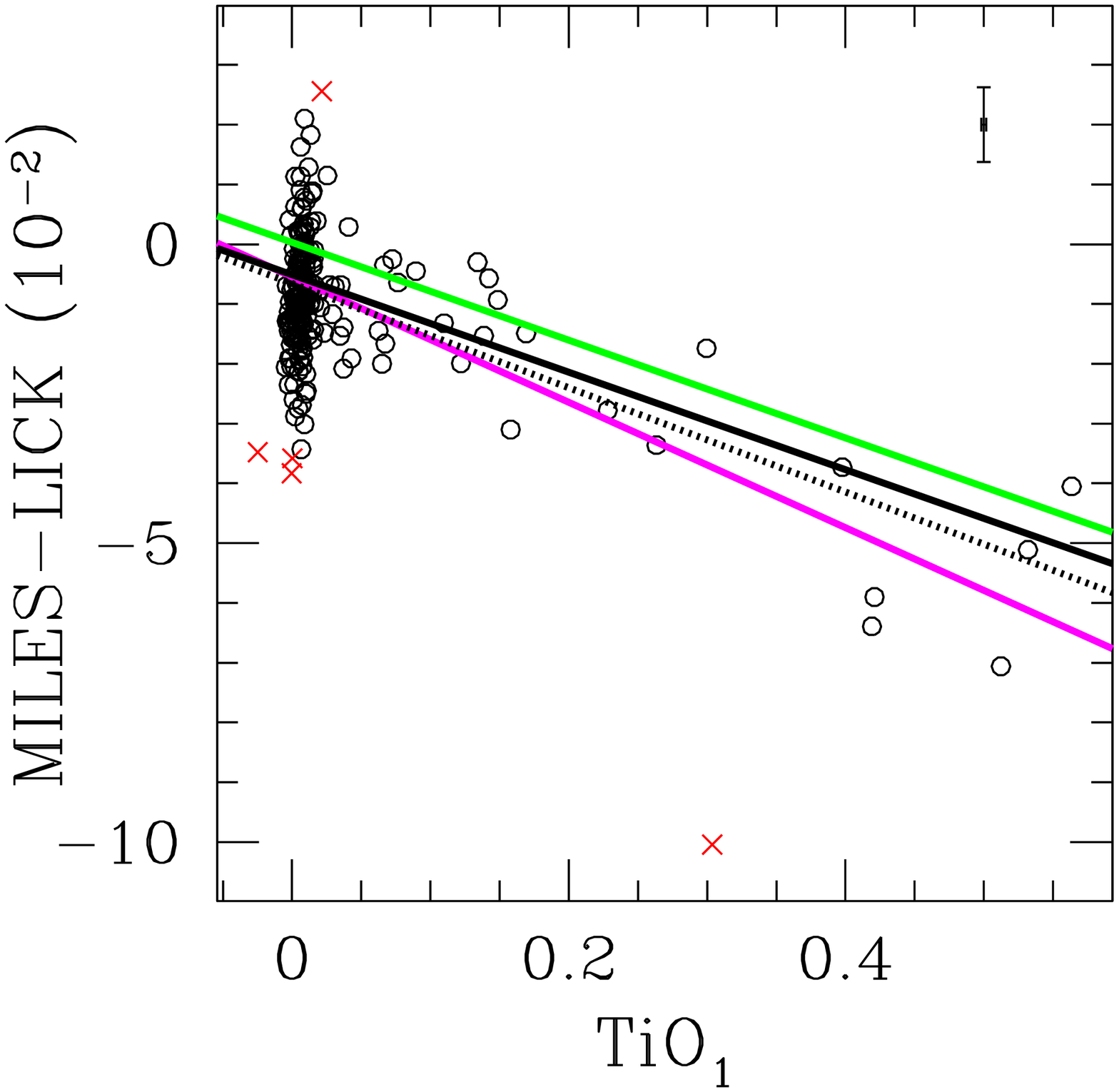}\\
\includegraphics[scale=0.2]{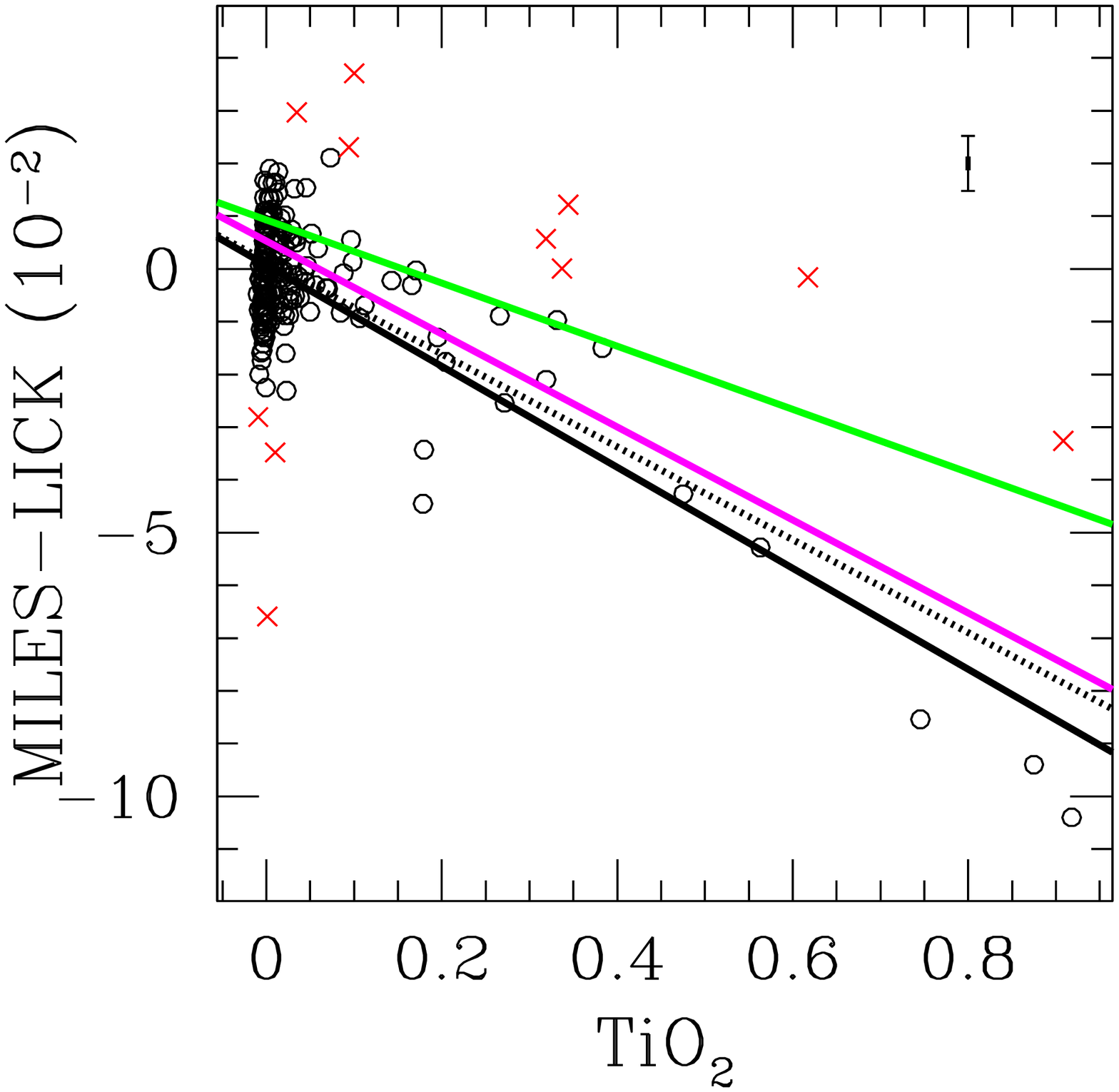}\\
\contcaption{}
\end{minipage}
\end{figure*}

\section{the \emph{MILES} STELLAR LIBRARY}
\label{stlib}

The \emph{MILES} library \citep{miles} consists of 985 stars with spectra in 
a wavelength range of 3525-7500 \AA, well covering the Lick indices, and 
with a spectral resolution of 2.3 \AA\ (see \citealt{miles} for further details). 
Important for the aim of this 
work is the careful flux-calibration of the \emph{MILES} spectra. 
Also, \citet{miles} selected the sample of stars to fill the gaps in stellar
parameter space covered by previous stellar libraries. This makes the \emph{MILES} library  
particularly suitable for modelling absorption line indices of stellar populations. 

Stellar parameter estimates in the literature show a scatter due to varying methods applied,
as discussed in \citet{maraston03} for [Fe/H].
The stellar parameters (T$_{\rm eff}$, $\log g$ or [Fe/H]) for the stars in the \emph{MILES} library 
are presented in \citet{cenarro07}, 
where estimates from the literature have been used and put on a homogeneous scale.
3 of 985 stars have no available estimates for none of the 
stellar parameters T$_{\rm eff}$, $\log g$ or [Fe/H]. 35 stars lack estimates only for [Fe/H] and are 
located in sparsely populated regions at the ends of the T$_{\rm eff}$ range. The stars have therefore been assigned 
a solar metallicity to increase the number of data points. 

\subsection{Empirical stellar Lick indices}
\label{absi}

Our aim was to produce fitting functions both for the resolution of the \emph{MILES} library (2.3 \AA) 
and for the resolution of the Lick/IDS library (8-11 \AA). We have therefore 
measured the 25 Lick indices directly on the original stellar spectra and on the spectra 
downgraded to the Lick/IDS resolution described by the curve presented in \citet{worthey97}. 
We have used the index definitions from \citet{trager98} and also from
\citet{worthey97} for the higher order Balmer lines (H$\delta_A$, H$\delta_F$, H$\gamma_A$ and H$\gamma_F$). 
Observational errors and offsets to the Lick/IDS library are described in the following paragraphs.

\subsubsection{Observational index errors}
\label{tierr}

We have derived typical 
observational index errors in order to evaluate the quality of our index measurements. 
To this end we have used pixel 1-$\sigma$ observational errors 
(P. S{\'a}nchez-Bl{\'a}szquez private communication) 
to perturb each stellar spectrum, both at \emph{MILES} resolution and Lick/IDS resolution,
through 600 Monte Carlo realizations. 
We have then measured the 25 Lick indices for each perturbed spectrum and 
determined 1-$\sigma$ errors for each index by using the spread in index measurements from the realizations. 
The index errors of the individual stellar spectra are used
for weighting the least square fits when deriving both the offsets to the Lick system (Section~\ref{loff}) and the 
fitting functions (Section~\ref{fitf}). 

Trends between the index errors and 
the atmospheric parameters or line-strength indices can in principle bias the fits, but we have found 
such trends not to affect the results. Only for the Balmer indices we find weak trends of increasing errors
with decreasing temperature and decreasing index strength. No trends with logg and Fe/H
are found for the Balmer indices. These weak trends can probably be explained with higher 
S/N for bright hot stars where the Balmer indices increase significantly in strength.
Since we compute the fitting functions in bins of temperature, these trends
have no significant effects on the final fitting functions.

The final 1-$\sigma$ typical index errors were determined by taking 
the median error
of the whole stellar library for each index. 
The typical index errors are presented in Table~\ref{offtable} both for \emph{MILES}
and Lick/IDS resolution. Compared to the typical
index errors for the Lick/IDS stellar library \citep{trager98}, also included in
Table~\ref{offtable},
we find the errors of the MILES library to have improved 
significantly. The stars of the Lick/IDS library were observed about thirty years before
the \emph{MILES} library. Considering the technical development in thirty years time, an 
improvement in the measured indices ought to be expected.

\subsubsection{Lick Index offsets}
\label{loff}

We have computed Lick index offsets between
the \emph{MILES} library and the Lick/IDS library using the stars in common between the 
two libraries. These offsets can be used for comparisons between models based on this work with models
based on the \citet{worthey94} and \citet{worthey97} fitting functions. 
The offsets are also used in 
Section ~\ref{comps} to compare the fitting functions of this work with the 
fitting functions of \citet{worthey94} and \citet{worthey97}. 

Fig.~\ref{offig} shows index by index comparisons for the residuals between the index 
measurements of the two libraries
as function of index strength. \citet{worthey97}, \citet{kuntschner01} and \citet{schiavon07} 
computed zero-point offsets to the Lick/IDS library, while \citet{puzia02} computed
their offsets as 2nd order least-square fits. For most indices we find index strength dependent residuals 
between the two libraries (Fig.~\ref{offig}). We have therefore computed the
offsets using a sigma-clipping linear least-square fitting routine,  
weighted with the individual index errors derived in Section~\ref{tierr}. 
The slope and
intercept of these fits are presented in Table~\ref{offtable} and also included
in Fig.~\ref{offig} (black solid lines). Sigma-clipped data points are indicated 
with red crosses in Fig.~\ref{offig} and the error bars are the 1-$\sigma$ index errors presented in 
Section~\ref{tierr}.
The error bars along the x-axes are represented by the index errors derived for the 
\emph{MILES} library, while the error bars along the y-axes are
represented by the combined errors of the \emph{MILES} and Lick/IDS libraries in quadrature.

Extreme outliers, 
i.e. data points that clearly showed strong deviating values compared to the bulk of data points,
were removed prior to running the fitting routine, 
in order to avoid stars with anomalous index strengths to affect the final fits.
For three indices (C$_2$4668, Fe5015 and Mgb) we found offsets at 
particularly high index strengths that deviated from the offset trends
for the majority of data points. The low number of data points at these index strengths and 
the absence of data points at intermediate index strengths induced a bias in the derived offsets. 
The data points at particularly high index strengths were therefore discarded when determining 
the final offsets. 

Offsets between the \emph{MILES} and the Lick/IDS library derived in \citet{sanchez09} are also
included in Fig.~\ref{offig} (dotted lines). These offsets and the offsets derived in this work for the 
\emph{MILES} library show in general very good agreement. Differences greater than the 1-$\sigma$ 
index errors are mainly found for Mgb, one of the indices for which we excluded data points at 
high index strengths due to deviations in offset trends. Noticeable offset differences, but still within 
the 1-$\sigma$ index errors, are also found for Fe5015, Fe5709, NaD, TiO$_1$ and TiO$_2$. 
Only small deviations are found between the offsets derived in \citet{sanchez09} and in this work for the rest of
the indices.

For comparison, and for the derivation of a possible universal offset between flux-calibrated 
system and the Lick/IDS system, we have also determined offsets to the Lick/IDS library for two other 
flux-calibrated stellar libraries, namely \emph{ELODIE} \citep{prugniel01} 
and \emph{STELIB} \citep{leBorgne03}. These offsets were
determined using the same procedure as described above for the \emph{MILES} library, except 
that no individual index errors were used as weights in the least-square fitting. For \emph{STELIB}
the lack of information did not allow for a computation of index errors, while the derived index errors for
\emph{ELODIE} were found to be unreliable as they showed unrealisticly small values. 
Since we only found small deviations in the offsets derived 
for the \emph{MILES} library when not weighting as compared to weighting the least-square fits, 
we compare the offsets derived for all three libraries.

In accordance with the \emph{MILES} library we found index strength dependent offsets also for the \emph{ELODIE} 
and \emph{STELIB} libraries. We found deviating offsets trends at high index values  
for the same indices as for the \emph{MILES} library (C$_2$4668, Fe5015 and Mgb).

The offsets derived for the \emph{ELODIE} and \emph{STELIB} libraries are also presented in 
Table~\ref{offtable} and Fig.~\ref{offig} (magenta and green lines, respectively). Clearly, deviations in the 
offsets are found between the 
libraries, especially for the \emph{STELIB} library compared to the other two libraries.
However, the \emph{STELIB} library is also the library having the least number of stars in 
common with the Lick/IDS library, giving a higher statistical uncertainty in the 
derived offsets. The \emph{STELIB} library only has 44 stars in common with the 
Lick/IDS library, while \emph{ELODIE} has 112 stars and the \emph{MILES} library has 
237 stars in common with the Lick/IDS library. Ca4227 showed particularly 
strange behaviour with index strength and the accuracy of the final offsets for this 
index could be questionable.

In Fig.~\ref{offig} we find agreements within the 1-$\sigma$ index errors between the 
offsets derived for all three libraries for H$\beta$, Mgb, Fe5270, Fe5335, Fe5406, Fe5709,
Fe5782 and NaD. This implies a better agreement between all libraries at wavelengths 
redder than $\sim$4800 \AA, with the exception
for the broader molecular indices Mg$_1$, Mg$_2$, TiO$_1$ and TiO$_2$ that show differences greater than 
the 1-$\sigma$ index errors, which is also found for Fe5015. Agreements between offsets derived for
\emph{MILES} and \emph{ELODIE} only, well within the 1-$\sigma$ index errors, 
are found for G4300, H$\gamma_A$, H$\gamma_F$, Fe4383, Ca4455, Fe4531 and 
C$_2$4668. This instead implies a worse agreement between \emph{MILES} and \emph{ELODIE} at 
wavelengths bluer than $\sim$4250 \AA\ (H$\delta_A$, H$\delta_F$, CN$_1$, CN$_2$ and Ca4227), 
where we in general find inconsistencies between all three libraries. 
The significant deviation in offset between the libraries for several indices hamper
the derivation of a universal offset between the Lick/IDS and flux-calibrated systems as 
described by these libraries.

This conclusion gets further support from the study of \citet{miles} who show that offsets exist
between the three flux-calibrated libraries \emph{MILES}, \emph{STELIB} and \emph{ELODIE}. 
These offsets are generally in good agreement with the individual Lick offsets found in this work.

\begin{figure*}
\begin{minipage}{17cm}
\centering
\includegraphics[scale=0.43]{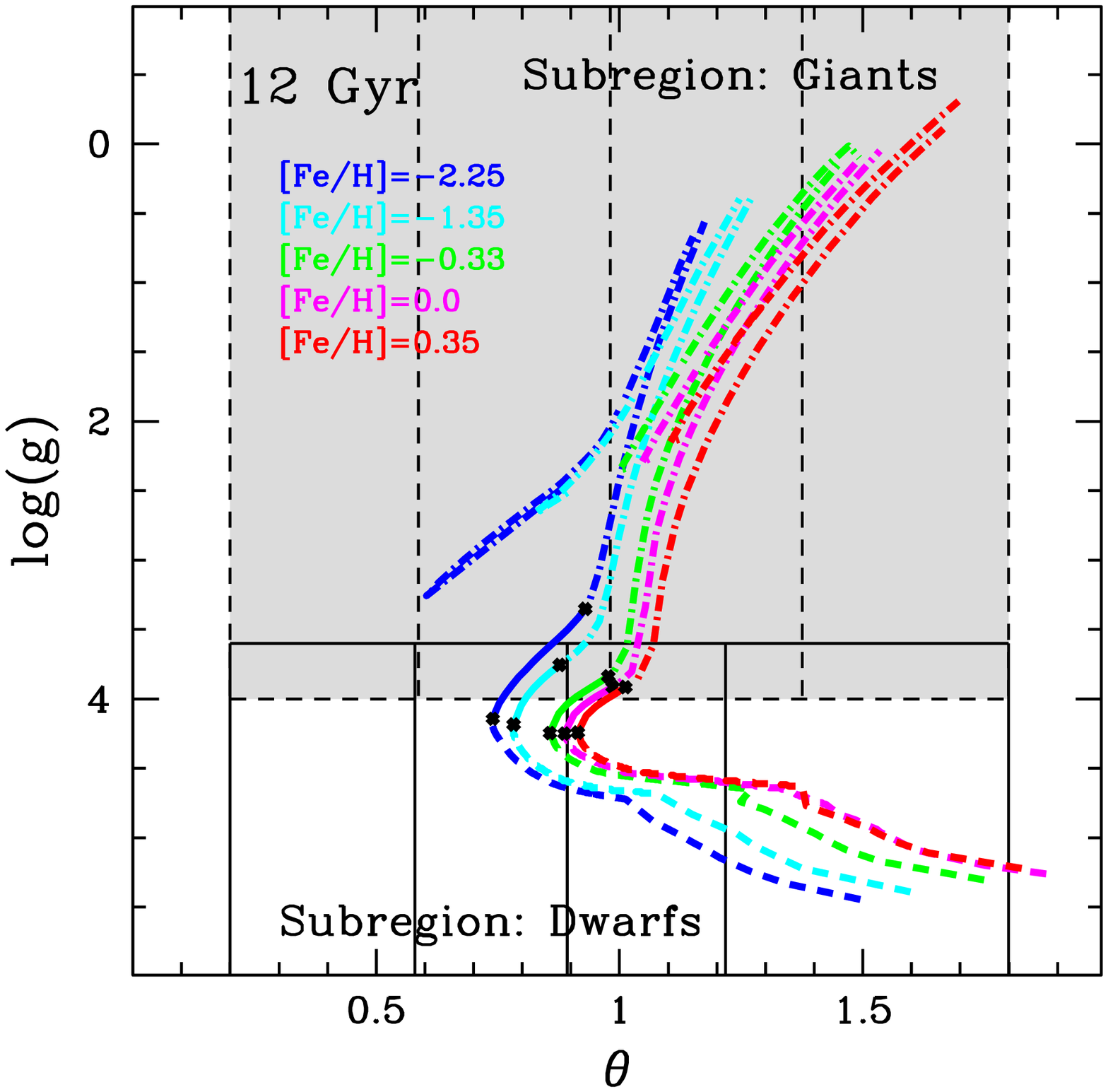}\includegraphics[scale=0.43]{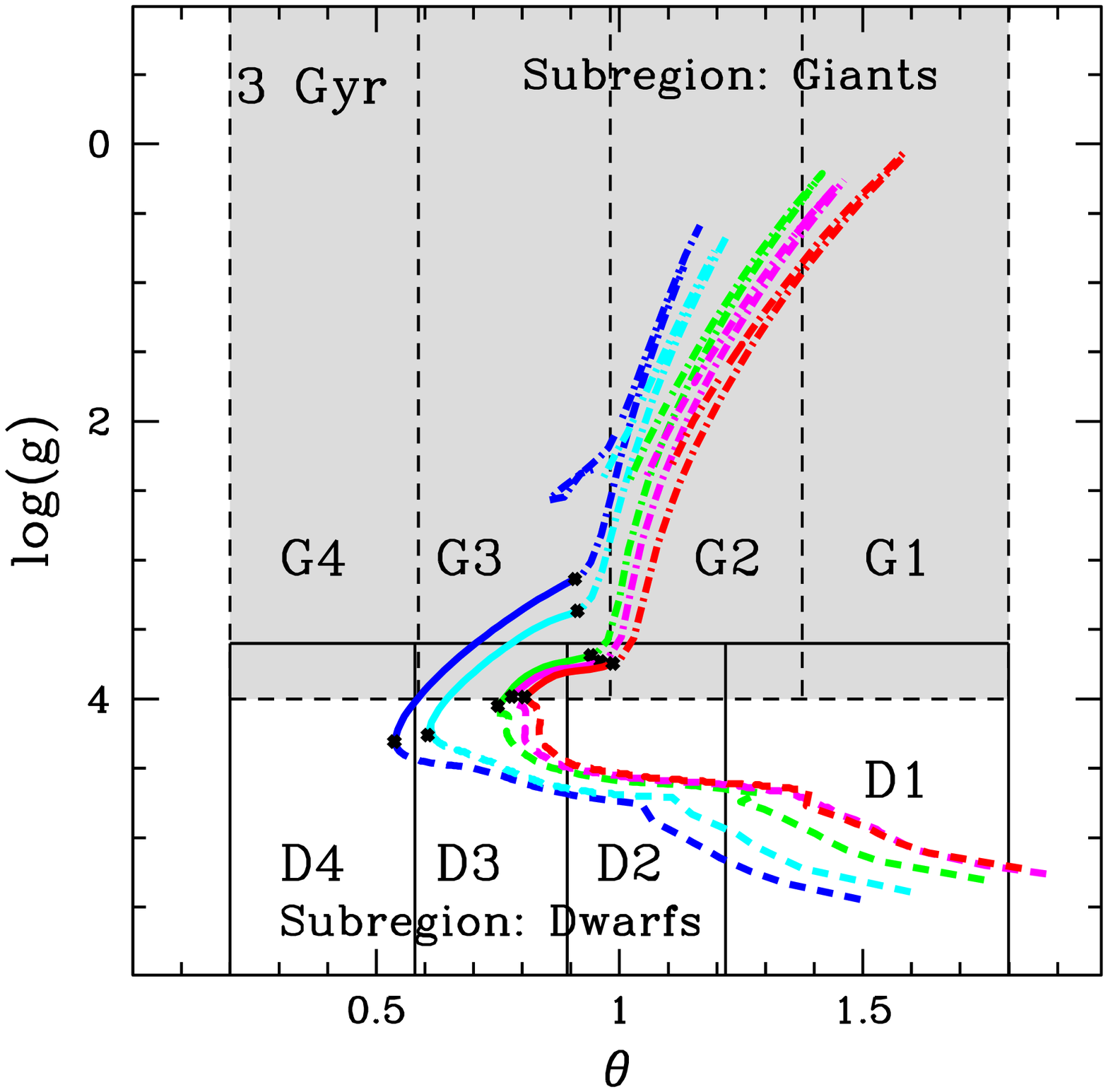}\\
\includegraphics[scale=0.43]{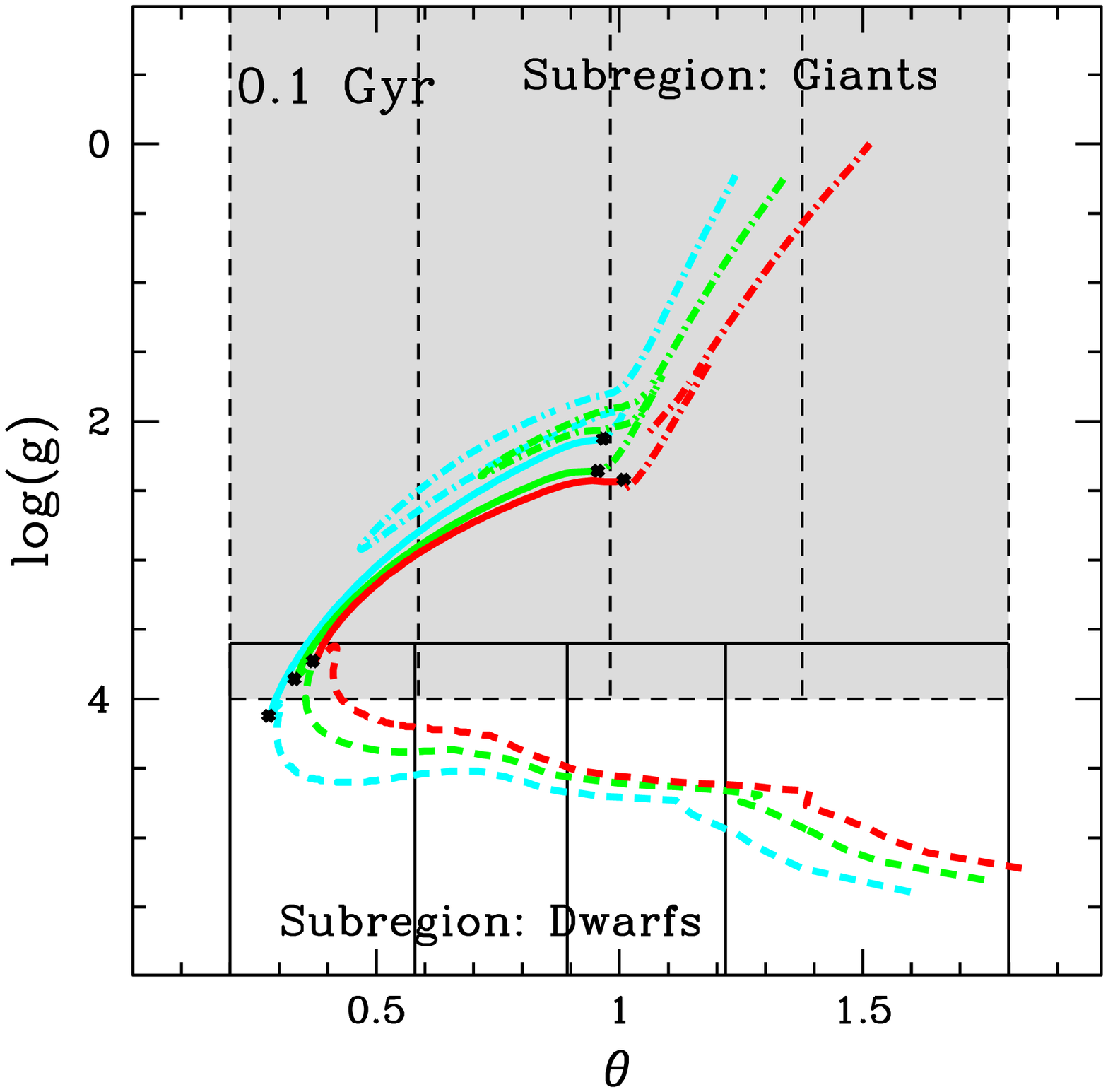}\includegraphics[scale=0.43]{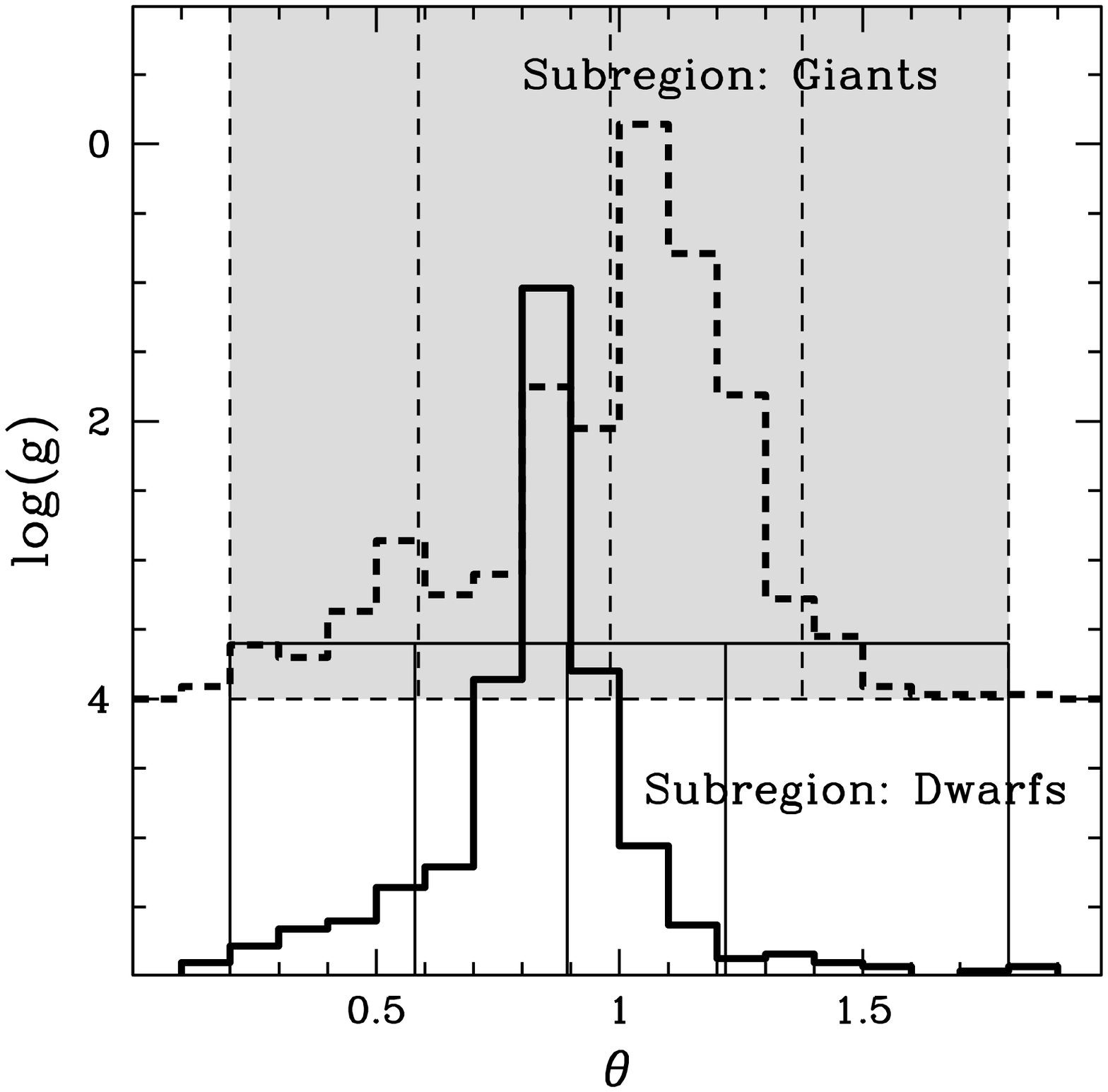}
\caption{The top panels and the lower left panel show the relationship between the chosen subregions and 
the analogous in the stellar population models of \citet{maraston05}. The three panels show models 
for different ages (as indicated)
and for each age the models are presented for varying metallicities as indicated by the corresponding 
colors. Each single model is divided into the main-sequence (dashed lines), sub-giant branch (solid line) and the rest of the
post-main-sequence phases (dash-dotted lines). The lower right panel shows the relationship between the chosen subregions and the distribution of 
data points as a function of $\theta$/T$_{\rm eff}$ for \emph{Dwarfs} ($\log g > 3.6$, solid lined histogram) 
and \emph{Giants} ($\log g > 4.0$, dashed line histogram) separately. 
In all panels the $\theta$/T$_{\rm eff}$ subregions for the \emph{Dwarf} subregion (solid lines) are indicated with D1-4 and for the \emph{Giant} subregion (shaded area and dashed lines) with 
G1-4.}
\label{fig_subreg}
\end{minipage}
\end{figure*}

\section{FITTING FUNCTIONS}
\label{fitf}

In order to produce empirical fitting functions for the
MILES library, we combine our measured Lick indices 
with the corresponding stellar atmospheric parameters (see Section ~\ref{stlib}). It is 
a complex task to find the
best relationship between indices and stellar atmospheric parameters, with several 
methods available in the literature. The method adopted in this work is 
presented in this section along with the derived fitting functions. 

A user friendly Fortran 90 code is available online at www.icg.port.ac.uk/$\sim$johanssj 
to make the implementation of our fitting functions easier.

\subsection{Fitting method}
\label{method}

The relationship between Lick index strengths and stellar parameters shows 
a complex behaviour, making it difficult to find one reliable empirical 
fitting function for the whole parameter space. To solve this problem the
parameter space must be divided into subregions where 
local fitting functions can be computed. 
However, it is desirable to find the simplest 
set of fitting functions and achieve a final representation of
the data that is as accurate as possible. Hence the limits of the
subregions have to be carefully chosen. It has also to be assured
that adjacent subregions overlap, making smooth transitions 
possible. For these transitions we have adopted cosine-weighted 
interpolations following \citet{cenarro02}. 
The choice of subregions are discussed in Section~\ref{ffres}.

Following the extensive number of published fitting functions in the literature 
\citep{worthey94,gorgas99,cenarro02,schiavon07,maraston09}, we use 
a linear least square fitting routine to 
determine the local relationships as polynomials in the following way

\begin{equation}\label{ffeq}
I(\theta,[Fe/H],log$g$)=\sum_{i}\beta_{i}\cdot\theta^{j}\cdot [Fe/H]^{k}\cdot log$g$^{l} 
\end{equation}

where $j,k,l\ge0$ and the atmospheric effective temperature is represented by 
$\theta =5040/T_{\rm eff}$. The representation of $T_{\rm eff}$ using $\theta$ is chosen
due to the wide range of spectral types in the stellar library.
The number of terms in Eq.~\ref{ffeq} can be made 
arbitrarily high. 
However, the goal is to find the best compromise between simplicity and accuracy
by discarding terms with higher order polynomials that are negligible 
or induce unphysical behaviours.
To this end several methods have been developed in the literature. 
\citet{worthey94} presented a method
to find the converging rms scatter by successively including terms and test if the 
rms scatter was significantly reduced by means of a F-test. \citet{gorgas99} and 
\citet{cenarro02} instead test if each term significantly differed from zero
through a T-test.
\citet{schiavon07} point out that both methods mentioned above are sensitive
to the coverage of parameter space. 
Therefore \citet{schiavon07} combine
the two methods by first successively removing statistically insignificant terms
and then interactively testing the remaining terms for unphysical behaviours
and their effect on the rms scatter.
 
In this work we adopt a mix of the above mentioned methods. 
We choose successive inclusion over successive removal of terms. The main 
reason for this choice is that the normal equations of the linear
least square routine run a high risk of becoming degenerate when terms
that respond similarly to the data are combined. By including terms
we can better control the degeneracy of the normal equations. If 
degenerate normal equations were reached after the inclusion of a new term,
this new term was discarded since a possible lower order term already 
responded to the data in a similar fashion. 

Finally, we determined the local fitting functions through an error
weighted linear least square routine
(for individual index errors see Section~\ref{tierr}). Terms were successively 
included following the procedure described in \citet{gorgas99}, 
by starting with the constant ($j,k,l=0$ in Eq. ~\ref{ffeq}) and then increasing
the sum of powers $j+k+l$ up to a maximum of $j+k+l=3$, including all
possible cross terms. However, since 
the effective temperature is the parameter showing the most complex behaviour 
we included polynomials of $\theta$ up to $j=5$. If the variance was not 
reduced at the inclusion of a new term the term was discarded. 
When a reduced variance was found the new term and all the previously 
included terms were tested by means of a T-test
to determine if the coefficients $\beta_{i}$ were statistically different from
zero (by using the coefficient errors following \citealt{gorgas99} and \citealt{cenarro02}). 
Terms with coefficients having a significance level $\alpha\le0.1$ was kept. 
 We then interactively studied the fitting
functions and removed terms inducing unphysical behaviours or not affecting the
rms scatter significantly. 
At the end of each run the sample was $\sigma$-clipped, by removing
data points deviating more than 3 $\sigma$, and the fitting redone on the new sample.

Extreme outliers that clearly deviated from the bulk of data points were discarded prior to running the
fitting routine. Hence to avoid stars with anomalous index strengths affecting the fitting functions.

\begin{table*}
\caption{Fe5335 fitting function coefficients for Lick/IDS resolution}
\label{fe5335table}
\begin{tabular}{|c|c|c|c|c|c|c|c|c|}
\hline
\multicolumn{9}{|c|}{\bf \small overall rms=0.2586}\\
  & \multicolumn{4}{|c|}{\small $\log g\le 4.0$ and $\theta$ limits:} & \multicolumn{4}{|c|}{\small $\log g\ge 3.6$ and $\theta$ limits:} \\
\small Term & \small $\le 0.58$ & \small $0.50-1.1$ & \small $0.95-1.5$ & \small $\ge 1.2$ & \small $\le 0.58$ & \small $0.50-1.0$ & \small $0.85-1.4$ & \small $\ge 1.2$ \\
\hline     
\scriptsize Const.               &\scriptsize -0.05682 &\scriptsize -1.257   &\scriptsize 125.1   &\scriptsize -279.0 &\scriptsize -0.8217 &\scriptsize -43.41 &\scriptsize 56.65   &\scriptsize 10.18 \\
\scriptsize $\theta$             &\scriptsize 0.4726   &\scriptsize 1.861    &\scriptsize -343.3  &\scriptsize 591.9  &\scriptsize 9.547   &\scriptsize 217.8  &\scriptsize -190.9  &\scriptsize -4.614 \\
\scriptsize [Fe/H]               &\scriptsize x        &\scriptsize -0.6719  &\scriptsize x       &\scriptsize x      &\scriptsize x       &\scriptsize 1.336  &\scriptsize -14.42  &\scriptsize 0.6270       \\
\scriptsize $\log g$             &\scriptsize x        &\scriptsize x        &\scriptsize x       &\scriptsize x      &\scriptsize x       &\scriptsize x      &\scriptsize 0.3445  &\scriptsize x       \\
\scriptsize $\theta^{2}$         &\scriptsize x        &\scriptsize 1.797    &\scriptsize 314.1   &\scriptsize -406.0 &\scriptsize -30.62  &\scriptsize -397.3 &\scriptsize 205.6   &\scriptsize x  \\
\scriptsize [Fe/H]$^{2}$         &\scriptsize x        &\scriptsize x        &\scriptsize x       &\scriptsize x      &\scriptsize x       &\scriptsize x      &\scriptsize 0.9821  &\scriptsize x                \\
\scriptsize $\theta$[Fe/H]       &\scriptsize x        &\scriptsize 1.808    &\scriptsize 1.048   &\scriptsize x      &\scriptsize x       &\scriptsize -4.202 &\scriptsize 31.08   &\scriptsize x                \\
\scriptsize $\theta^{3}$         &\scriptsize x        &\scriptsize x        &\scriptsize -93.45  &\scriptsize 90.94  &\scriptsize 30.84   &\scriptsize 313.2  &\scriptsize -69.81  &\scriptsize x \\
\scriptsize [Fe/H]$^{3}$         &\scriptsize x        &\scriptsize -0.05781 &\scriptsize -0.1268 &\scriptsize x      &\scriptsize x       &\scriptsize x      &\scriptsize x       &\scriptsize x                \\
\scriptsize $\theta^{2}$[Fe/H]   &\scriptsize x        &\scriptsize x        &\scriptsize x       &\scriptsize x      &\scriptsize x       &\scriptsize 4.535  &\scriptsize -15.09  &\scriptsize x                \\
\scriptsize $\theta$[Fe/H]$^{2}$ &\scriptsize x        &\scriptsize x        &\scriptsize -0.2159 &\scriptsize x      &\scriptsize x       &\scriptsize 0.2707 &\scriptsize -0.7390 &\scriptsize x                \\
\scriptsize $\theta^{4}$         &\scriptsize x        &\scriptsize x        &\scriptsize x       &\scriptsize x      &\scriptsize x       &\scriptsize -87.19 &\scriptsize x       &\scriptsize x \\
\hline
\small rms                       &\small 0.1111        &\small 0.2352        &\small 0.3446       &\small 0.7921      &\small 0.08168      &\small 0.1348      &\small 0.1879      &\small 0.7221                \\     
\small N                         &\small 81            &\small 358           &\small 365          &\small 113         &\small 51           &\small 349         &\small 207         &\small 17                \\     
\hline
\end{tabular} 
\end{table*}

\begin{table*}
\caption{Fe5335 fitting function coefficient errors for Lick/IDS resolution}
\label{fe5335errtable}
\begin{tabular}{|c|c|c|c|c|c|c|c|c|}
\hline
  & \multicolumn{4}{|c|}{\small $\log g\le 4.0$ and $\theta$ limits:} & \multicolumn{4}{|c|}{\small $\log g\ge 3.6$ and $\theta$ limits:} \\
\small Term & \small $\le 0.58$ & \small $0.50-1.1$ & \small $0.95-1.5$ & \small $\ge 1.2$ & \small $\le 0.58$ & \small $0.50-1.0$ & \small $0.85-1.4$ & \small $\ge 1.2$ \\
\hline     
\scriptsize Const.               &\scriptsize 0.01305 &\scriptsize 0.04702  &\scriptsize 1.237    &\scriptsize 2.860 &\scriptsize 0.1784 &\scriptsize 2.914    &\scriptsize 2.000   &\scriptsize 0.06966\\
\scriptsize $\theta$             &\scriptsize 0.03082 &\scriptsize 0.1169   &\scriptsize 3.120    &\scriptsize 5.928 &\scriptsize 1.544  &\scriptsize 15.82    &\scriptsize 5.800   &\scriptsize 0.04788 \\
\scriptsize [Fe/H]               &\scriptsize x       &\scriptsize 0.01732  &\scriptsize x        &\scriptsize x     &\scriptsize x      &\scriptsize 0.1617   &\scriptsize 0.3727  &\scriptsize 0.01636       \\
\scriptsize $\log g$             &\scriptsize x       &\scriptsize x        &\scriptsize x        &\scriptsize x     &\scriptsize x      &\scriptsize x        &\scriptsize 0.01056 &\scriptsize x       \\
\scriptsize $\theta^{2}$         &\scriptsize x       &\scriptsize 0.07110  &\scriptsize 2.603    &\scriptsize 4.059 &\scriptsize 4.200  &\scriptsize 31.77    &\scriptsize 5.571   &\scriptsize x  \\
\scriptsize [Fe/H]$^{2}$         &\scriptsize x       &\scriptsize x        &\scriptsize x        &\scriptsize x     &\scriptsize x      &\scriptsize x        &\scriptsize 0.08236 &\scriptsize x                \\
\scriptsize $\theta$[Fe/H]       &\scriptsize x       &\scriptsize 0.01959  &\scriptsize 0.008170 &\scriptsize x     &\scriptsize x      &\scriptsize 0.4293   &\scriptsize 0.7186  &\scriptsize x                \\
\scriptsize $\theta^{3}$         &\scriptsize x       &\scriptsize x        &\scriptsize 0.7181   &\scriptsize x     &\scriptsize 3.624  &\scriptsize 27.99    &\scriptsize 1.772   &\scriptsize x \\
\scriptsize [Fe/H]$^{3}$         &\scriptsize x       &\scriptsize 0.001251 &\scriptsize 0.003640 &\scriptsize x     &\scriptsize x      &\scriptsize x        &\scriptsize x       &\scriptsize x                \\
\scriptsize $\theta^{2}$[Fe/H]   &\scriptsize x       &\scriptsize x        &\scriptsize x        &\scriptsize x     &\scriptsize x      &\scriptsize 0.2782   &\scriptsize 0.3495  &\scriptsize x                \\
\scriptsize $\theta$[Fe/H]$^{2}$ &\scriptsize x       &\scriptsize x        &\scriptsize 0.01012  &\scriptsize x     &\scriptsize x      &\scriptsize 0.005307 &\scriptsize 0.08761 &\scriptsize x                \\
\scriptsize $\theta^{4}$         &\scriptsize x       &\scriptsize x        &\scriptsize x        &\scriptsize x     &\scriptsize x      &\scriptsize 9.138    &\scriptsize x       &\scriptsize x \\
\hline
\end{tabular} 
\end{table*}

\begin{figure*}
\begin{minipage}{17cm}
\centering
\includegraphics[scale=0.4]{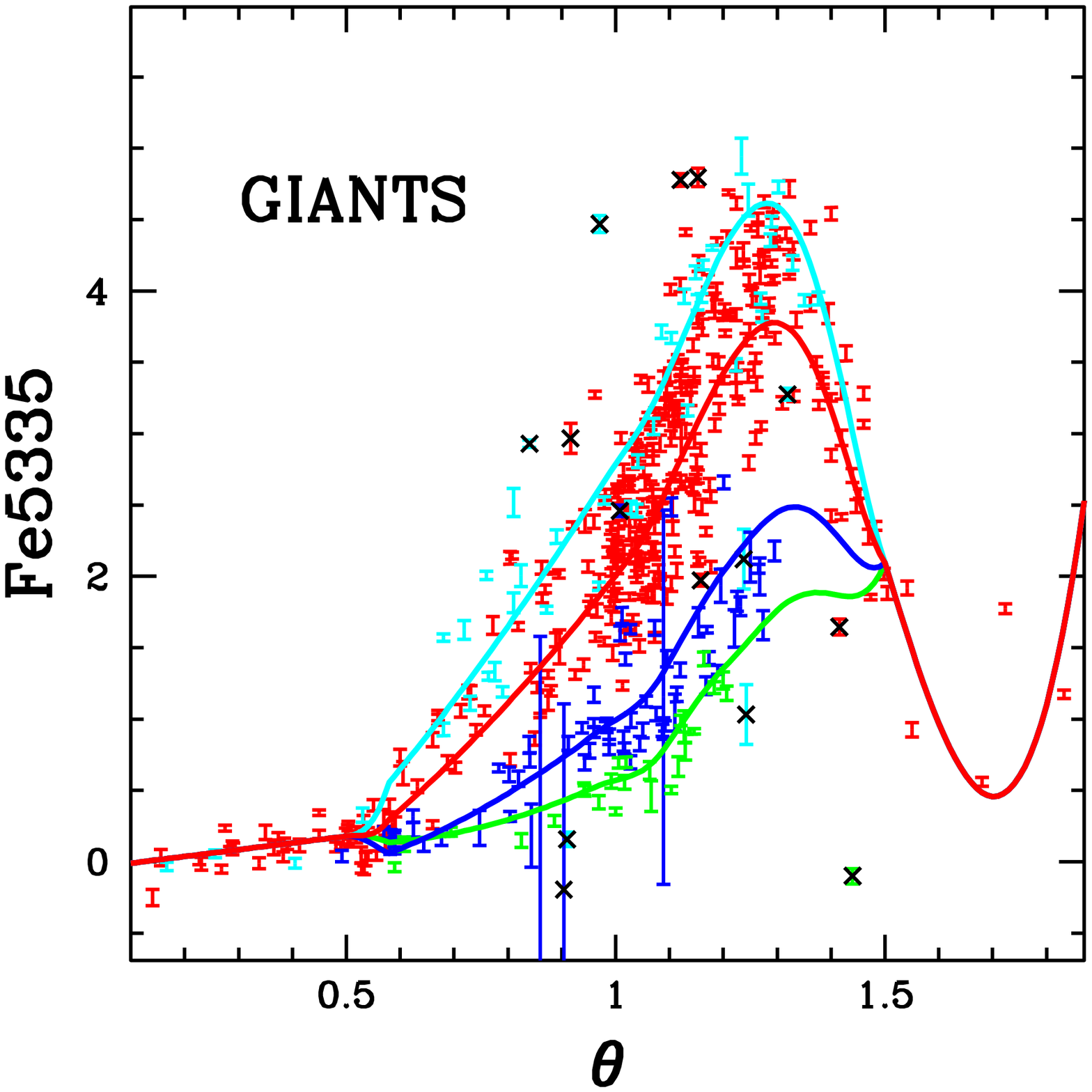}\includegraphics[scale=0.4]{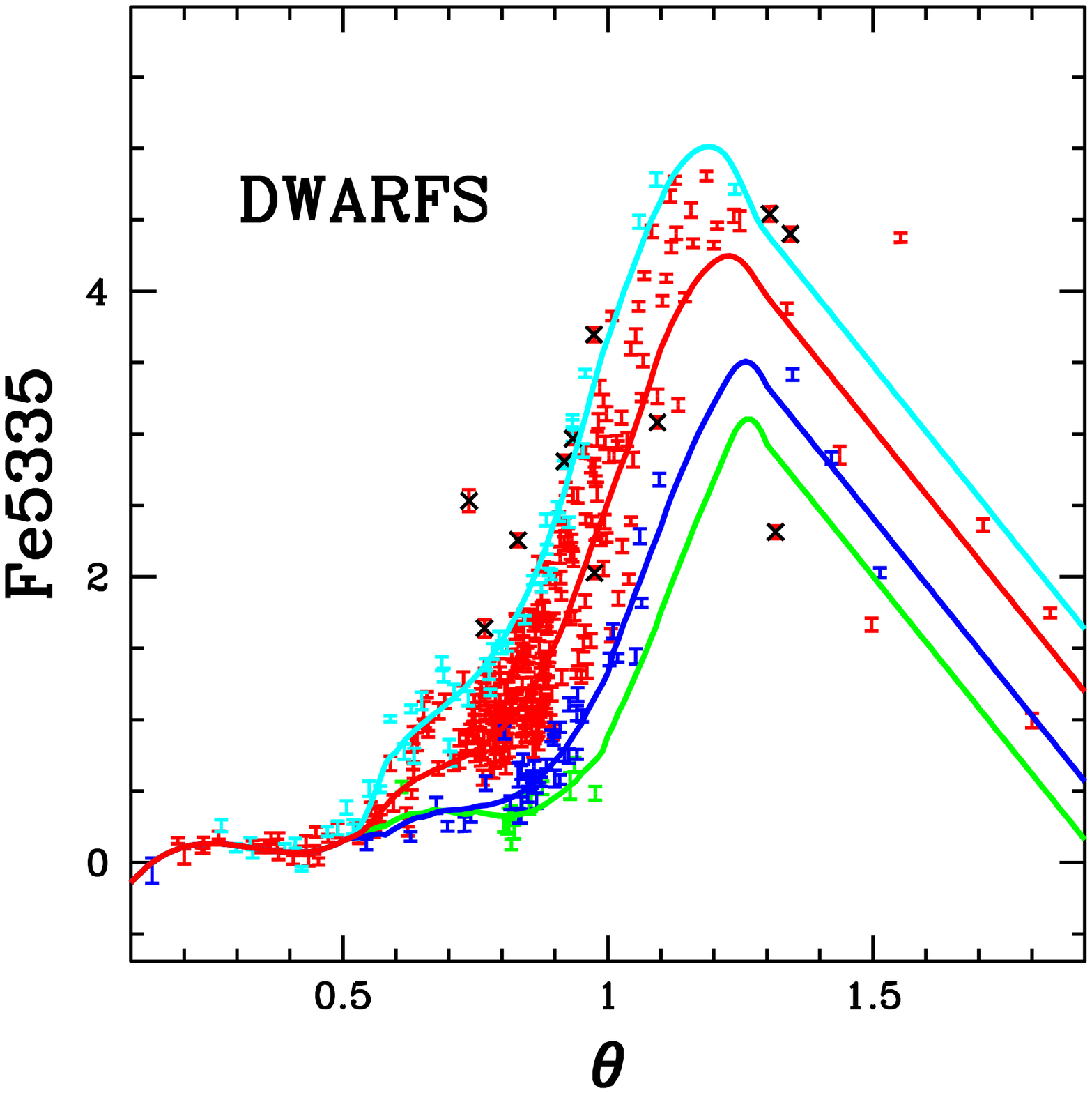}\\
\includegraphics[scale=0.4]{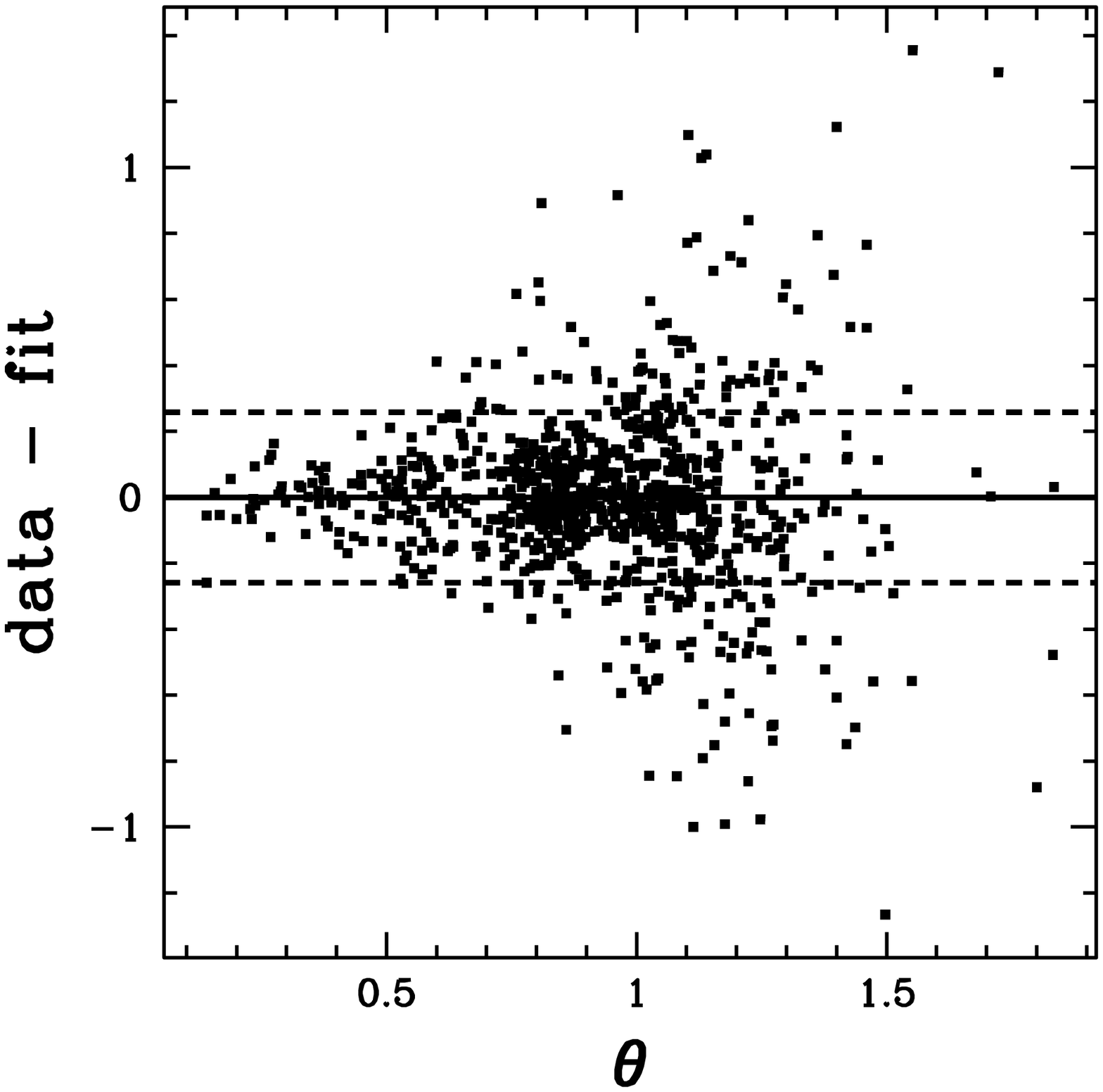}\includegraphics[scale=0.4]{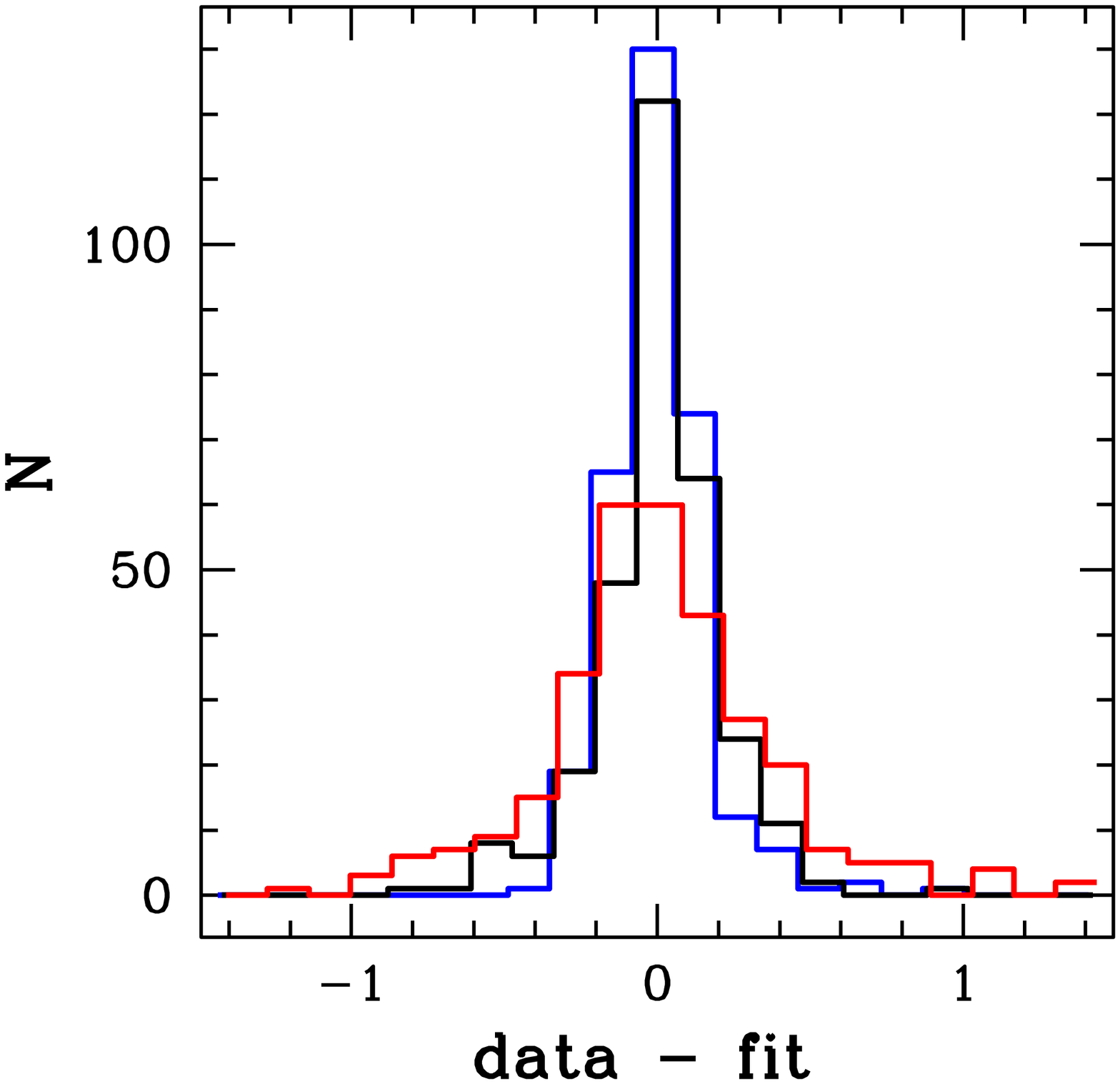}\\
\caption{The fitting functions for Fe5335 (Lick resolution) are shown in the  
upper panels for various metallicities and over-plotted on data in corresponding 
metallicity bins. The error bars on the data are observational index errors (see 
Section~\ref{tierr}). The colors correspond to [Fe/H]$=-2.0$ (green) , [Fe/H]$=-1.35$ (blue), [Fe/H]$=-0.35$ (red) 
and [Fe/H]$=0.35$ (cyan) for the fitting functions and [Fe/H]$<-1.8$ (green) , $-1.8<$[Fe/H]$<-1.0$ (blue), 
$-1.0<$[Fe/H]$<0.2$ (red) and [Fe/H]$>0.2$ (cyan) for the data. The left and right upper panels show
Giants ($\log g<3.6$) and Dwarfs ($\log g>3.6$) , respectively, for the average $\log g$ of the data in bins of 
$\Delta\theta=0.1$ at steps of $\theta=0.01$. Fixed $\log g$ values are used at the ends of the $\theta$/T$_{\rm eff}$ range, with 
$\log g=1.0,2.0$ (cold,warm end) and $\log g=4.6,4.0$ (cold,warm end) for Giants and Dwarfs, respectively. Data points with black crosses have been sigma clipped by the least-square fitting routine.
The lower left panel shows the residuals between the data and the fitting functions
as a function of $\theta$ and the dashed lines represents the overall rms value for the fitting functions. 
The lower right panel shows the 
distribution of the residuals for three $\theta$/T$_{\rm eff}$ bins, indicated by different colors where blue have $\theta<0.841$ 
black $0.841\ge\theta<1.045$ and red $\theta\ge1.045$.}
\label{fe5335ff}
\end{minipage}
\end{figure*}

\subsection{Definition of subregions in parameter space}
\label{ffres}

Thanks to the good coverage of stellar parameters the \emph{MILES} 
library show a complex behaviour of the relationship between the Lick
indices and the stellar parameters. We have therefore divided parameter space 
into several subregions. 

The relationship 
between the Lick indices and the stellar parameters show a bimodality 
between high and low gravity stars (i.e. \emph{Giants} and \emph{Dwarfs}). 
The first major subregions that we have chosen are therefore in high and low values of $\log g$ space (from now on referred
to as the \emph{Dwarf} and \emph{Giant} subregion, respectively), 
in accordance with \citet{gorgas99}, \citet{cenarro02} and \citet{schiavon07}. 
The same $\log g$ subregion limits have been used for all indices.
The lower limit for the \emph{Dwarf} subregion was set to $\log g=3.6$, while 
the upper limit for the \emph{Giant} subregion was set to $\log g=4.0$, giving an overlap region of $\Delta\log g=0.4$. In
Fig~\ref{fig_subreg} the subregions are shown together with the analogous 
in the stellar population models of \citet{maraston05}, for $\log g$ as a function of $\theta$.
The different evolutionary phases for the models are indicated in Fig~\ref{fig_subreg}. 
This shows that the choice of limits for the \emph{Dwarf} and \emph{Giant} subregions coincides 
very well with the division into the main-sequence and the post-main-sequence, as the $\log g$ overlap region
mainly covers the sub-giant branch (SGB).

To fully recover the detailed behaviour 
within the $\log g$ subregions we divided the full $\theta$/T$_{\rm eff}$ range into
four subregions. The choice of 
the limits for $\theta$/T$_{\rm eff}$ subregions follow the behaviour of the models 
and the distribution of stars as a function $\theta$/T$_{\rm eff}$. This can be seen in Fig~\ref{fig_subreg}
where the limits of the $\theta$/T$_{\rm eff}$ subregions are represented by the midpoints in the overlap
regions, averaged over all indices. The $\theta$/T$_{\rm eff}$ subregions 
are discussed in the following bullet points, by referring to the $\theta$/T$_{\rm eff}$ subregions using the 
names (D1-4 and G1-4) in Fig~\ref{fig_subreg}, first for the \emph{Giant} subregion
\begin{itemize}
\item Only the tip of the RGB for high metallicities fall within G1 (Fig~\ref{fig_subreg}). 
The lower limit (in $\theta$) for 
this subregion coincides with the strong drop-off in the distribution of data points (Fig~\ref{fig_subreg}). 
With the weak dependency on metallicity for this subregion and the low number of data points we fit this
subregion independently of metallicity. 
\item G2 and G3 clearly separates out the red-giant branch (RGB) to be fitted mainly in G2 
(Fig~\ref{fig_subreg}). 
\item Most indices show a distinct change in the behaviour of the index strengths as a function of 
the stellar parameters for hot A-type stars, around $\theta=0.5-0.6$, see Fig.~\ref{fe5335ff} and 
Fig. A1-A24. The overlap regions between G3 and G4 are therefore 
located around this range in $\theta$.
\end{itemize}
and then for the \emph{Dwarf} subregion
\begin{itemize}
\item The lowest part of the main-sequence fall within D1 (Fig.~\ref{fig_subreg}). As for the \emph{Giant} 
subregion, the lower limit (in $\theta$) for this subregion coincides with the strong drop-off in 
the distribution of data points.  
\item The division of $\theta$/T$_{\rm eff}$ space into D2 and D3 were found to improve the fits in terms of a significantly 
reduced rms scatter.
\item As for the \emph{Giant} subregion, most indices show a distinct change in the behaviour of the index strengths as a function of 
the stellar parameters for hot A-type stars, around $\theta=0.5-0.6$, see Fig.~\ref{fe5335ff} and 
Fig. A1-A24. The overlap regions 
between D3 and D4 are therefore located around this range in $\theta$.
\end{itemize}

The number of the $\theta$/T$_{\rm eff}$ subregions is the same for all indices. With the exceptions for TiO$_1$
and TiO$_2$ that show a much simpler behaviour and we have 
therefore used less $\theta/T_{\rm eff}$ subregions (see Fig. A23-A24, Table A23-A24 and Table B24-B25). 
Since the different indices 
show a varying dependence on the stellar parameters, the limits for the subregions have been adjusted 
for each index individually to reduce the rms scatter.

The choice of subregions in $\log g$ and $\theta$/T$_{\rm eff}$ space make up the base for our 
fitting functions. On top of these, metallicity space had to be divided into two subregions 
for 10 indices (CN$_1$, CN$_2$, Ca4227, 
G4300, Fe4383, Fe5015, Mg$_1$, Mg$_2$, Mgb and NaD) in order to fully reproduce the 
metal-poor end, but only in the low gravity subregion and in the specific temperature range 
around $1.0<\theta<1.4$ ($5040<T_{\rm eff}<3600$). 
We have therefore independently fitted metal-rich and metal-poor stars, divided at [Fe/H]$\sim-1.0$
for the affected temperatures in the low gravity subregion for the 12 indices. 

Even though the \emph{MILES} library covers an extensive range of stellar parameter space, the 
very ends are obviously still sparsely populated. Therefore, 
the fitting functions are not
valid beyond $\theta>1.8$ ($T_{\rm eff}<2800$) and $\theta<0.2$ ($T_{\rm eff}>25200$). The dwarf 
main-sequence that extends to very low temperatures is well covered within these limits (Fig.~\ref{fig_subreg}). 
Very hot young stars with temperatures greater than 25200 K do not have strong indices in the visual parts of their spectra.

\subsection{[$\alpha$/Fe] trends}
\label{alpha-bias}

Globular cluster stars are significantly [$\alpha$/Fe]-enhanced with respect to solar values 
\citep[$\sim0.3$, ][]{carney96}. The [$\alpha$/Fe]-trend of field stars in the solar neighborhood instead show increasing 
[$\alpha$/Fe]-enhancements with decreasing metallicity down to [Fe/H]$\sim-1.0$ \citep*{edvardsson93,fuhrmann98,milone09}. 
It is first at this metallicity that the field stars reach globular cluster [$\alpha$/Fe]-values. 
Having globular cluster stars for [Fe/H]$>-1.0$ can therefore induce [$\alpha$/Fe] trends biased towards 
globular cluster values in stellar libraries dominated by field stars. 
The globular cluster M71 has a metallicity of [Fe/H]$=-0.84$ and is represented by a significant number of 28 stars in the 
\emph{MILES} library, which could possibly induce such a bias. The stars from this globular cluster were therefore discarded
when computing the final fitting functions, since the \emph{MILES} library is reasonably well 
populated with field stars around the metallicity of M71.

The [$\alpha$/Fe]-bias of the 
solar neighborhood must be taken into account when deriving stellar population
models based on empirical stellar libraries, as discussed in \citet{maraston03}. Model adjustments are 
therefore needed when adopting the fitting functions of this work. Such adjustments are described in \citet{trip95,TMB03,thomas04,korn05,thomas05}.

\subsection{Spectral resolution}
\label{spec_res}

We have computed fitting functions for both the \emph{MILES} and Lick/IDS resolutions (see Section~\ref{absi}).
The same final set of terms were used for both resolutions.
Coefficients and coefficient errors for the 
fitting functions are presented in Appendix A for 
Lick resolution and Appendix B for \emph{MILES} resolution. The sigma clipped number of data points (N) for the local
fitting functions are also included in the coefficient Tables, along with the rms 
of the residuals between the data and the final fitting functions,
both local and overall. The visual behaviours, residuals and 
distribution of residuals of the fitting functions are shown for Lick resolution in Appendix A.
An example is presented for Fe5335 and Lick resolution in Table~\ref{fe5335table} and Table~\ref{fe5335errtable} for coefficients and coefficient errors, respectively. 
The visual behaviours of the fitting functions for Fe5335 are shown in Fig.~\ref{fe5335ff}, 
where they are presented for the \emph{Dwarf} and \emph{Giant} subregions separately 
and for varying metallicity. In Appendix A the visual behaviour of fitting functions for several 
$\log g$ values at fixed $\theta$ are also presented for indices showing strong 
$\log g$ dependencies within the $\log g$ subregions.

\subsection{Errors}
\label{errs}

In this section we briefly discuss possible error sources affecting the final fitting functions.
Such error sources include the index measurements of the \emph{MILES} spectra, 
but these show very high quality, in terms of typical observational index errors, as discussed in 
Section~\ref{tierr}.
However, the overall rms of the final fitting functions (see Section~\ref{spec_res}) are considerably  
larger than the typical observational index errors (see Section~\ref{tierr}). 
Possible error sources for this scatter are instead uncertainties in the stellar parameter estimates 
and intrinsic scatter in the index strengths.

The residuals between the final fitting functions and the data, presented in the lower left panels of
Fig.~\ref{fe5335ff} and Fig. A1-A24 as a function of $\theta$, 
show typically larger scatter for cooler temperatures where index values exhibit strong sensitivities to 
T$_{\rm eff}$. The source of this correlation is probably, at least partly, uncertainties in the stellar 
parameters, since these will have a larger effect when the index strengths show strong dependencies on 
the stellar parameters, 
i.e. $\theta/T_{\rm eff}$ uncertainties will have less effect when the index strengths show 
weaker dependencies on $\theta/T_{\rm eff}$.

\section{COMPARISONS WITH THE LITERATURE}
\label{comps}

In this section we compare the fitting functions derived in this work with 
fitting functions in the literature derived for stellar libraries other than \emph{MILES}. 
We search for differences in various parameter regimes. 
Comparisons are made with the classical 
and extensively adopted fitting functions of \citet{worthey94} and \citet{worthey97} (from now on $WFF$), 
shifted with the offsets derived in Section~\ref{loff}, and
with the more recent fitting functions of \citet{schiavon07} (from now on $SFF$) which were based on the
\emph{JONES} library \citep{jones99}.

We have performed the comparisons in different regions of parameter space 
to find the regimes where major differences roam.
The comparisons have been divided into three $\theta/T_{\rm eff}$ bins, 
referred to as \emph{Cold}, \emph{Intermediate} and \emph{Warm} temperatures, with $\theta/T_{\rm eff}$ limits presented
in Table~\ref{complimits}. Each of these bins have been 
further divided into two $\log g$ bins with $\log g=2.0$ (referred to as \emph{Giants}) and $\log g=4.5$ 
(referred to as \emph{Dwarfs}) to make up a total of six bins. The average
residuals between the fitting functions were computed in each bin at [Fe/H] 
steps of 0.5 in the range $-2\le$[Fe/H]$\le0.5$ and presented in 
Fig.~\ref{WFFcomp} and Fig.~\ref{SFFcomp} as a function of metallicity for the comparisons with
WFF and SFF, respectively. 

The comparisons have only been made within the parameter limits for which
the fitting functions are applicable, described in \citet{worthey94} (WFF), \citet{schiavon07} (SFF) 
and Section~\ref{ffres} (this work), resulting in the limits of the $\theta$/T$_{\rm eff}$ bins presented
in Table~\ref{complimits}. Due to the limitations of the SFF we can not make
comparisons for the \emph{Warm} $Giant$ regime, while the \emph{Intermediate} $Giant$ regime have a varying 
lower $\theta$ limit (see \citealt{schiavon07} for individual index limits). 

The overall rms of the final fitting functions (see Section~\ref{spec_res}) 
are shown in Fig.~\ref{WFFcomp}~-~\ref{SFFcomp}
as grey shaded areas (1rms dark grey and 2rms light grey).
This gives a reference to the differences found between the libraries. 

Overall there is good agreement between fitting functions within the rms.
We find the biggest residuals to occur at the ends of parameter space, i.e. at
the metallicity and temperature ends (see Fig.~\ref{WFFcomp}~-~\ref{SFFcomp}). This was
expected since the number of data points decrease towards the ends of parameter space, 
resulting in larger uncertainties of the fitting functions.
In the rest of this Section we discuss the comparisons for individual indices in terms of  
stellar parameter regions that show differences beyond the 1rms and 2rms levels.\\

\begin{table}
\centering
\caption{Limits for the different bins of $\theta/T_{\rm eff}$ space used in the fitting function comparisons. 
SFF-G and SFF-D correspond to the fitting functions of \citet{schiavon07} for \emph{Giants} and \emph{Dwarfs}, respectively.}
\label{complimits}
\begin{tabular}{cccc}
\hline
\bf FF & \bf \emph{COLD} (T$_{\rm eff}$) & \bf \emph{INTER.} (T$_{\rm eff}$) & \bf \emph{WARM} (T$_{\rm eff}$)\\
\hline
WFF &  2800-4582 & 4582-7200 & 7200-12263\\
SFF-G & 2800-4582 & 4582-($\sim$)6300 & - \\
SFF-D & 3220-4582 & 4582-7200 & 7200-18000 \\
\hline
\end{tabular} 
\end{table}

\begin{flushleft}\textbf{H$\mathbf{\delta_A}$}\\\end{flushleft}
\emph{WFF comparison (Fig.~\ref{WFFcomp}):} \emph{Warm} \emph{Giants} extend well beyond the 2rms level where
this work show much weaker indices. We find both \emph{Warm} and \emph{Cold} \emph{Dwarfs} to show 
stronger indices for this work, even extending beyond the 2rms level for the metal-poor and metal-rich ends, respectively. 
Otherwise, this work show slightly weaker indices extending to the 1rms level.\\
\emph{SFF comparison (Fig.~\ref{SFFcomp}):} \emph{Cold} \emph{Dwarfs} show weaker indices for this work, beyond the 1rms level.
\emph{Warm} and \emph{Intermediate} temperature \emph{Dwarfs} show stronger indices for this work out to the 2rms level in the metal-poor
regime.
\emph{Intermediate} temperature \emph{Giants} show stronger indices out to the 2rms level at the ends of the metallicity scale.
Otherwise are mainly differences within the 1rms level found.\\

\begin{flushleft}\textbf{H$\mathbf{\delta_F}$}\\\end{flushleft}
\emph{WFF comparison (Fig.~\ref{WFFcomp}):} The most obvious difference is found for \emph{Warm} \emph{Giants} where
this work show much weaker indices, extending well beyond the 2rms level. Otherwise are differences
mainly within the 1rms level, except for the metal-rich end of \emph{Cold} and \emph{Warm} \emph{Dwarfs} 
that show stronger indices for this work beyond the 2rms level.\\
\emph{SFF comparison (Fig.~\ref{SFFcomp}):} This work shows in general stronger indices in the
metal-poor regime, beyond the 1 rms level for \emph{Intermediate} temperature and \emph{Warm} \emph{Dwarfs} and beyond the
2rms level for \emph{Intermediate} temperature \emph{Giants}. In the metal-rich regime we instead find weaker for this work, out
to the 2rms level for \emph{Intermediate} temperature and \emph{Cold} \emph{Giants}.\\

\begin{figure*}
\begin{minipage}{17cm}
\centering
\includegraphics[scale=0.21,angle=-90]{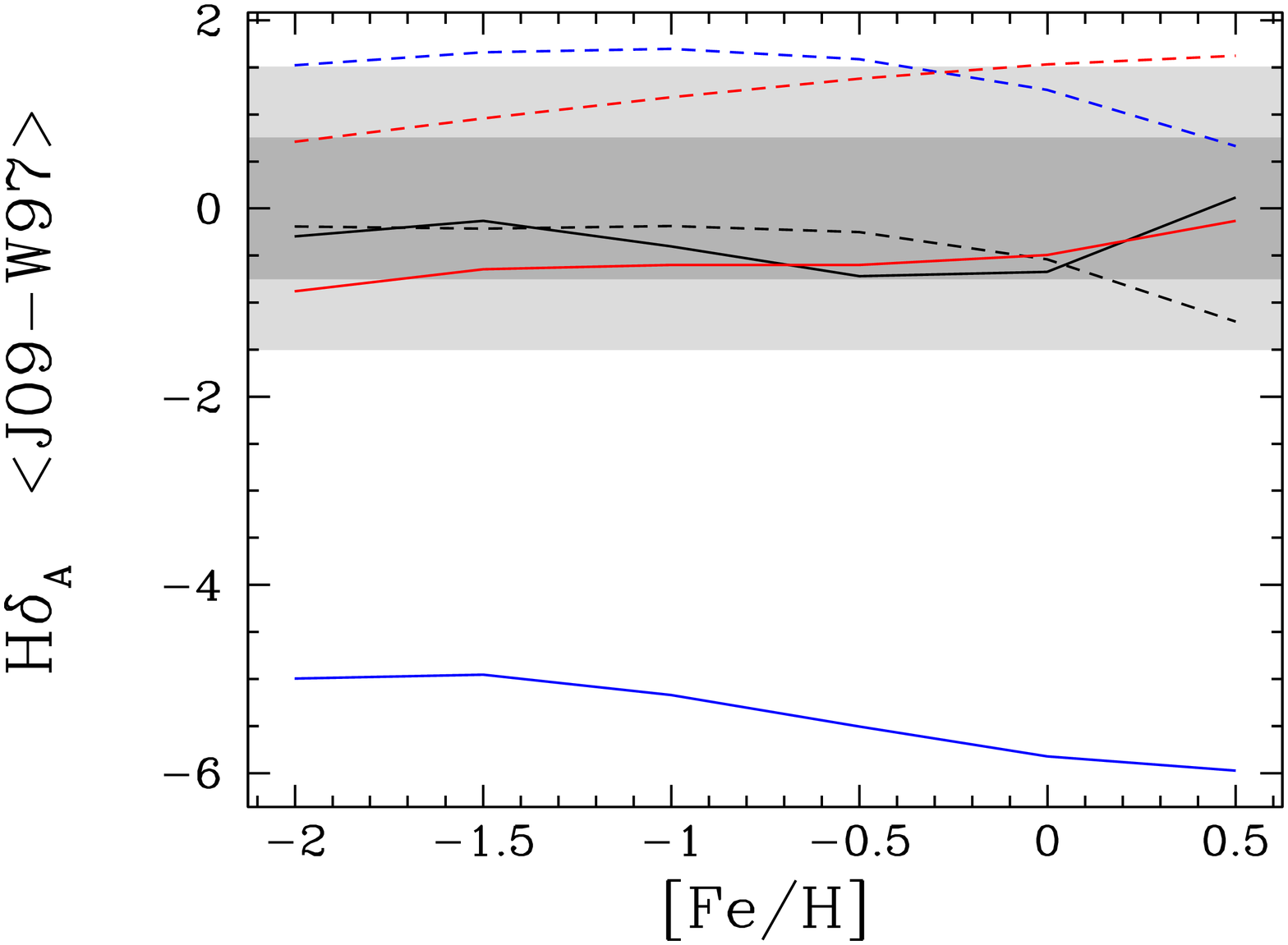}\includegraphics[scale=0.21,angle=-90]{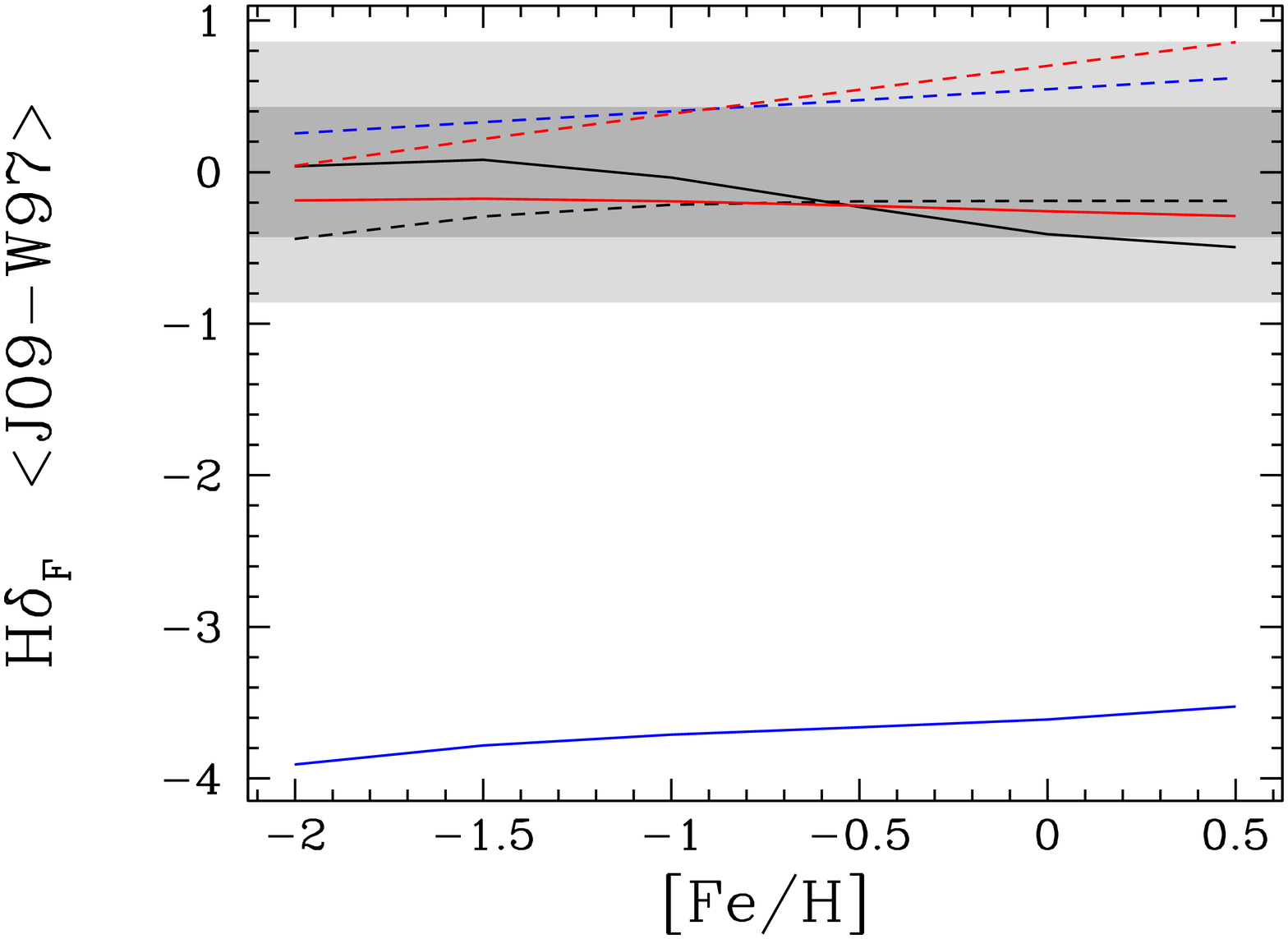}\includegraphics[scale=0.21,angle=-90]{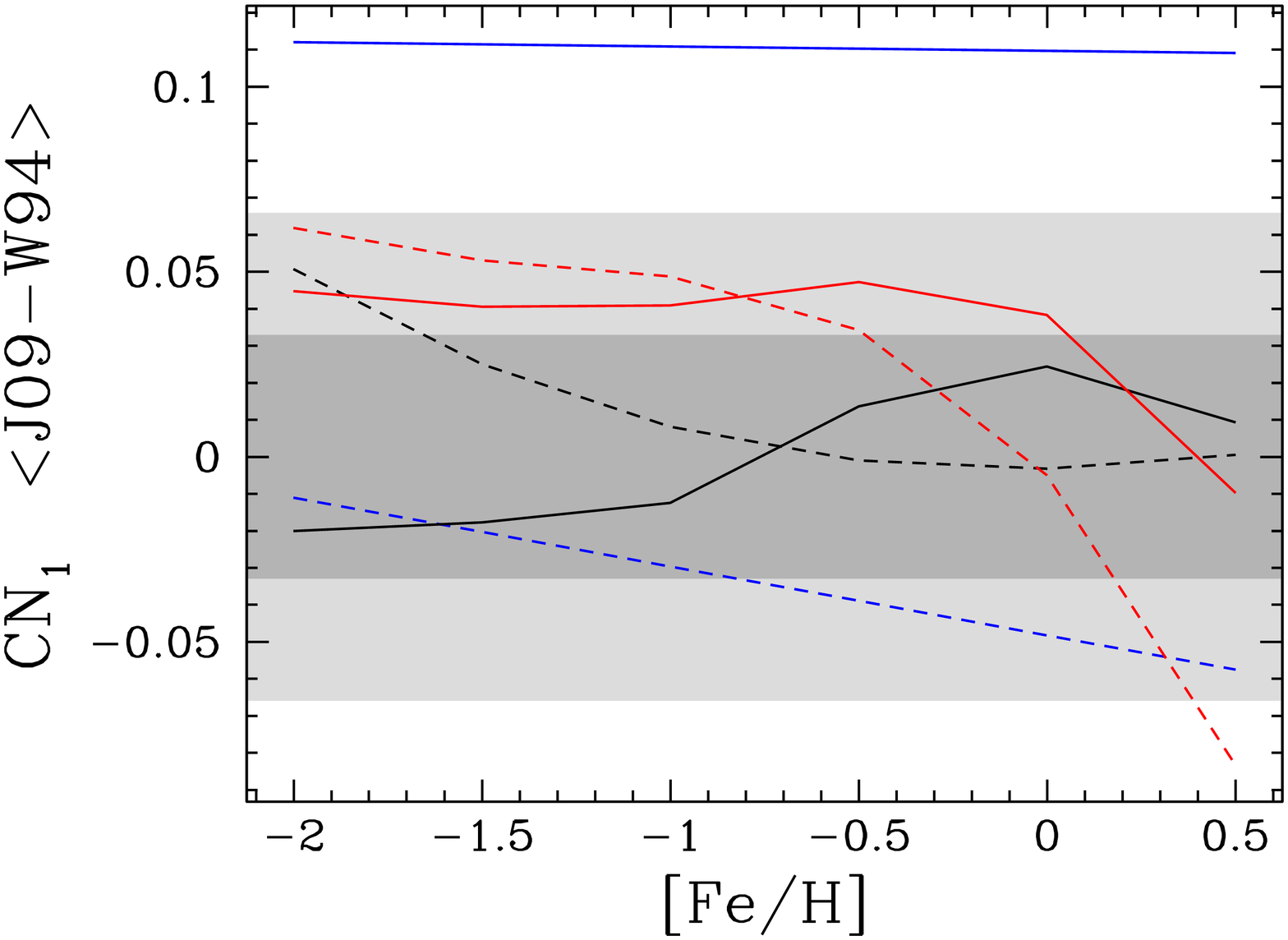}\\
\includegraphics[scale=0.21,angle=-90]{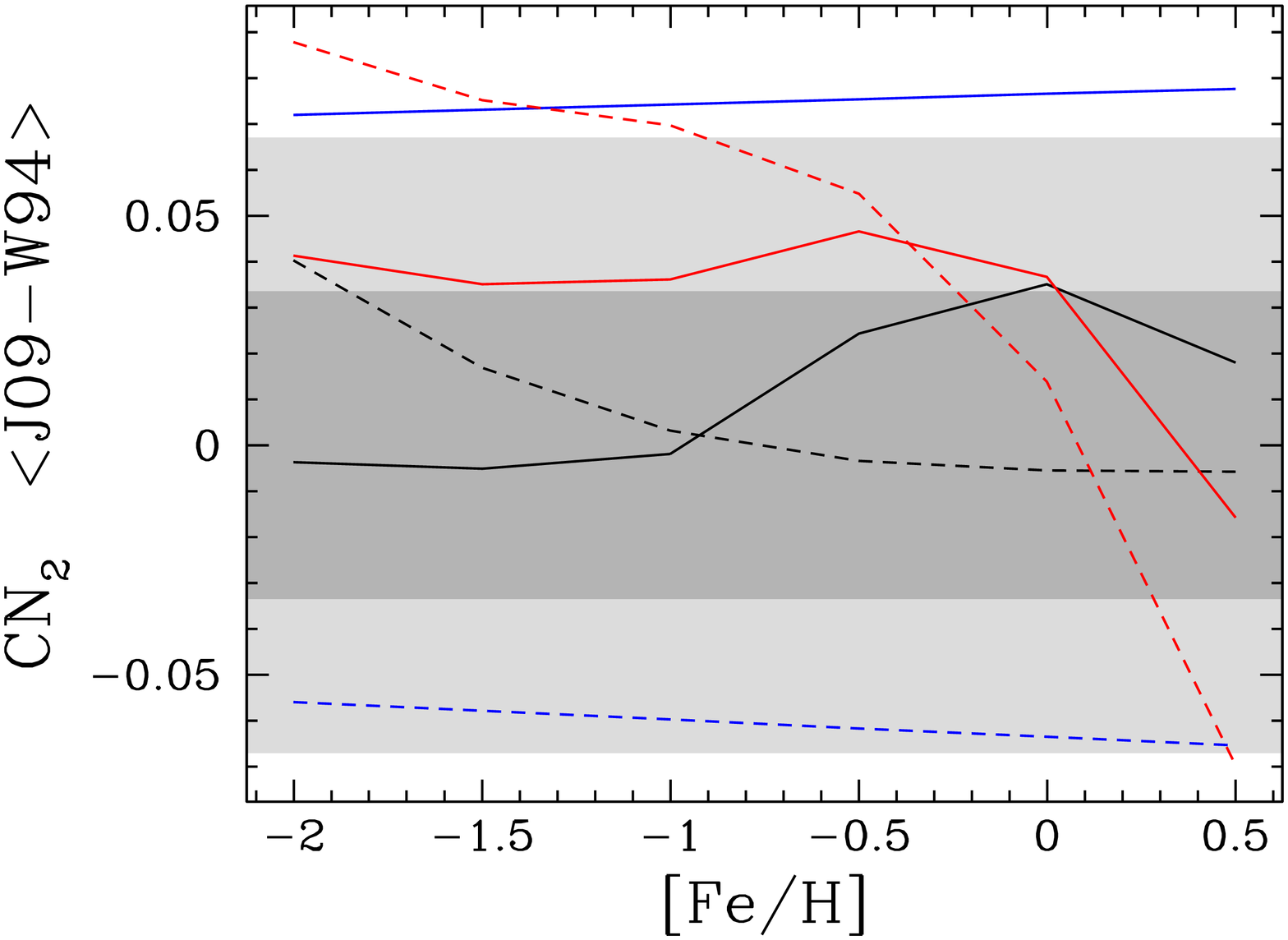}\includegraphics[scale=0.21,angle=-90]{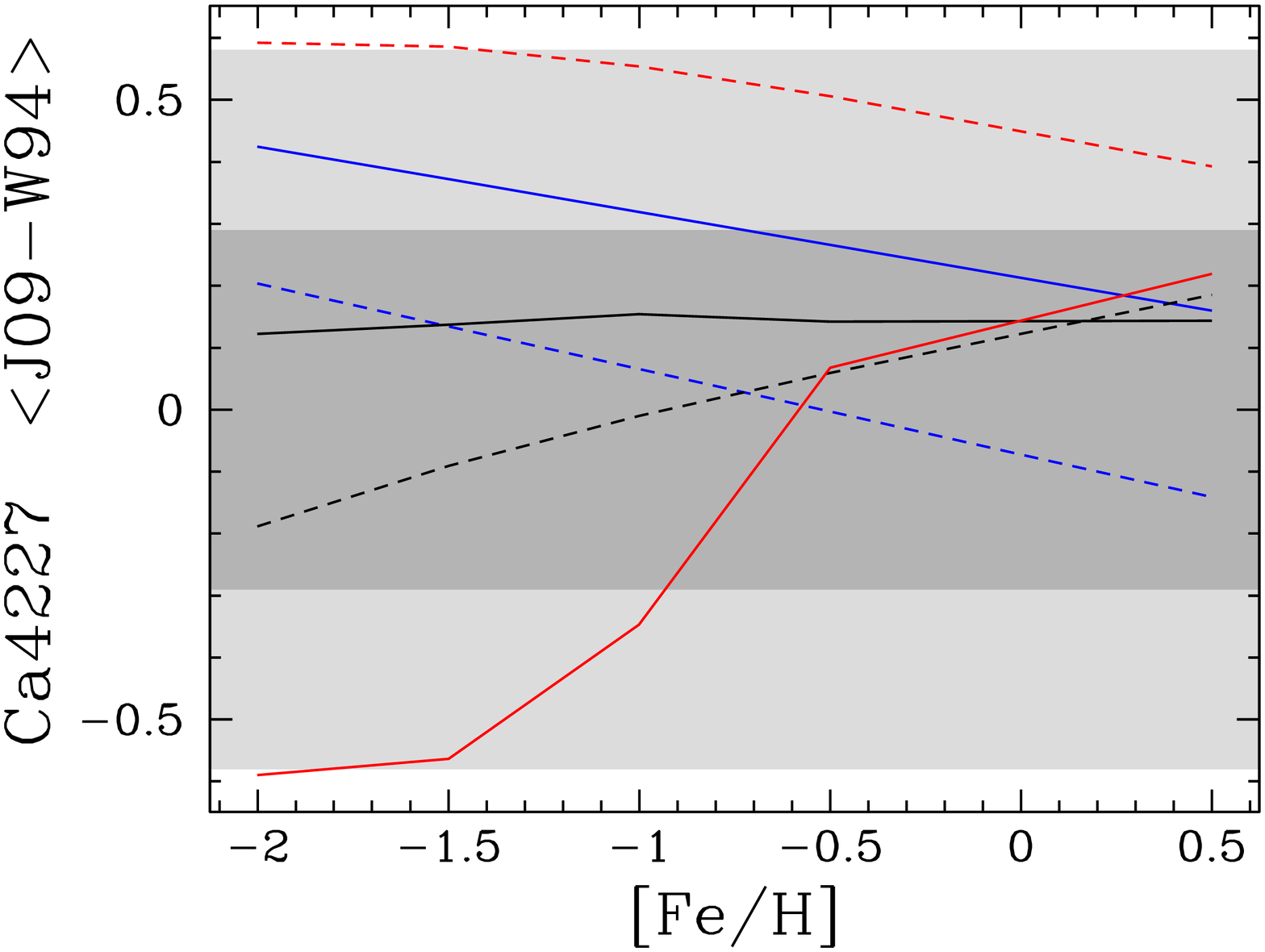}\includegraphics[scale=0.21,angle=-90]{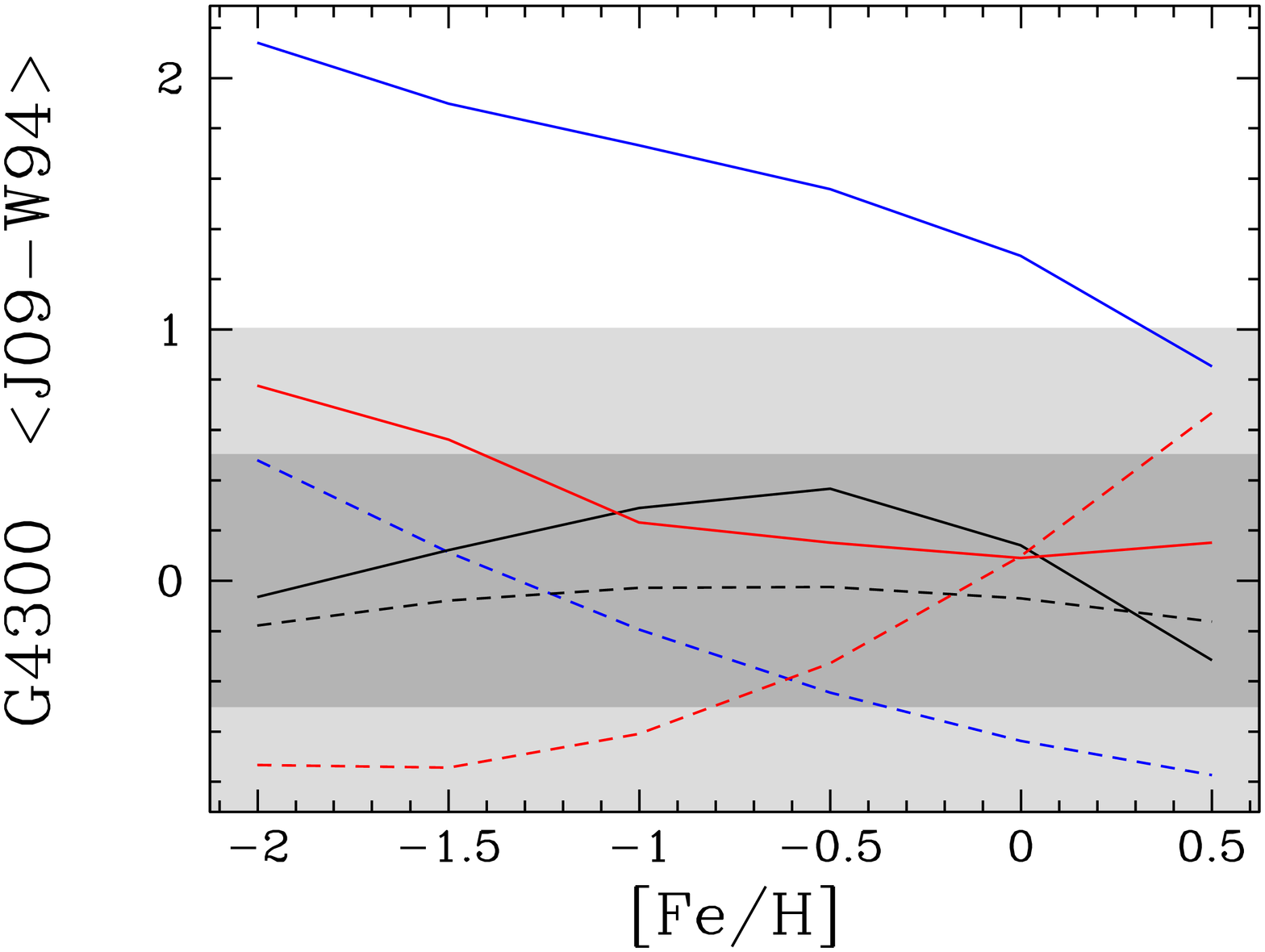}\\
\includegraphics[scale=0.21,angle=-90]{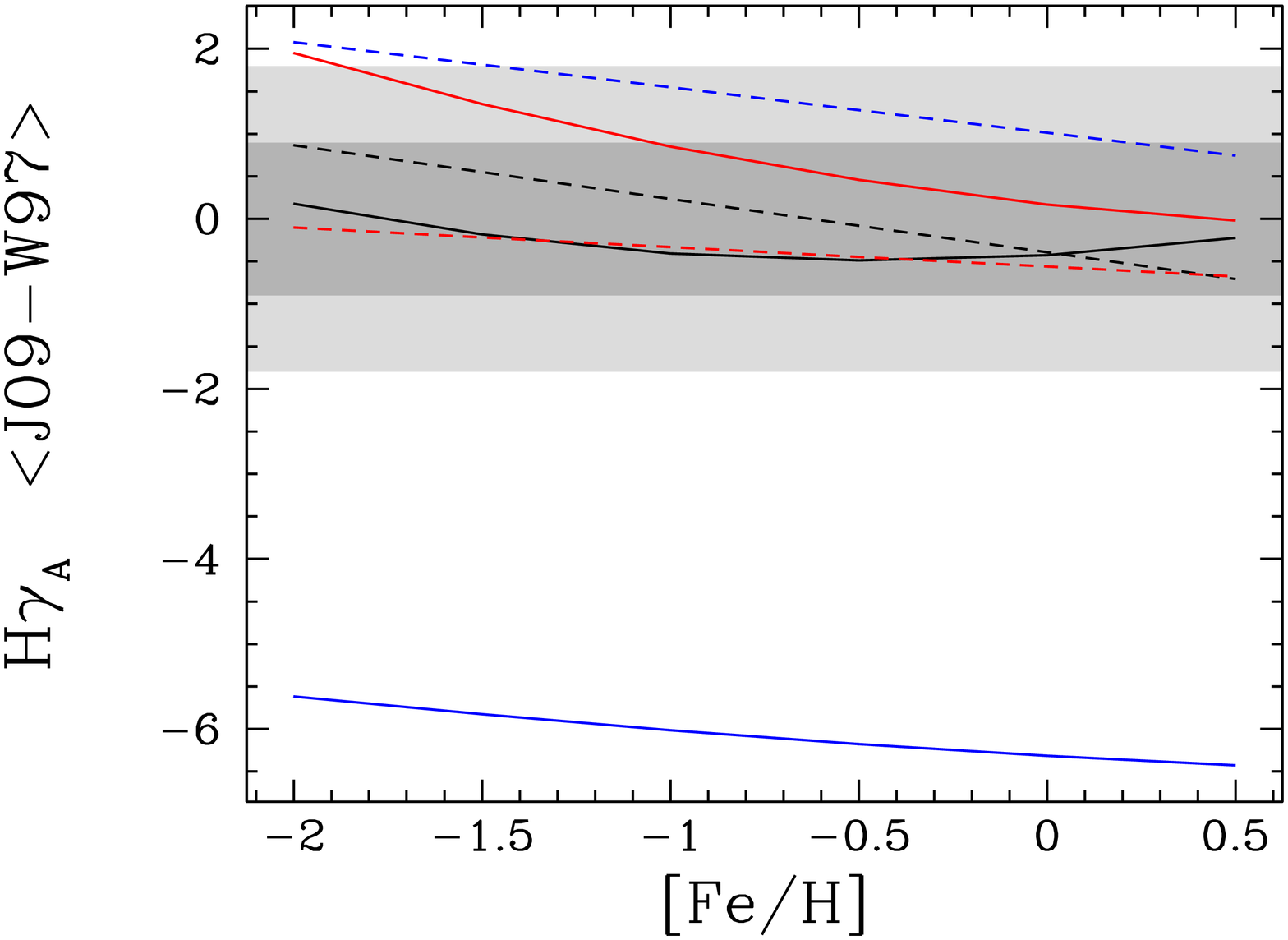}\includegraphics[scale=0.21,angle=-90]{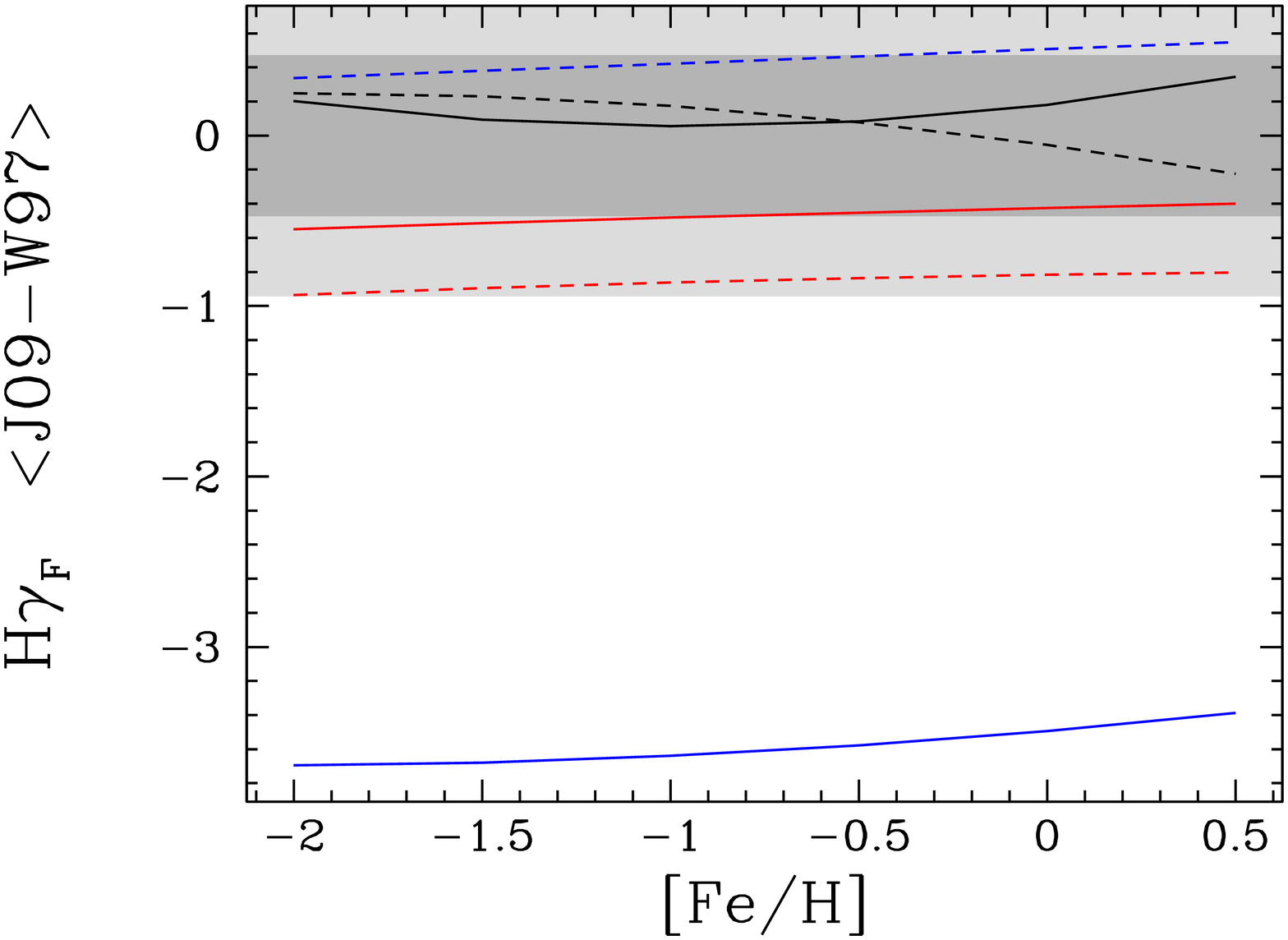}\includegraphics[scale=0.21,angle=-90]{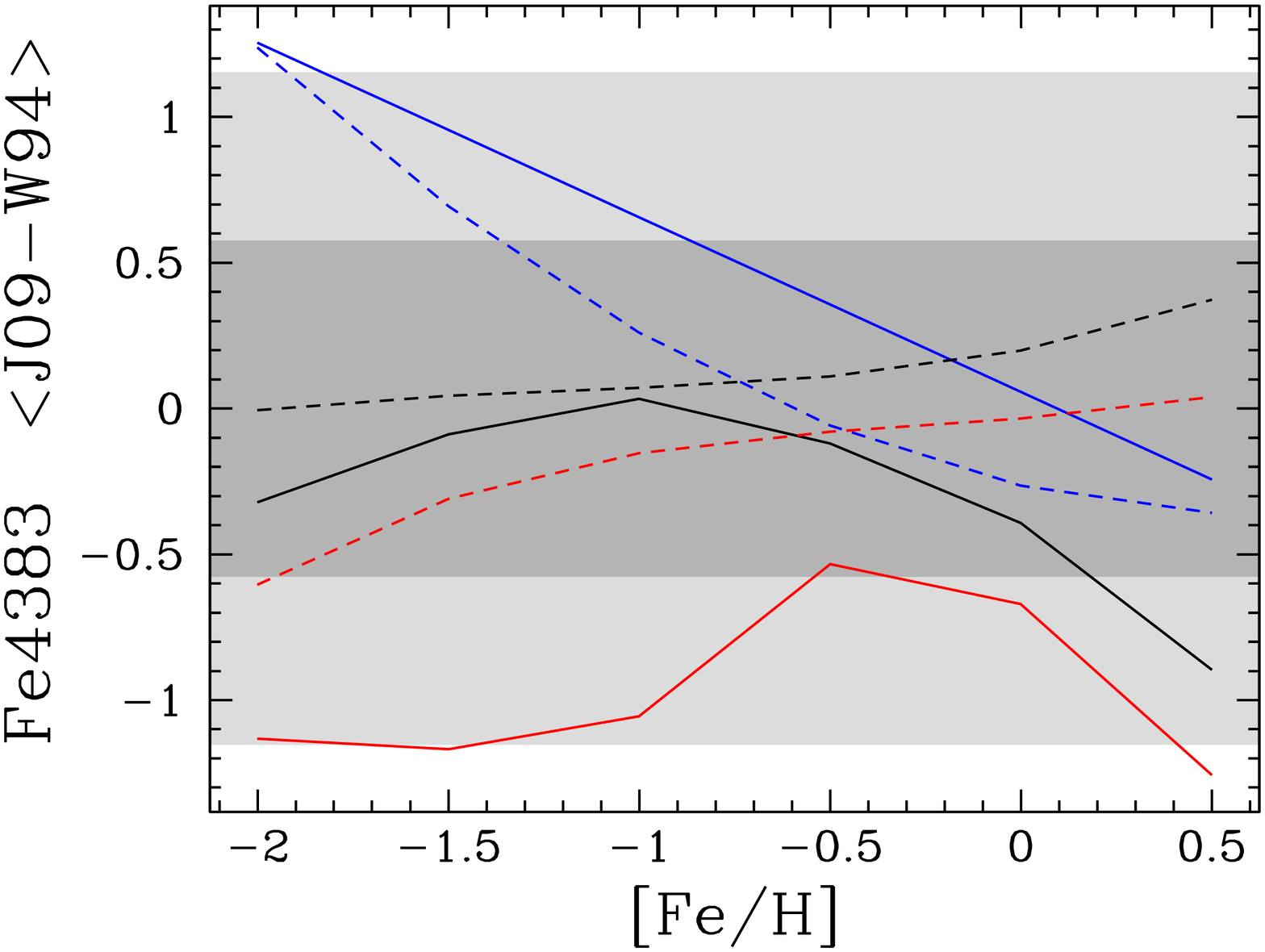}\\
\includegraphics[scale=0.21,angle=-90]{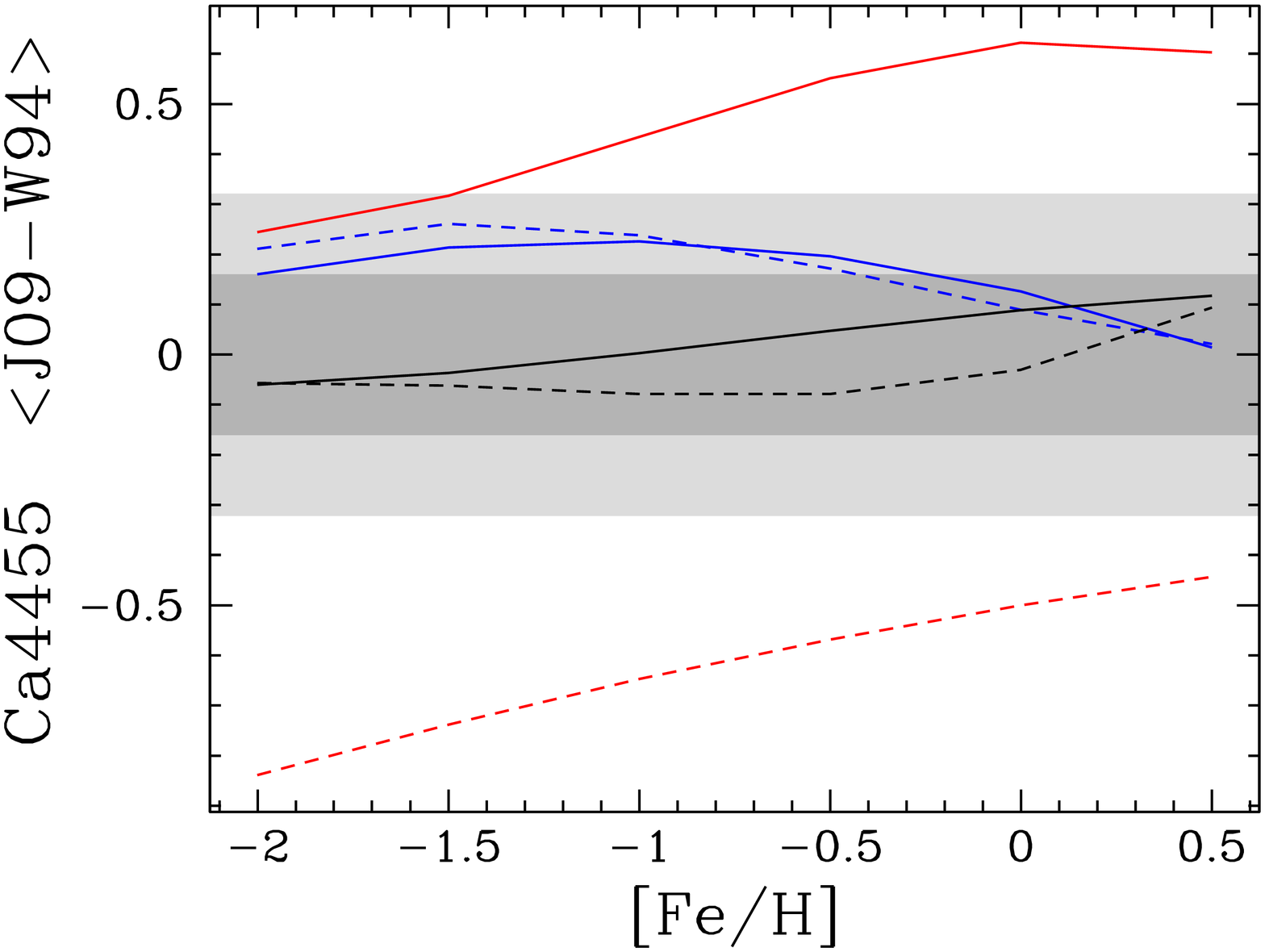}\includegraphics[scale=0.21,angle=-90]{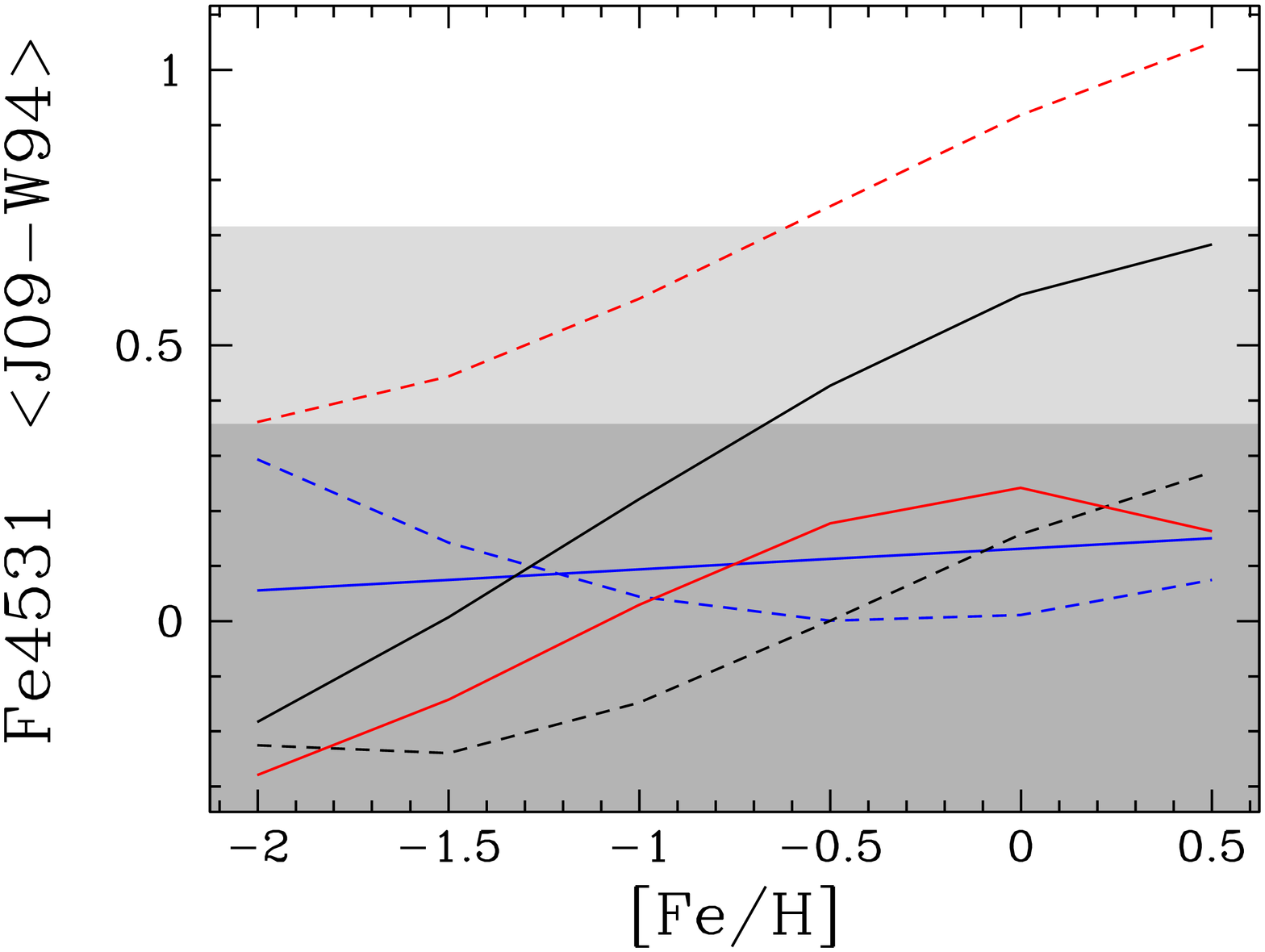}\includegraphics[scale=0.21,angle=-90]{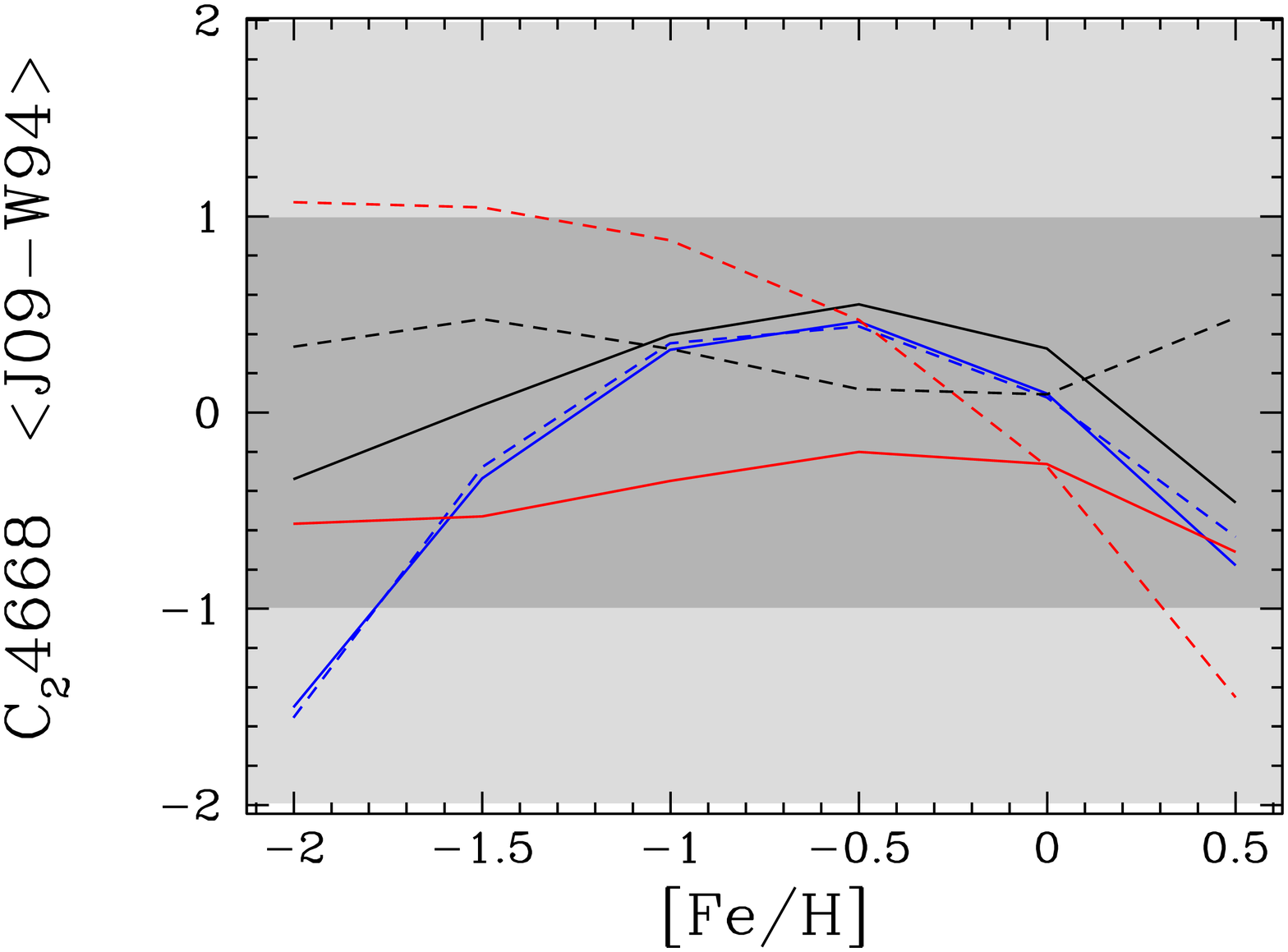}\\
\caption{Comparisons between the fitting functions of this work (referred to as J09) and \citet{worthey94} (W94), and with \citet{worthey97} (W97) for H$\delta_A$, H$\delta_F$, H$\gamma_A$ and H$\gamma_F$. The panels show the difference J09-W94/W97 as a function of metallicity for each Lick index. The comparisons are made for \emph{Giants} ($\log g=2.0$, solid lines) and \emph{Dwarfs} ($\log g=4.5$, dashed lines). The different colors correspond to 
the different bins of $\theta/T_{\rm eff}$ space, with limits stated in Table~\ref{complimits}, where the average 
difference has been computed, blue for the \emph{Warm}, black for the \emph{Intermediate} and red for the \emph{Cold} temperature bin. 
Fitting function residuals in terms of 1rms (dark grey shaded areas) and 2rms levels (light grey shaded areas) are indicated. 
The errors are represented by the combined 
errors of the \emph{MILES} and Lick/IDS libraries in quadrature (for more on the errors see Section~\ref{tierr}). }
\label{WFFcomp}
\end{minipage}
\end{figure*}
\begin{figure*}
\begin{minipage}{17cm}
\centering
\includegraphics[scale=0.21,angle=-90]{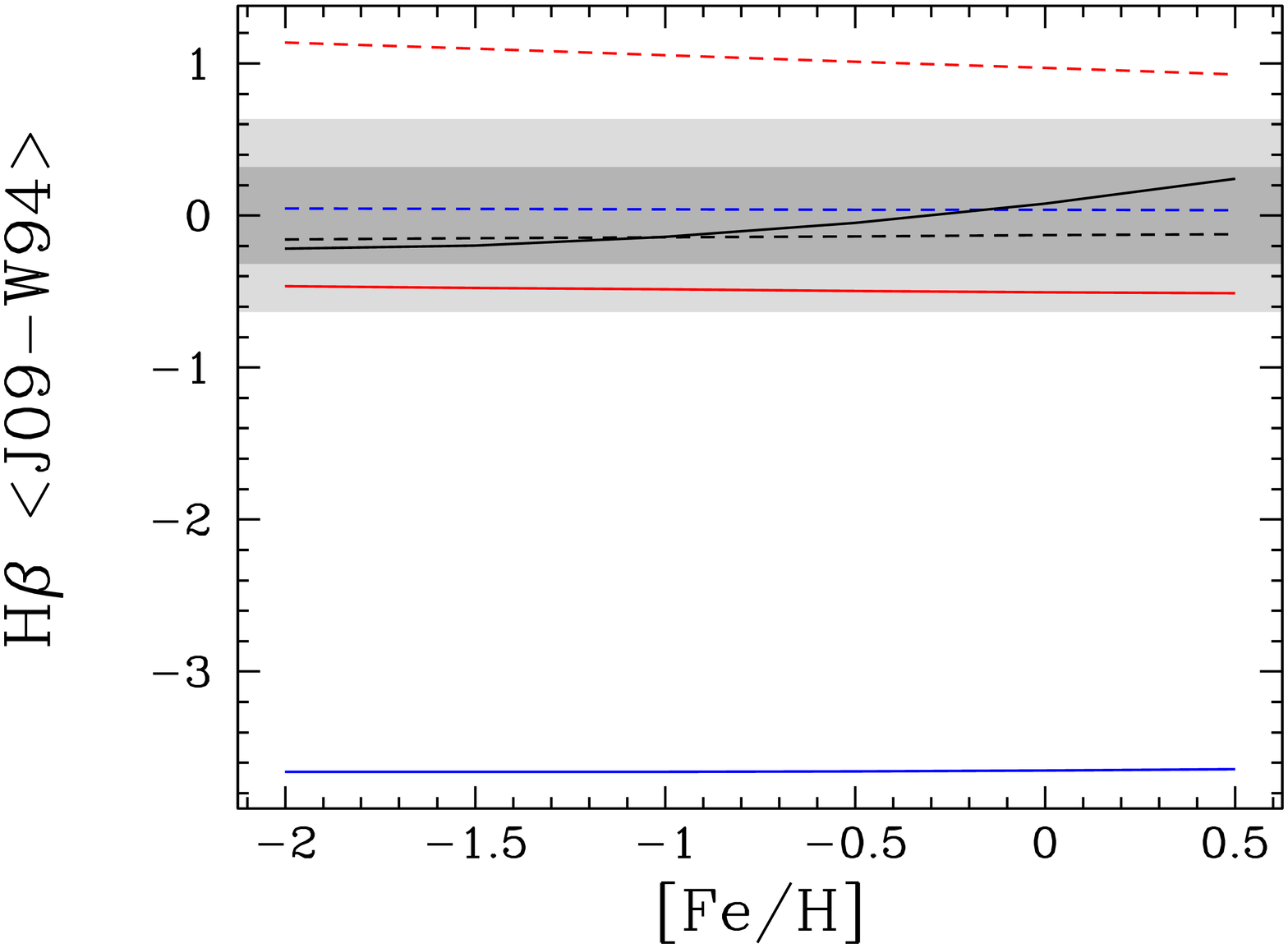}\includegraphics[scale=0.21,angle=-90]{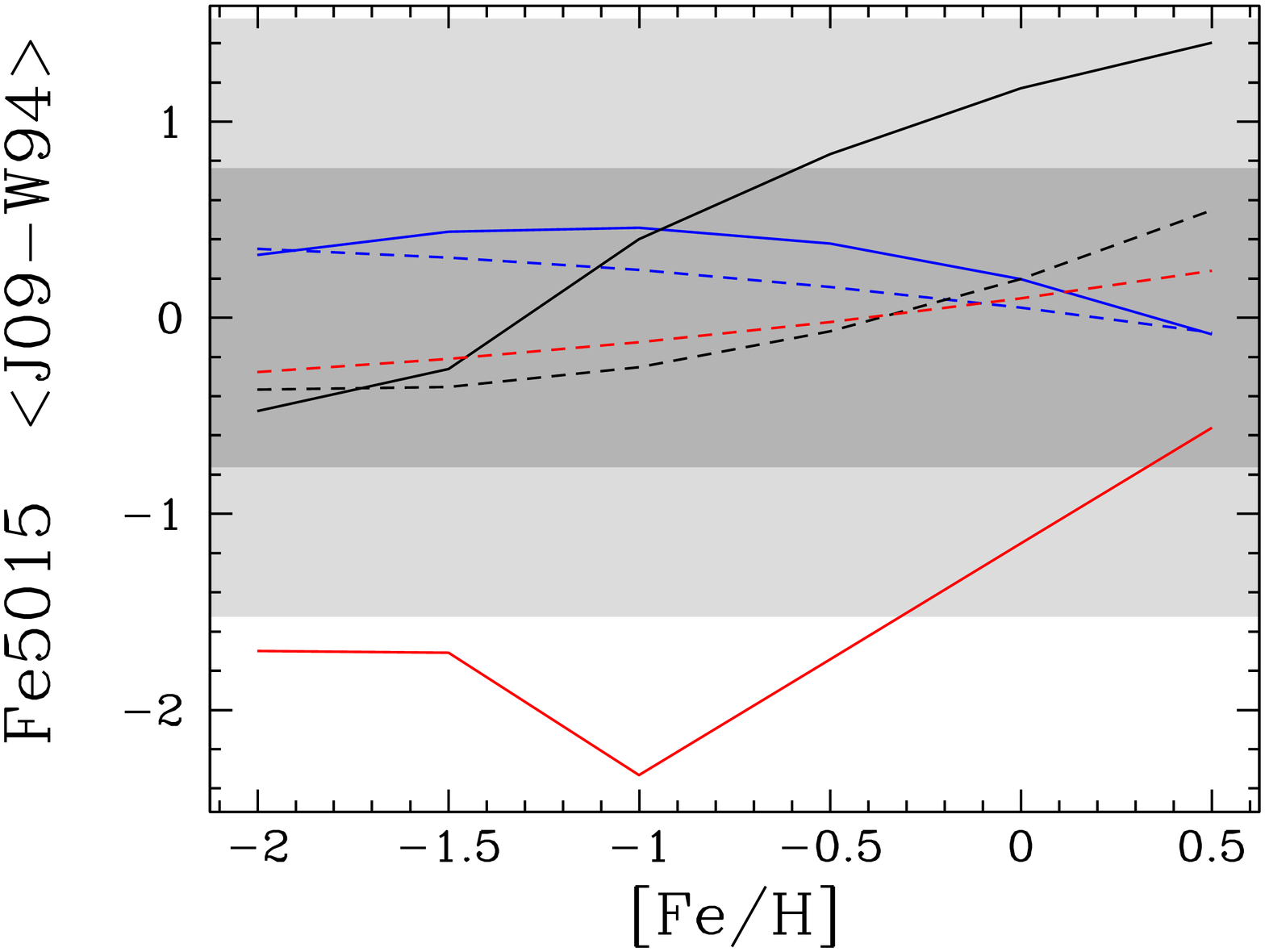}\includegraphics[scale=0.21,angle=-90]{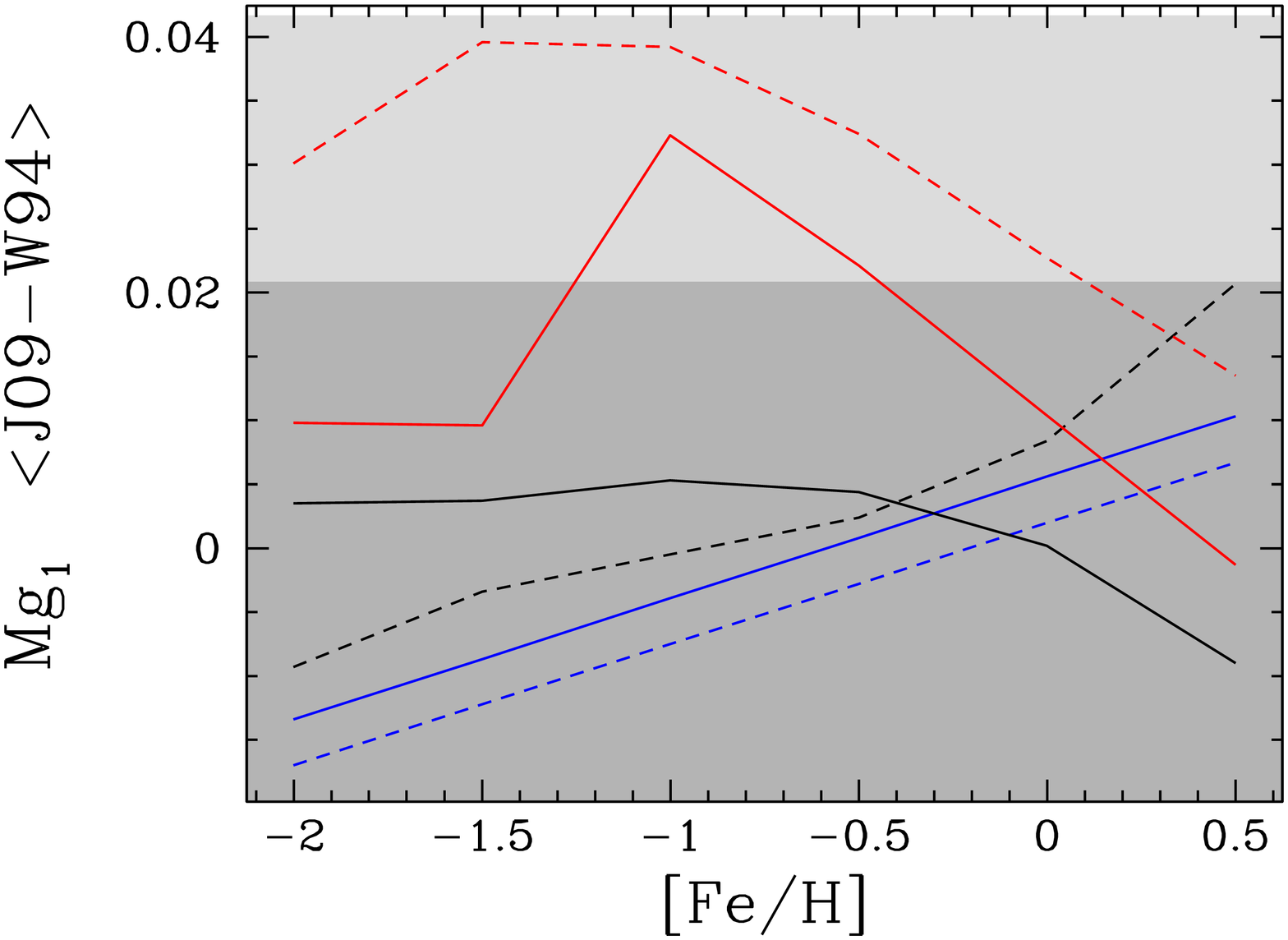}\\
\includegraphics[scale=0.21,angle=-90]{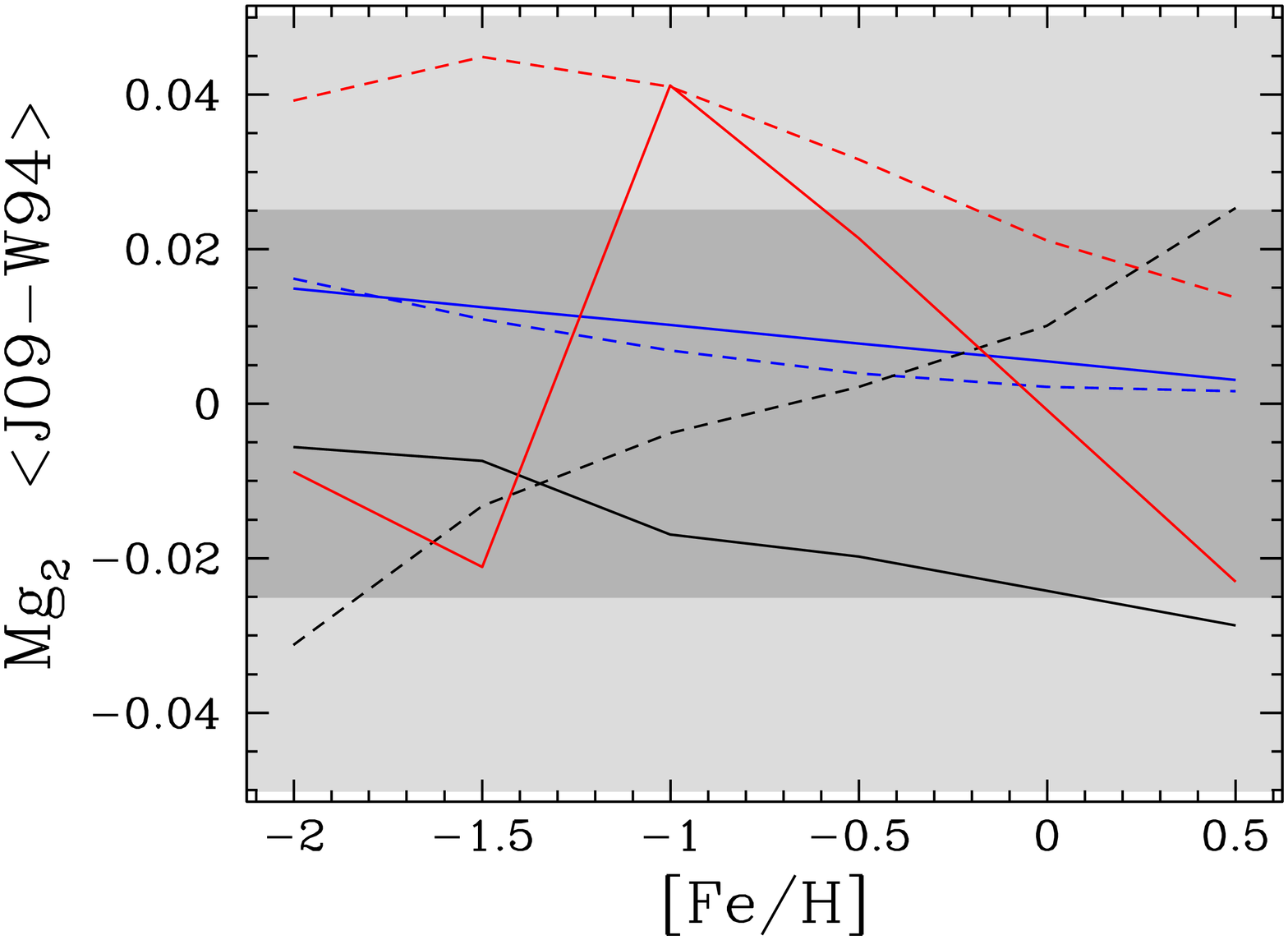}\includegraphics[scale=0.21,angle=-90]{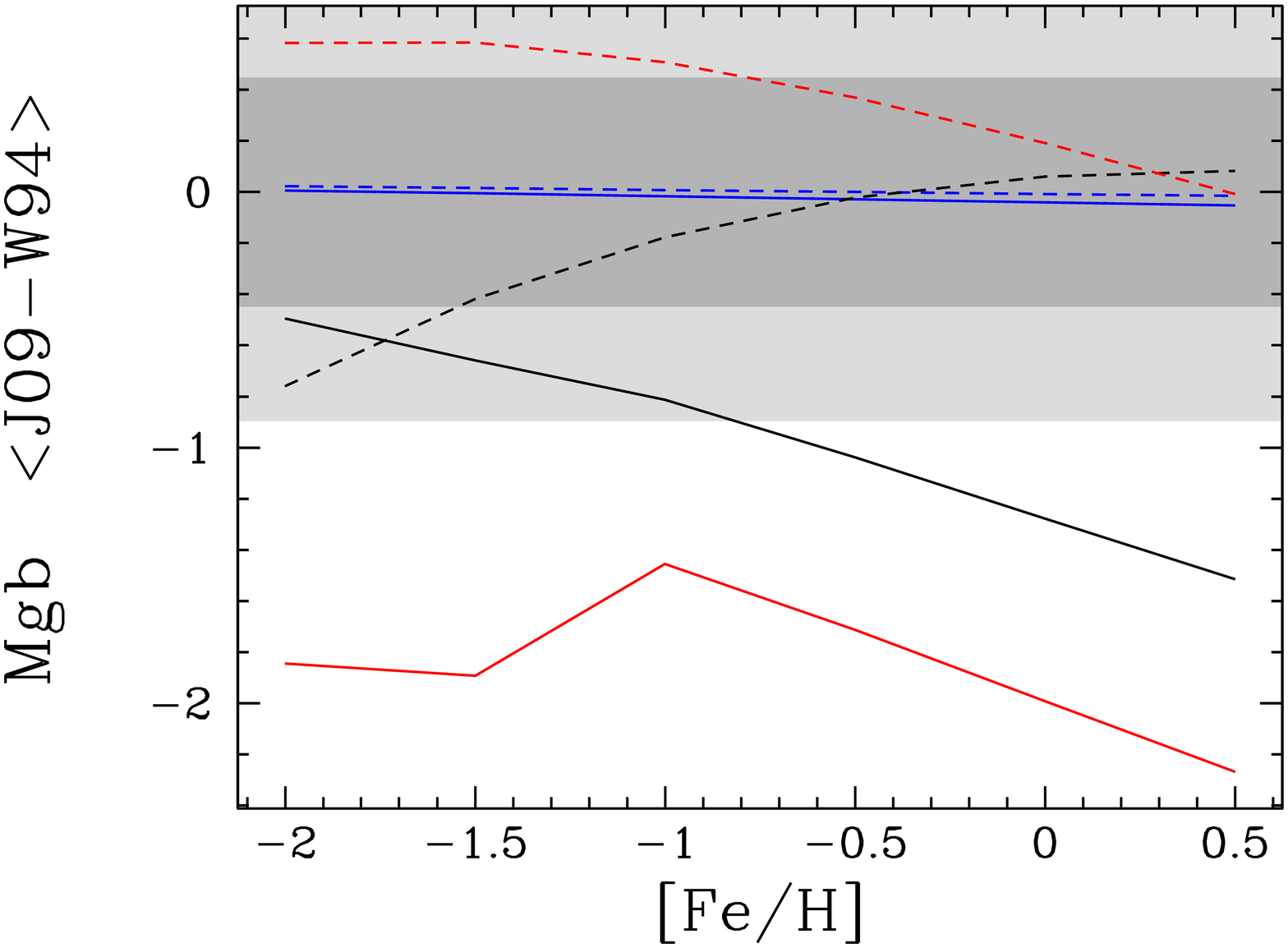}\includegraphics[scale=0.21,angle=-90]{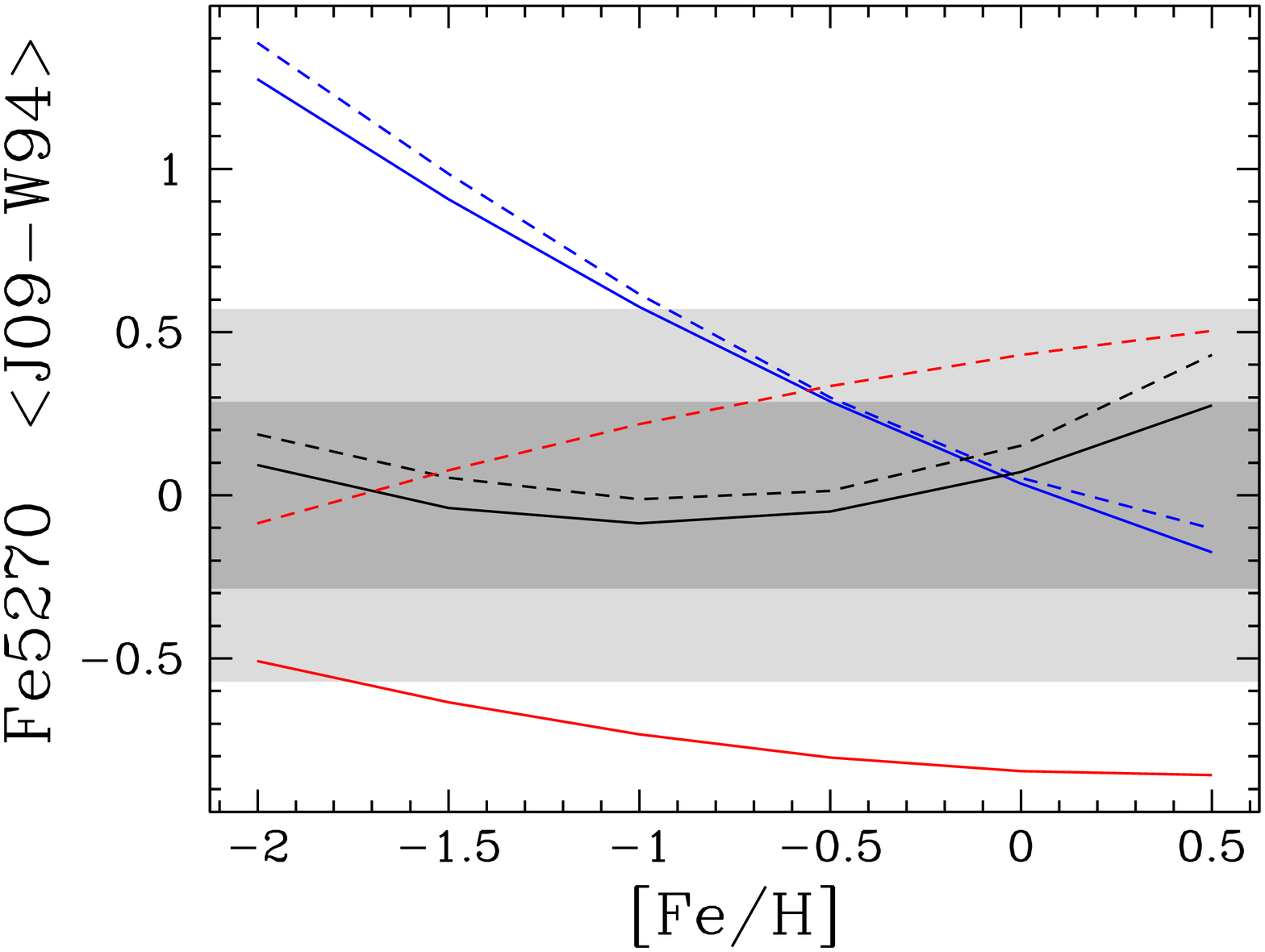}\\
\includegraphics[scale=0.21,angle=-90]{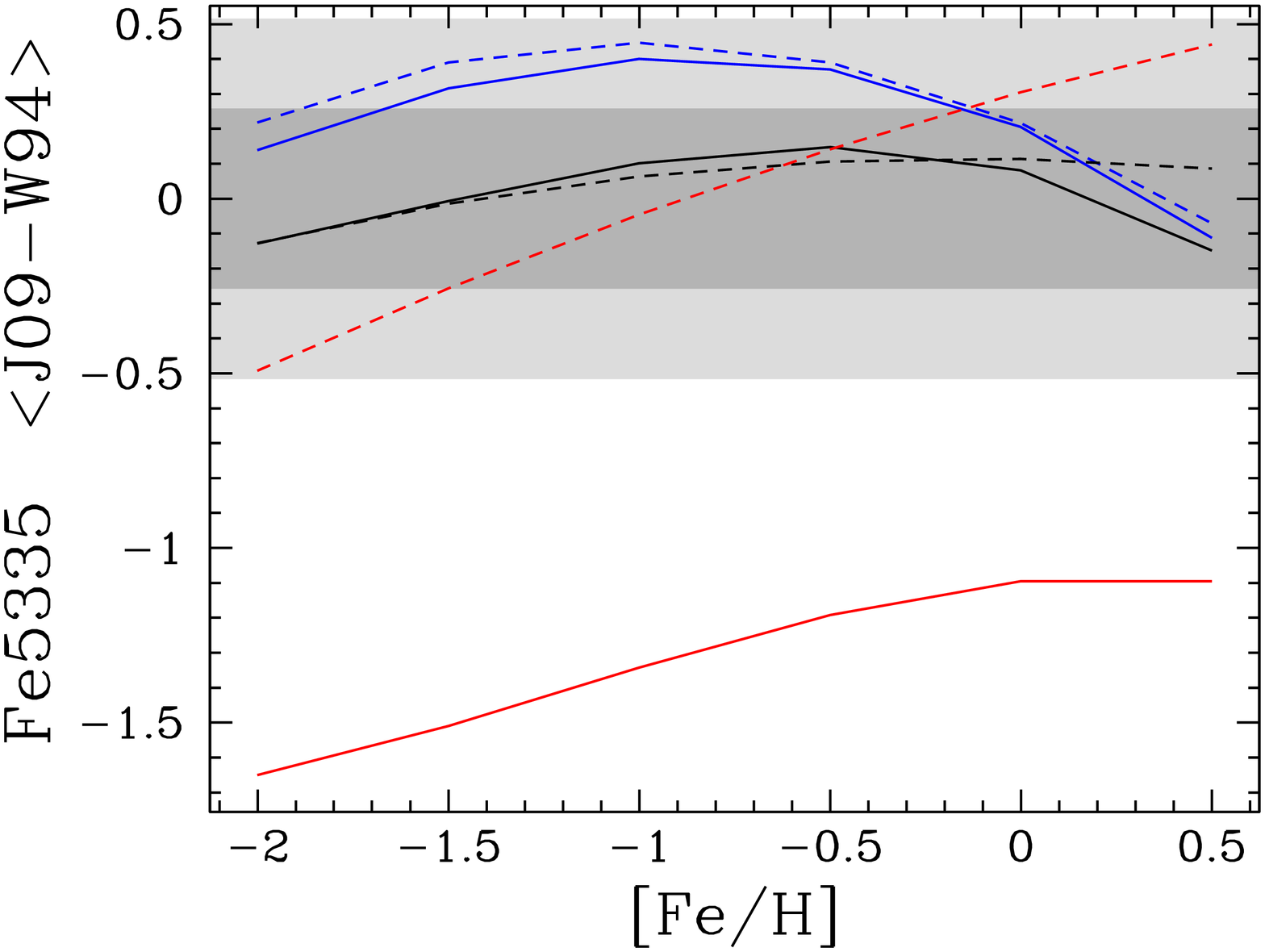}\includegraphics[scale=0.21,angle=-90]{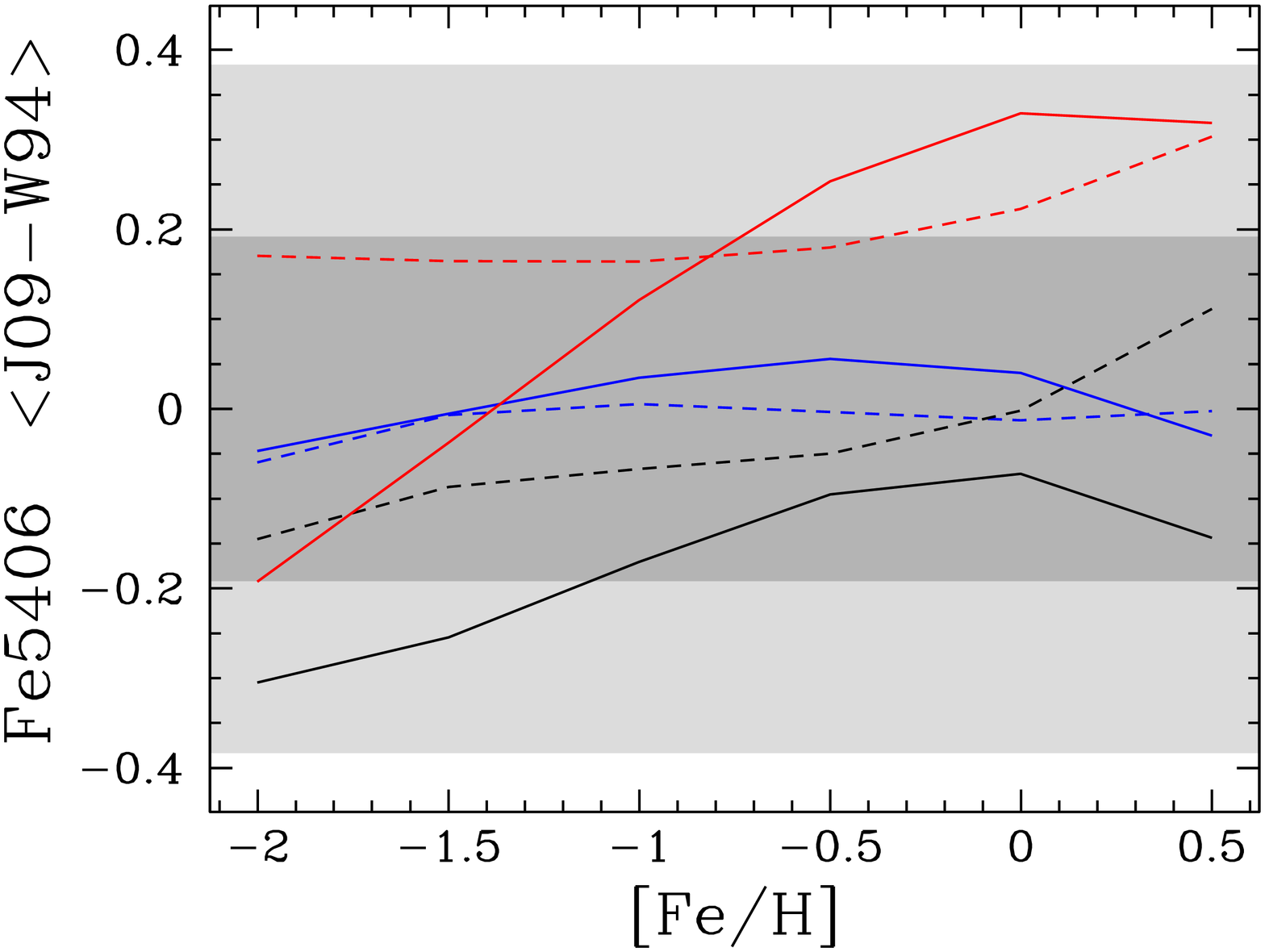}\includegraphics[scale=0.21,angle=-90]{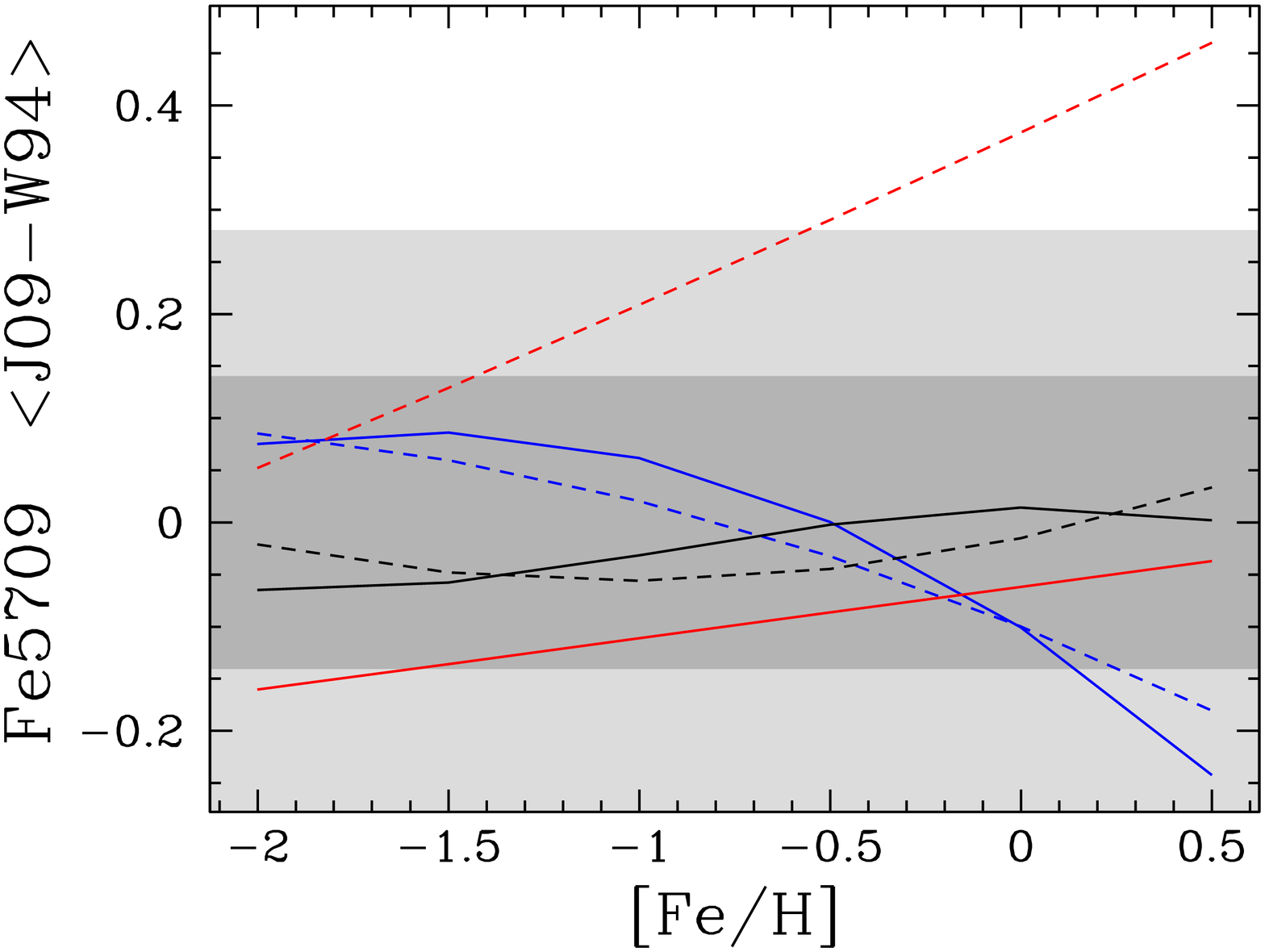}\\
\includegraphics[scale=0.21,angle=-90]{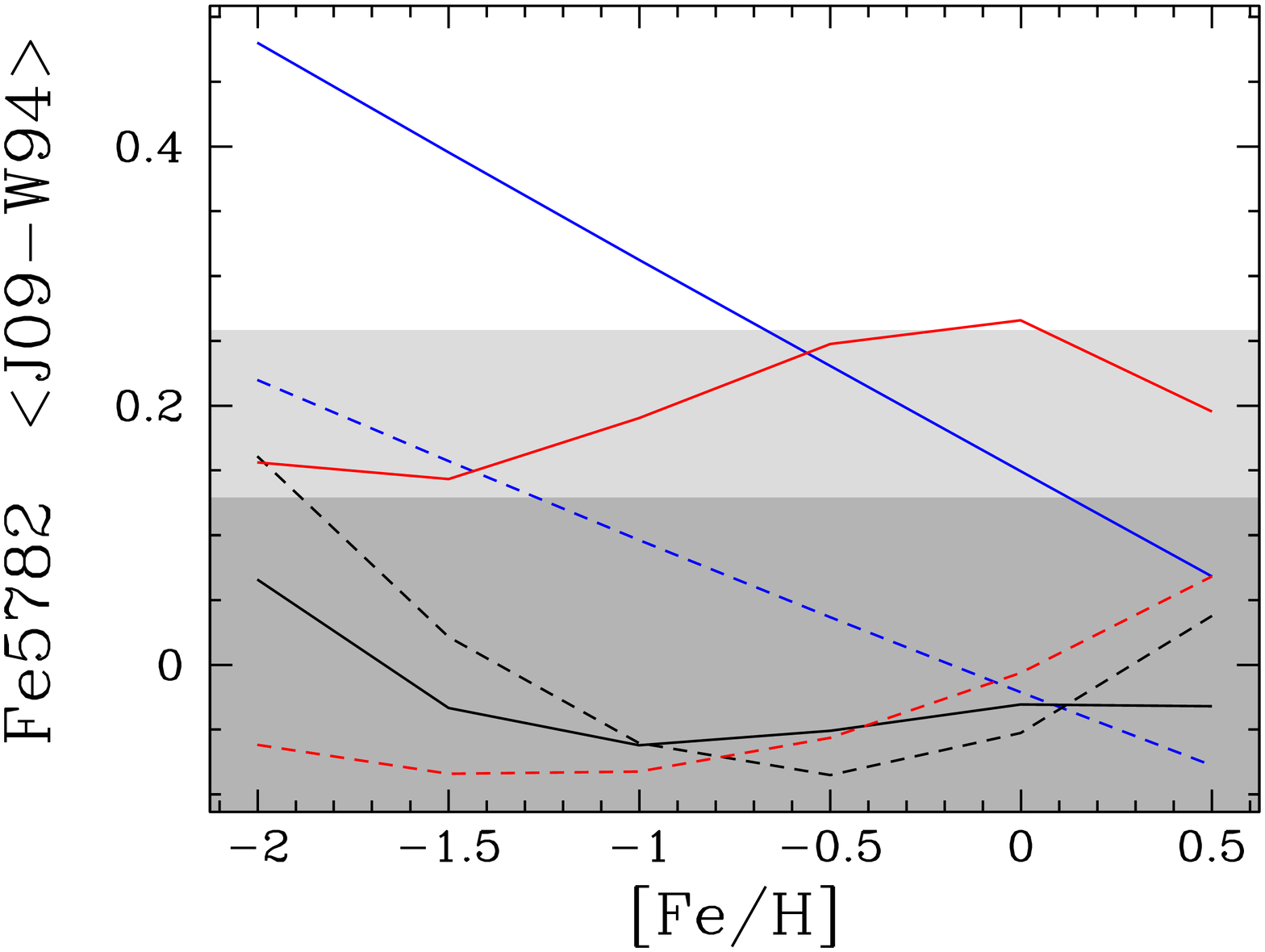}\includegraphics[scale=0.21,angle=-90]{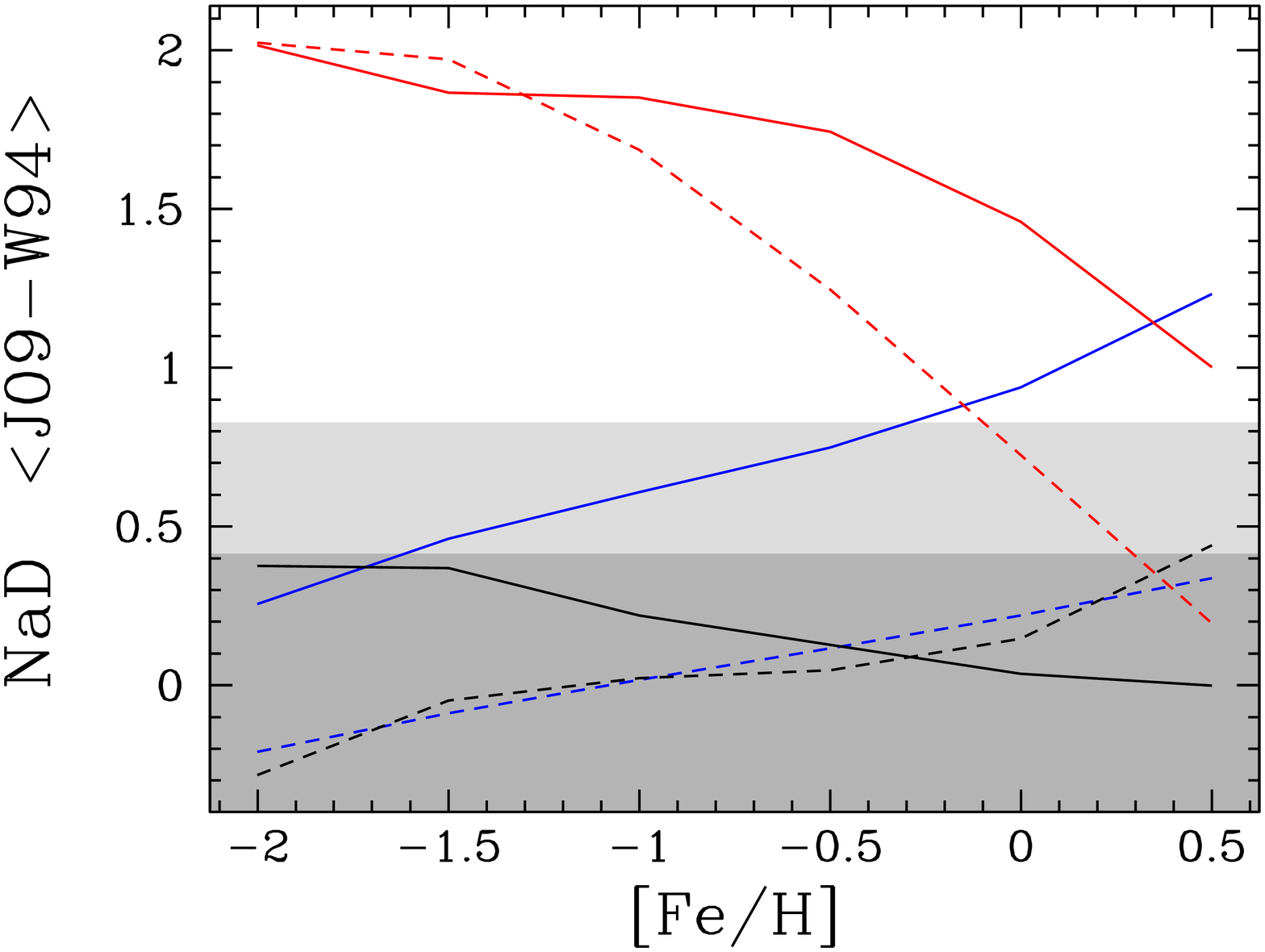}\includegraphics[scale=0.21,angle=-90]{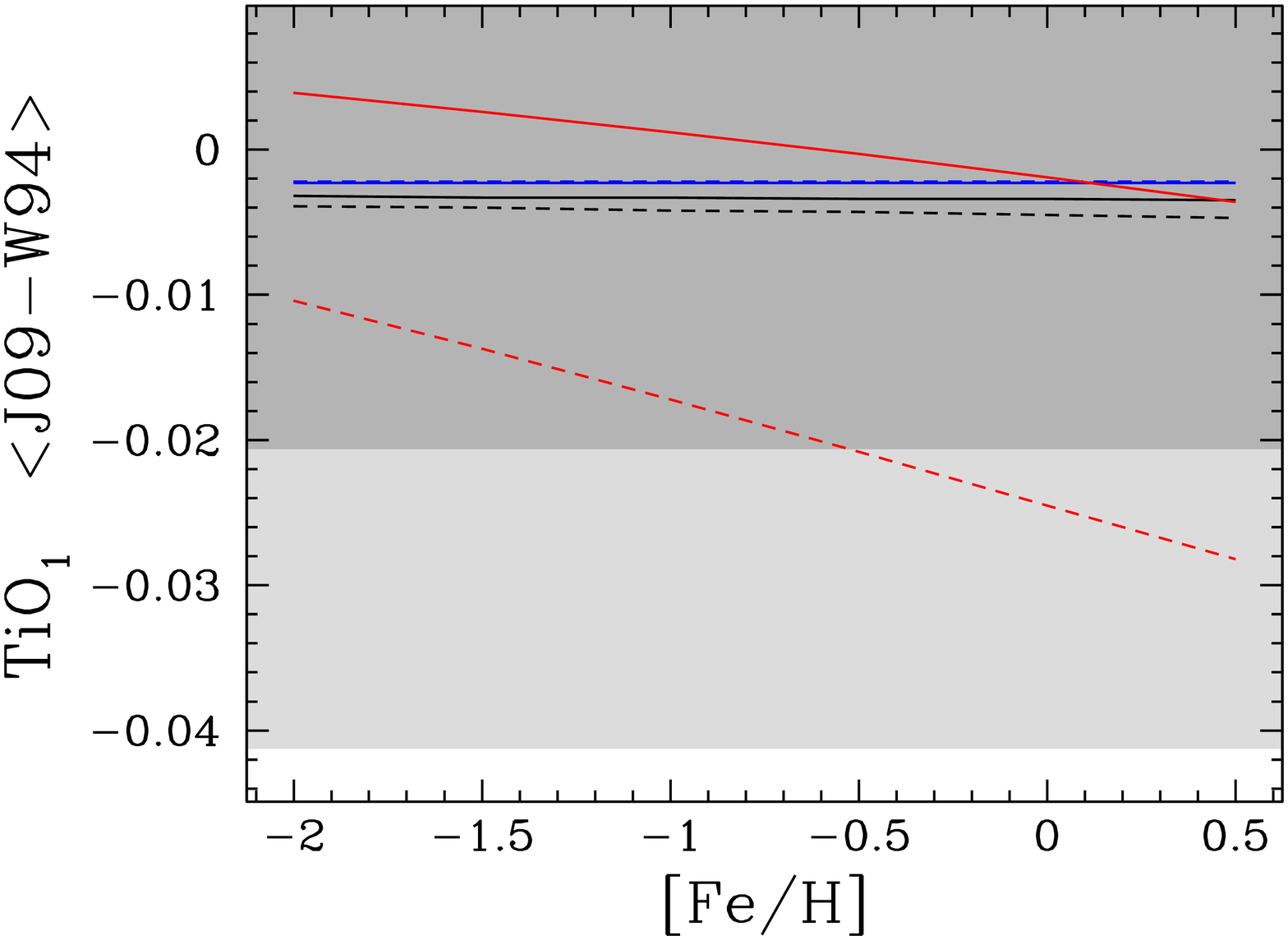}\\
\includegraphics[scale=0.21,angle=-90]{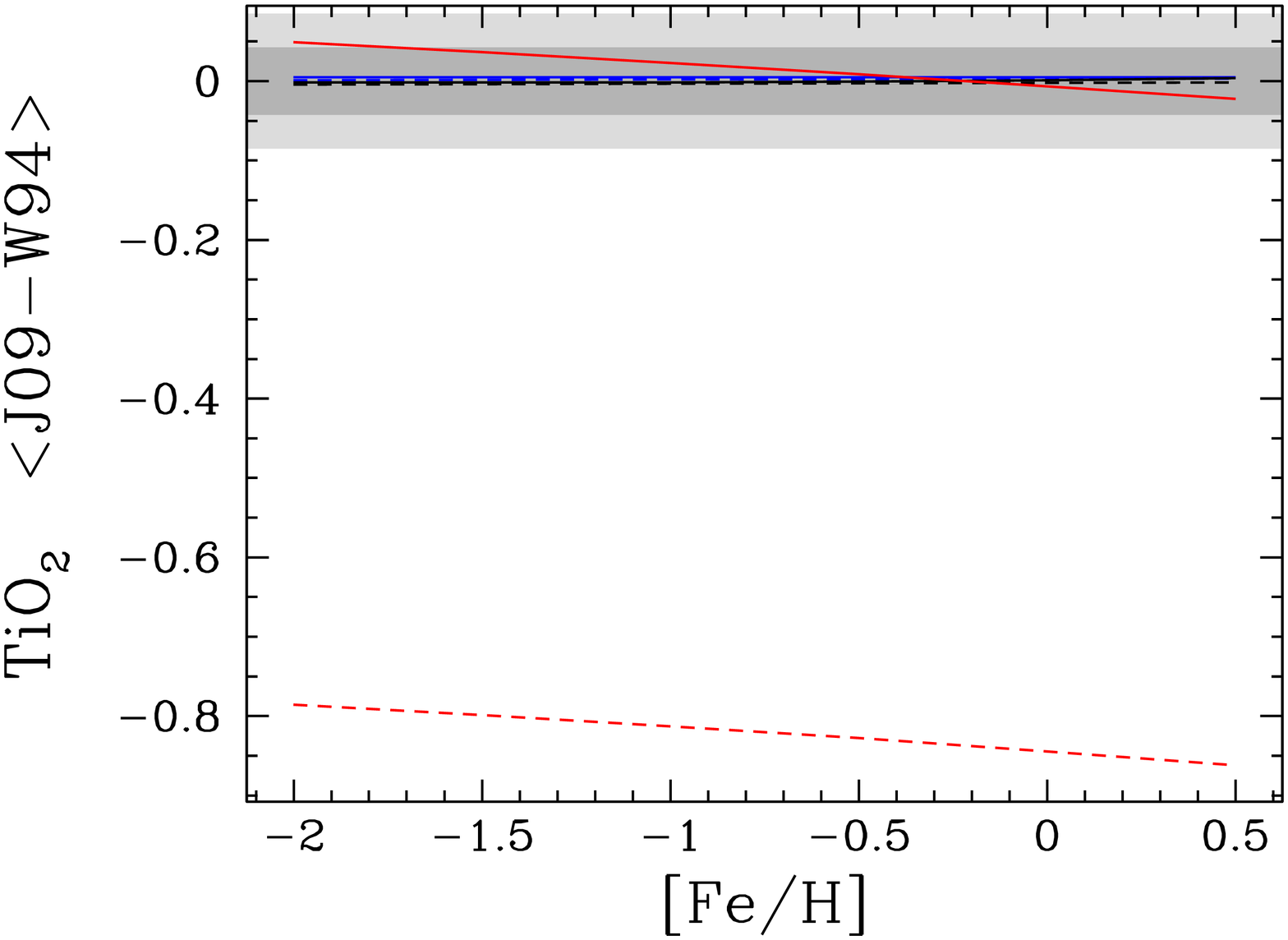}\\
\contcaption{}
\end{minipage}
\end{figure*}

\begin{figure*}
\begin{minipage}{17cm}
\centering
\includegraphics[scale=0.21,angle=-90]{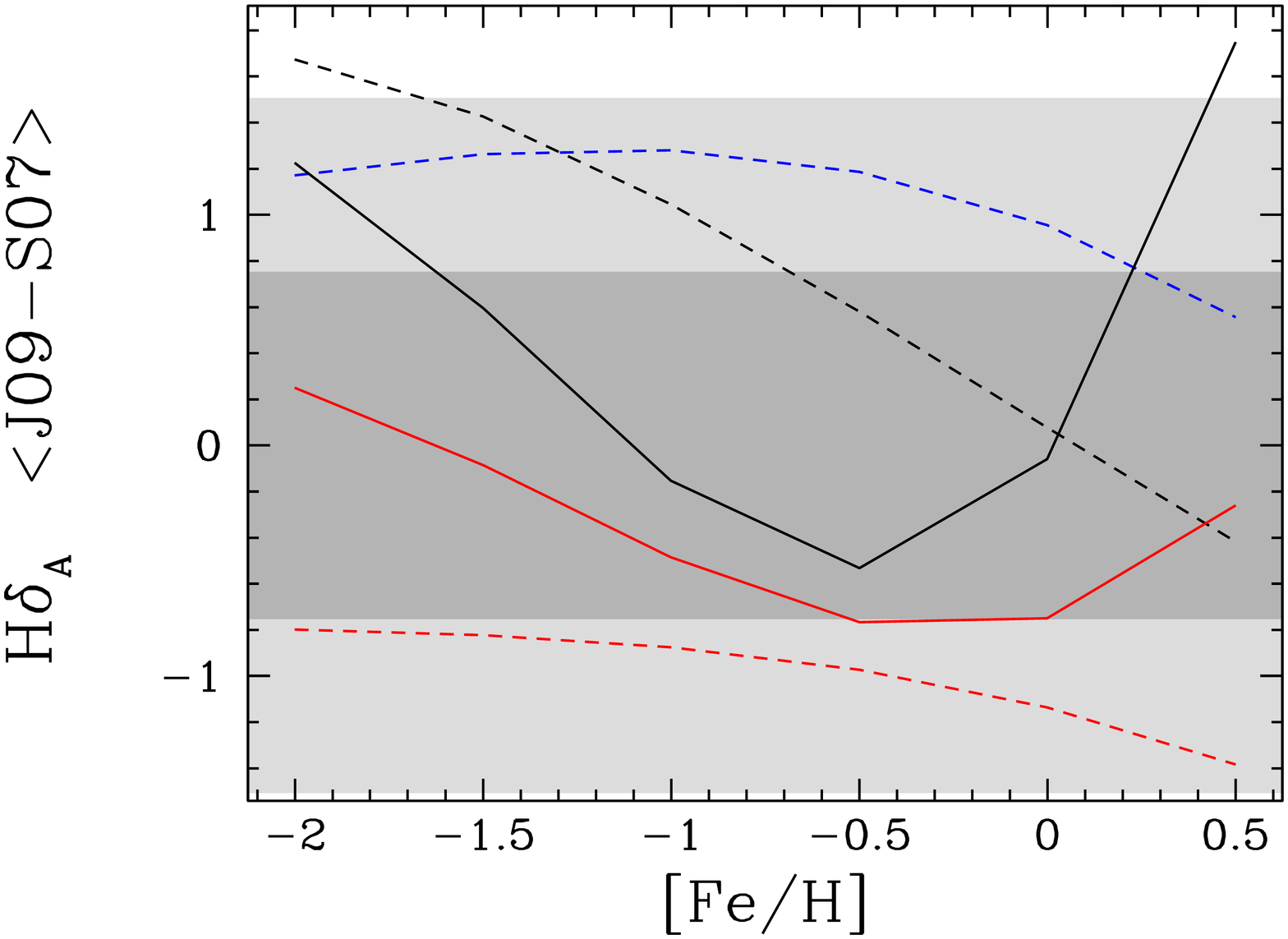}\includegraphics[scale=0.21,angle=-90]{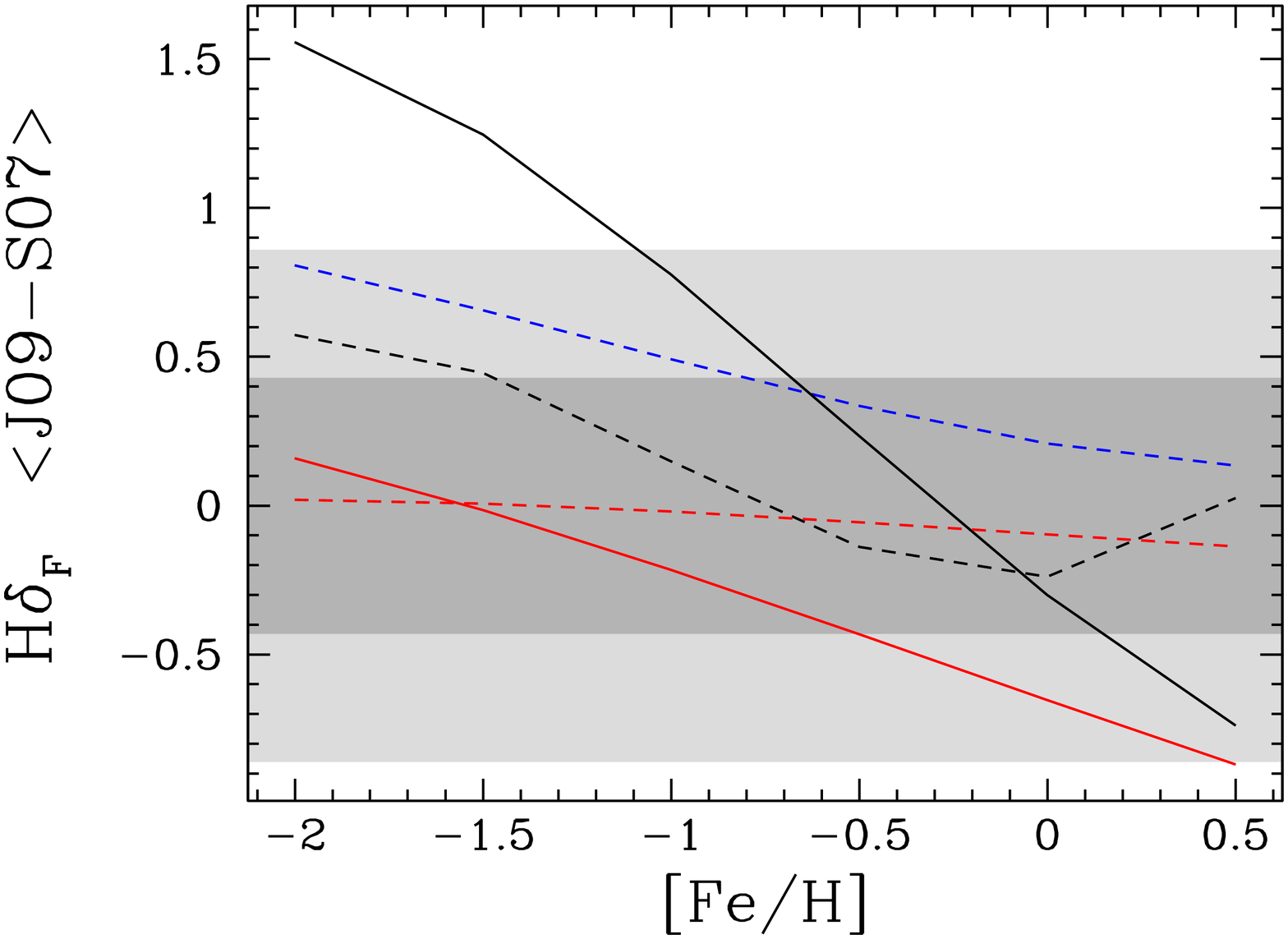}\includegraphics[scale=0.21,angle=-90]{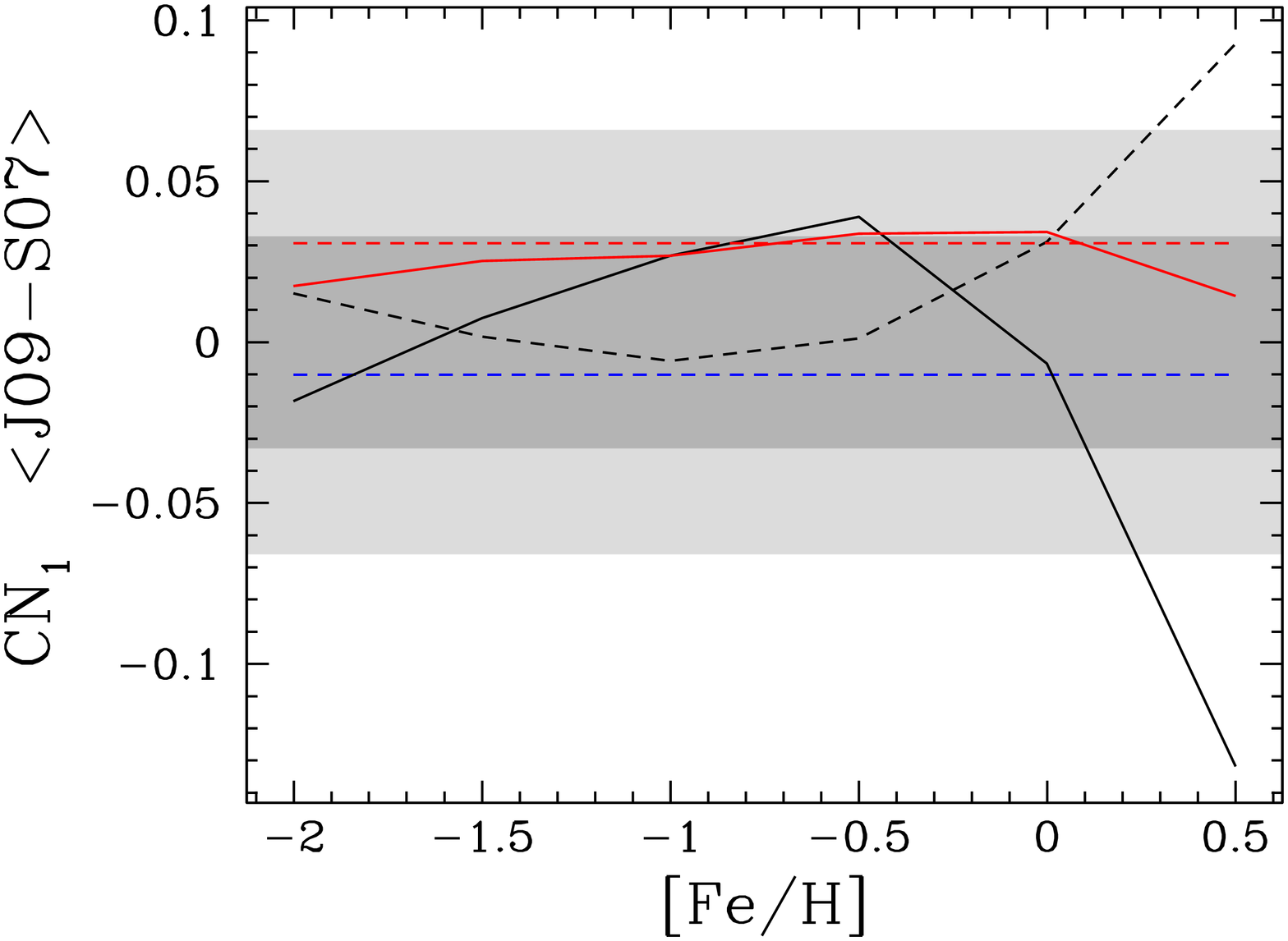}\\
\includegraphics[scale=0.21,angle=-90]{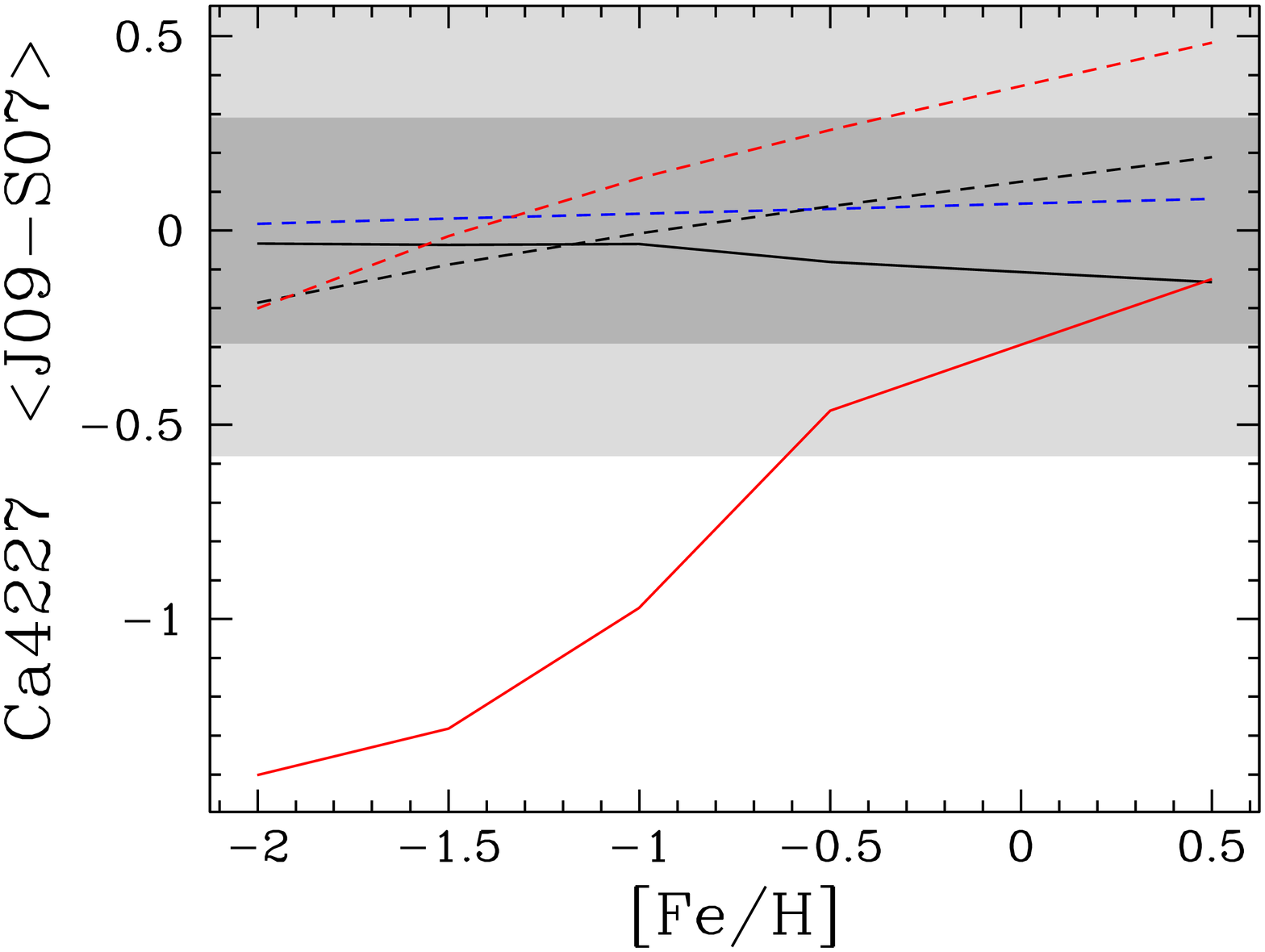}\includegraphics[scale=0.21,angle=-90]{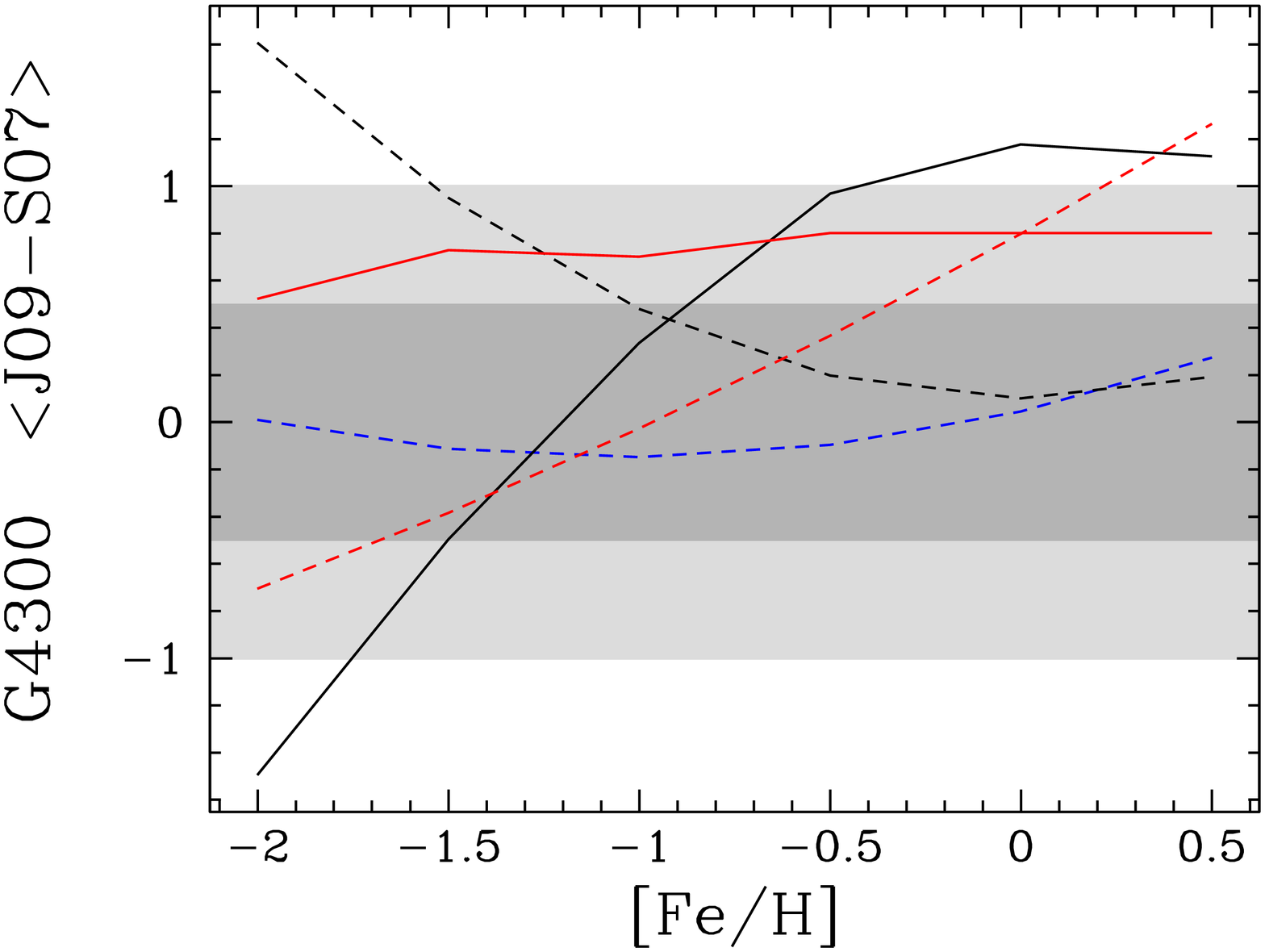}\includegraphics[scale=0.21,angle=-90]{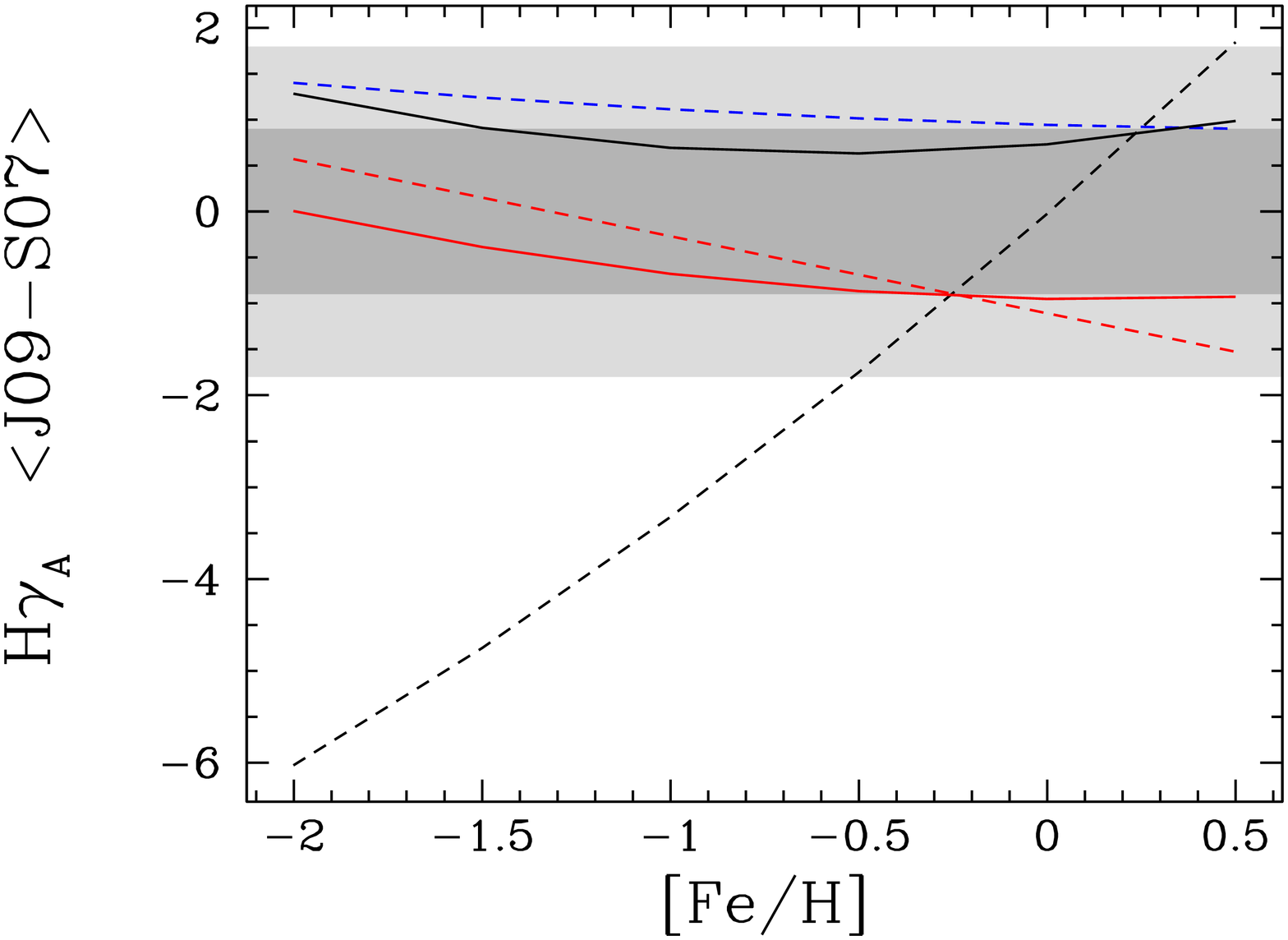}\\
\includegraphics[scale=0.21,angle=-90]{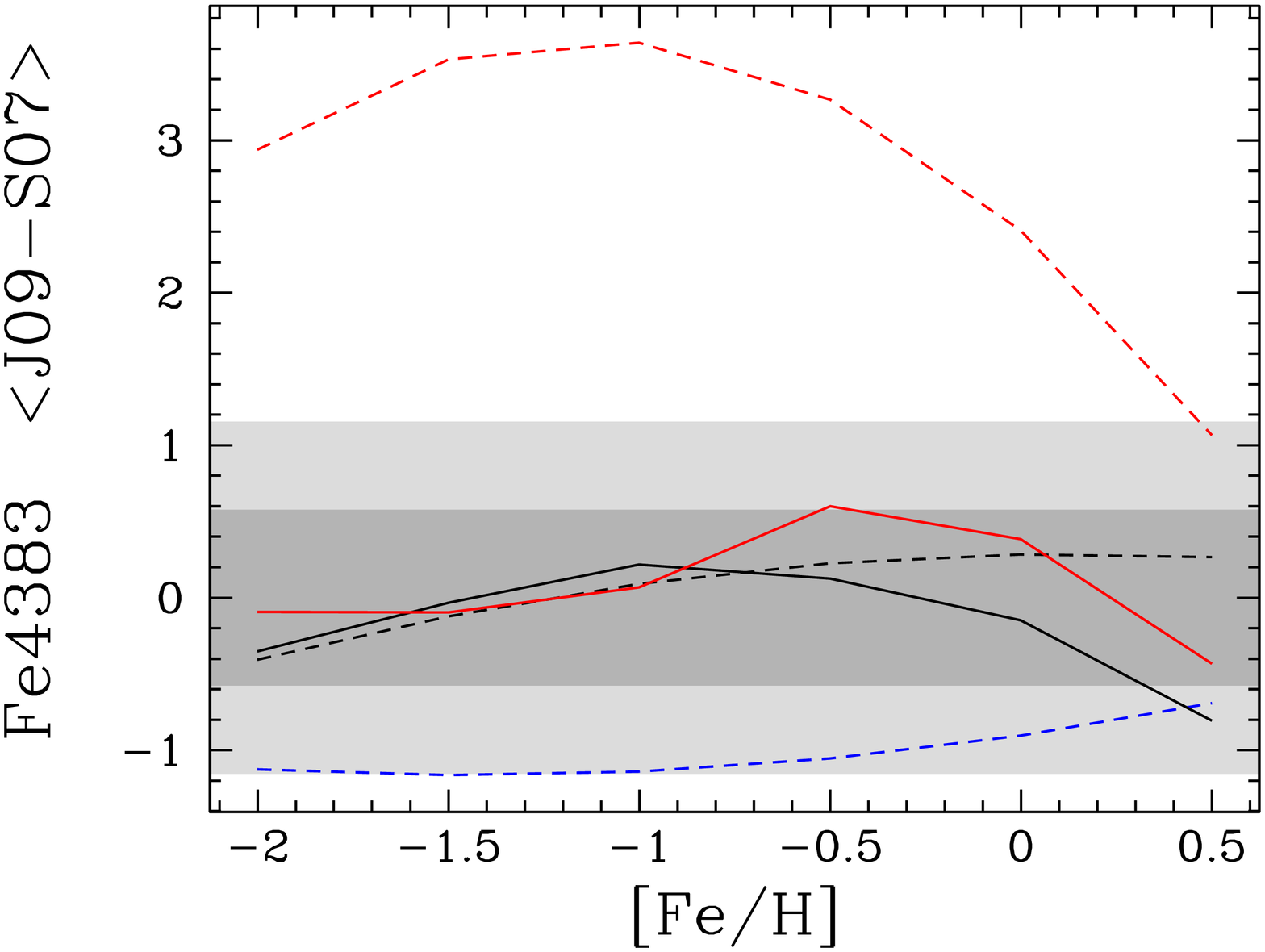}\includegraphics[scale=0.21,angle=-90]{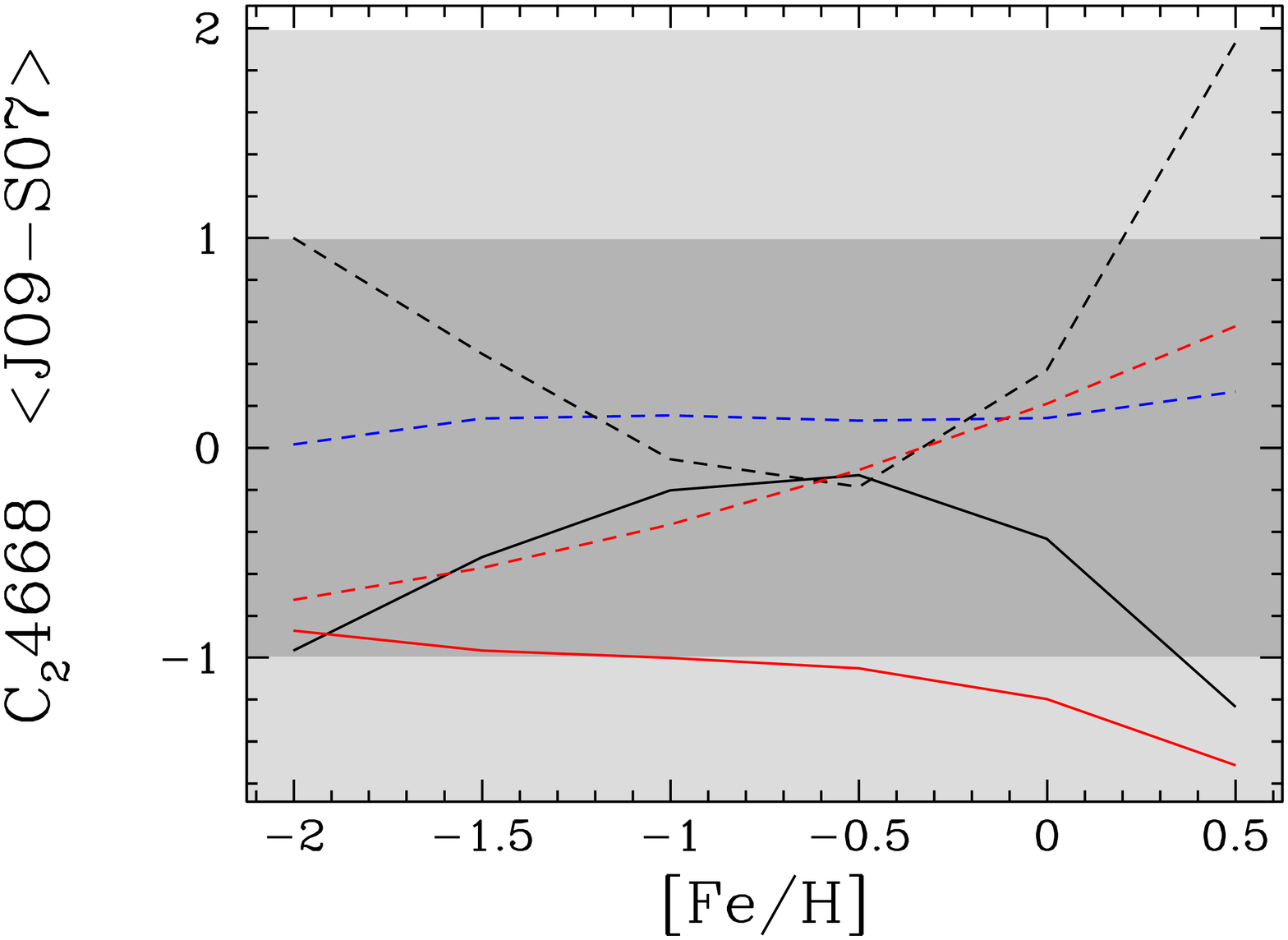}\includegraphics[scale=0.21,angle=-90]{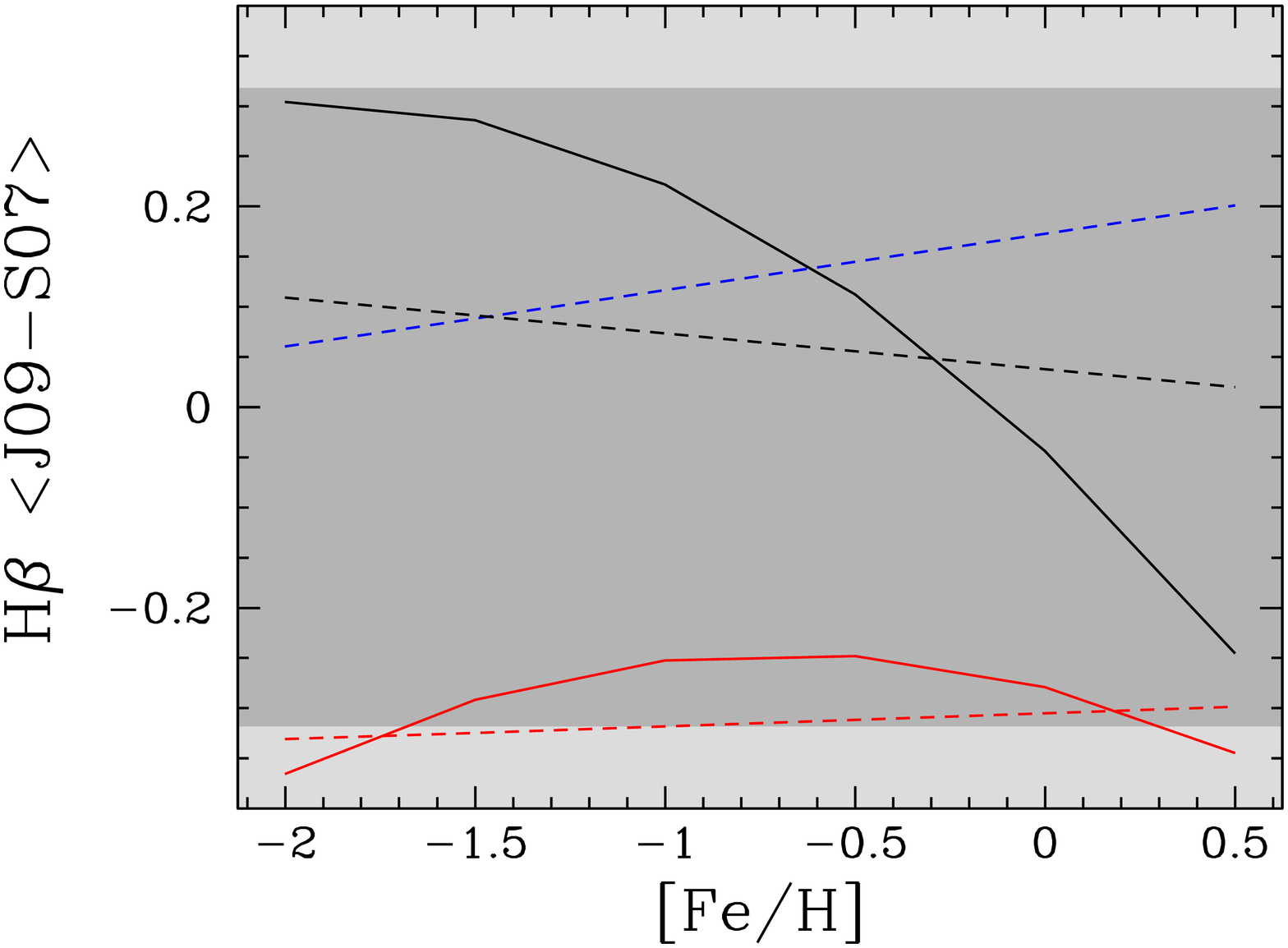}\\
\caption{Same as Fig.~\ref{WFFcomp}, but for the comparison between this work (J09) and \citet{schiavon07} (S07).
The errors are represented by index errors of the \emph{MILES} library (see Section~\ref{tierr}). }
\label{SFFcomp}
\end{minipage}
\end{figure*}

\begin{flushleft}\textbf{CN$\mathbf{_1}$}\\\end{flushleft}
\emph{WFF comparison (Fig.~\ref{WFFcomp}):} The \emph{Warm} end for \emph{Giants} show 
significantly stronger indices for this work, extending well beyond the 2rms level. 
Otherwise, this work show in general stronger indices
at the metal-poor end and weaker indices at the metal-rich end, out to the 2rms level in both cases.\\
\emph{SFF comparison (Fig.~\ref{SFFcomp}):} \emph{Intermediate} temperature \emph{Giants} and \emph{Dwarfs} show
weaker and stronger indices for this work, respectively, at the metal-rich end. Otherwise are agreements within the
1rms level mainly found.\\ 

\begin{flushleft}\textbf{CN$\mathbf{_2}$}\\\end{flushleft}
\emph{WFF comparison (Fig.~\ref{WFFcomp}):} Similar to the previous index, 
but \emph{Warm} \emph{Dwarfs} show weaker indices for the entire metallicity scale for this work, out to the 2rms level.\\
\emph{SFF comparison (Fig.~\ref{SFFcomp}):} Due to problems with implementing
the SFFs we can not make a reliable comparison.\\ 

\begin{flushleft}\textbf{Ca4227}\\\end{flushleft}
\emph{WFF comparison (Fig.~\ref{WFFcomp}):} \emph{Cold} \emph{Dwarfs} show stronger indices for this work, extending out to the
2rms level in the metal-poor regime, while \emph{Cold} \emph{Giants} instead show weaker indices for this work
out to the 2rms level at the metal-poor end. \emph{Warm} \emph{Giants} show stronger indices for this work
beyond the 1rms level at the metal-poor end.  
\emph{Cold} \emph{Dwarfs} show stronger indices for this work, even extending beyond the 2rms level at the metal-poor end.\\
\emph{SFF comparison (Fig.~\ref{SFFcomp}):} The most prominent difference
is found for \emph{Cold} \emph{Giants} in the metal-poor regime, extending well beyond the 2rms level. 
\emph{Cold} \emph{Dwarfs} show stronger indices for this work at the metal-rich end, beyond the 1rms level. 
Otherwise are differences within the 1rms level found.\\ 

\begin{flushleft}\textbf{G4300}\\\end{flushleft}
\emph{WFF comparison (Fig.~\ref{WFFcomp}):} \emph{Warm} \emph{Giants} extend well beyond the 2rms level with
stronger indices for this work. Metal-poor \emph{Cold} \emph{Giants} show stronger indices for this work, 
extending to the 2rms level. \emph{Cold} metal-poor \emph{Dwarfs} extend beyond the 1rms level, 
showing weaker indices for this work. \emph{Cold} and \emph{Warm} metal-rich \emph{Dwarfs} show stronger and weaker indices
for this work, respectively, beyond the 1rms level.\\
\emph{SFF comparison (Fig.~\ref{SFFcomp}):} Differences beyond the 1rms and 2rms levels are found in
several regimes, strongest at the ends of the metallicity scale.\\ 

\begin{flushleft}\textbf{H$\mathbf{\gamma_A}$}\\\end{flushleft}
\emph{WFF comparison (Fig.~\ref{WFFcomp}):} \emph{Warm} \emph{Giants} show significantly
weaker indices for this work, well beyond the 2rms level. 
\emph{Warm} \emph{Dwarfs} and \emph{Cold} \emph{Giants} show stronger indices for this work out to the 2rms level.
Otherwise are mainly differences within the 1rms level found.\\
\emph{SFF comparison (Fig.~\ref{SFFcomp}):} The most significant difference
is found for \emph{Intermediate} temperature \emph{Dwarfs}, showing weaker indices
for this work in the metal-poor regime well beyond the 2rms level. Otherwise are differences mainly found
around the 1rms level.\\

\begin{flushleft}\textbf{H$\mathbf{\gamma_F}$}\\\end{flushleft}
\emph{WFF comparison (Fig.~\ref{WFFcomp}):} Weaker indices are found for this work for \emph{Warm} \emph{Giants} 
and \emph{Cold} \emph{Dwarfs} beyond the 2rms and 1rms level, respectively. 
Otherwise are mainly differences within the 1rms level found.\\
\emph{SFF comparison (Fig.~\ref{SFFcomp}):} Due to problems with implementing
the SFFs we can not make a reliable comparison.\\ 

\begin{flushleft}\textbf{Fe4383}\\\end{flushleft}
\emph{WFF comparison (Fig.~\ref{WFFcomp}):} \emph{Warm} \emph{Giants} and \emph{Warm} \emph{Dwarfs} show stronger
indices out to the 2rms level at the metal-poor end. \emph{Cold} \emph{Giants} show weaker
indices, out to the 2rms level at the metal-poor and metal-rich ends. 
Otherwise are mainly differences within the 1rms level found.\\
\emph{SFF comparison (Fig.~\ref{SFFcomp}):} \emph{Cold} \emph{Dwarfs} show significantly 
stronger indices for this work, well beyond the 2rms level. \emph{Warm} \emph{Dwarfs} instead show weaker indices, out to
the 2rms level. Otherwise are mainly differences within the 1rms found.\\ 

\begin{figure*}
\begin{minipage}{17cm}
\centering
\includegraphics[scale=0.21,angle=-90]{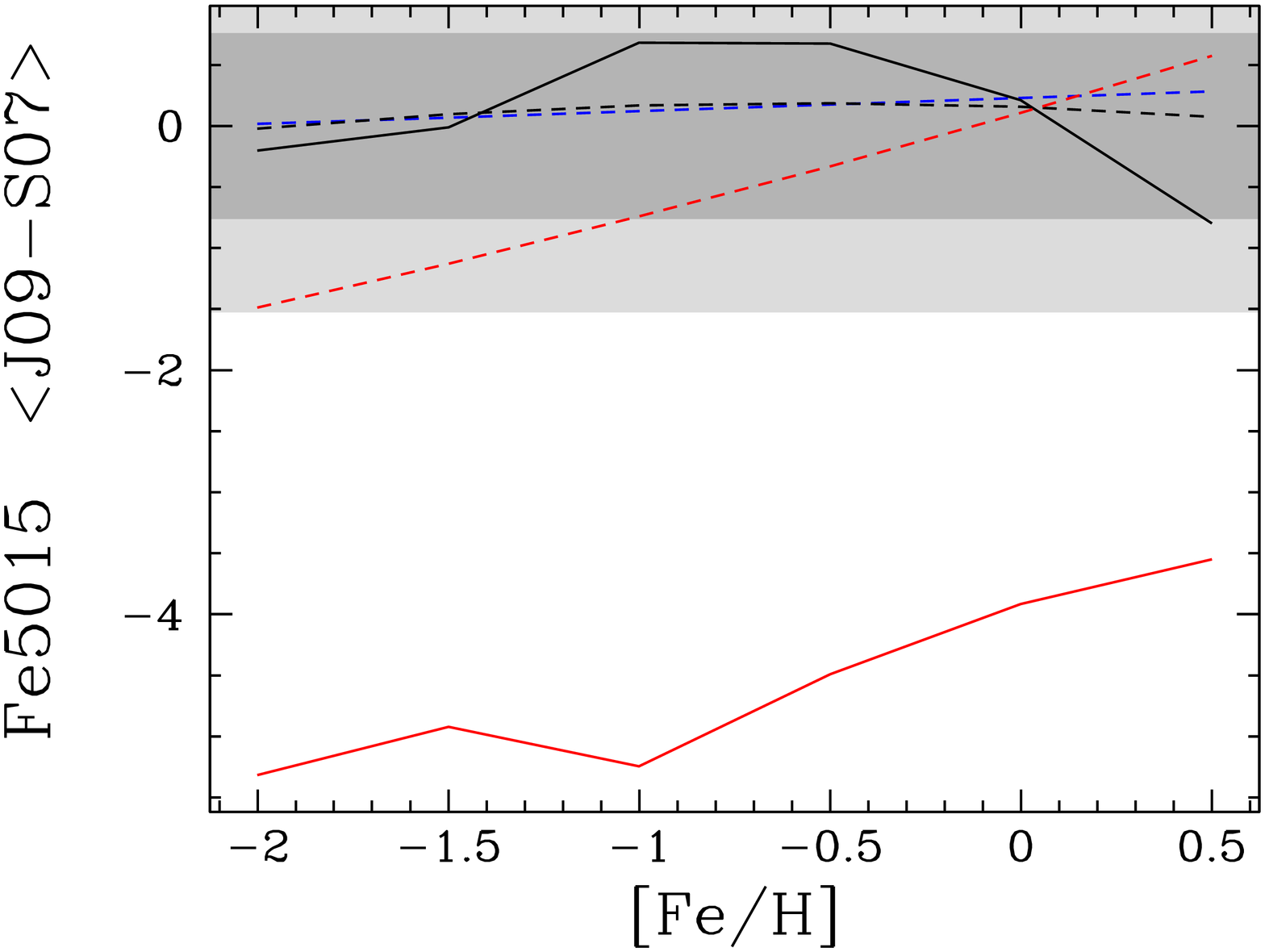}\includegraphics[scale=0.21,angle=-90]{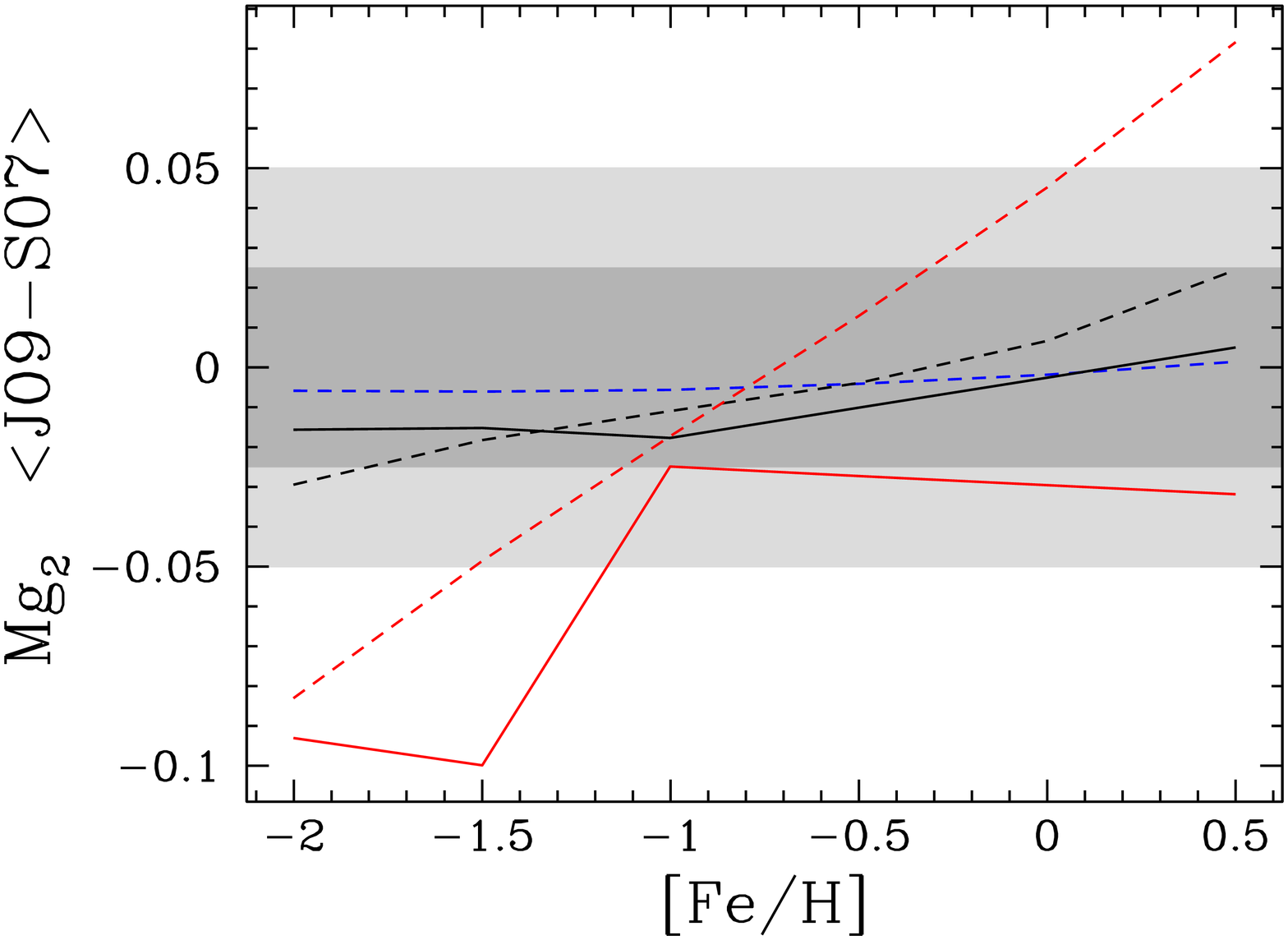}\includegraphics[scale=0.21,angle=-90]{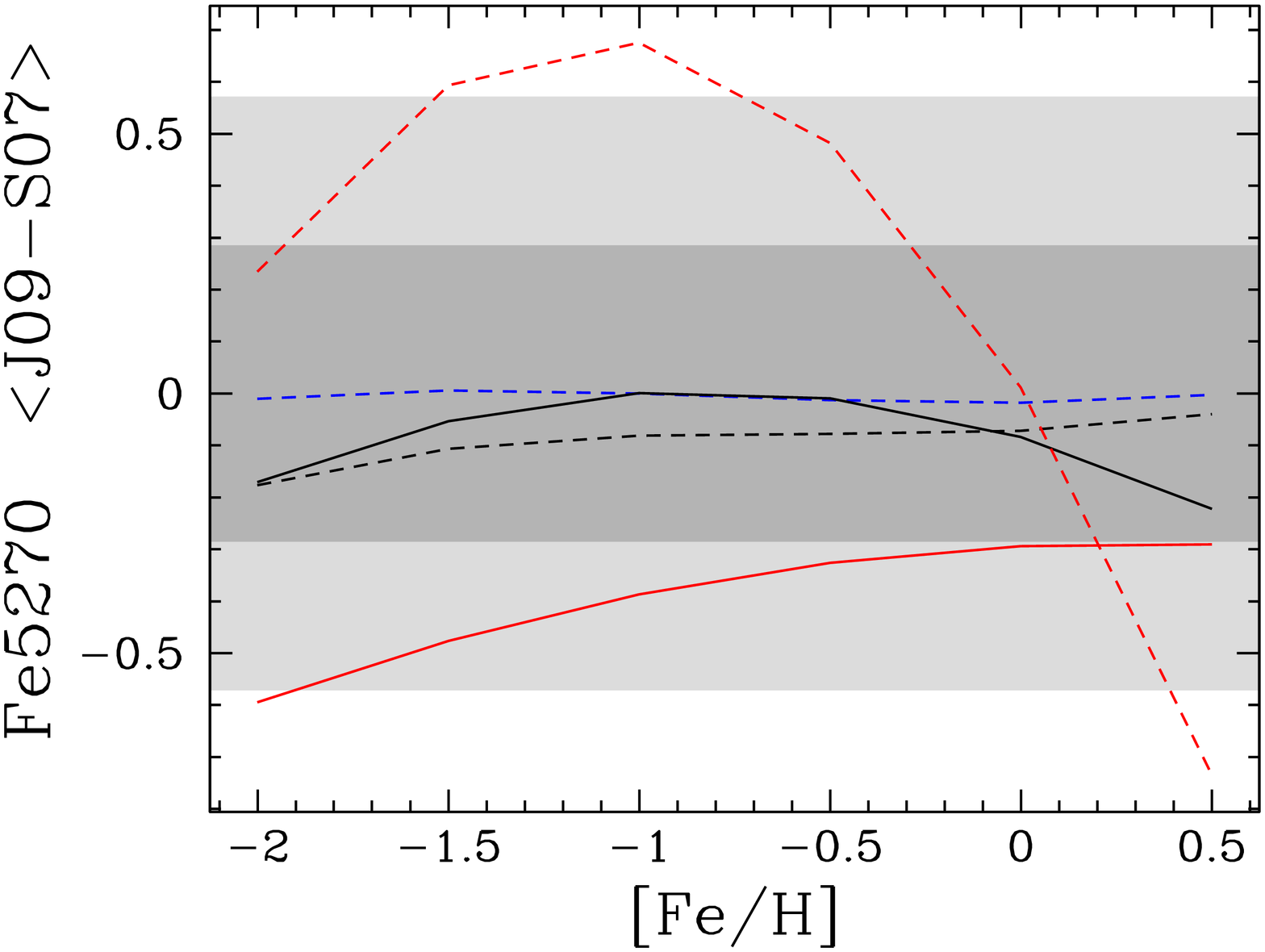}\\
\includegraphics[scale=0.21,angle=-90]{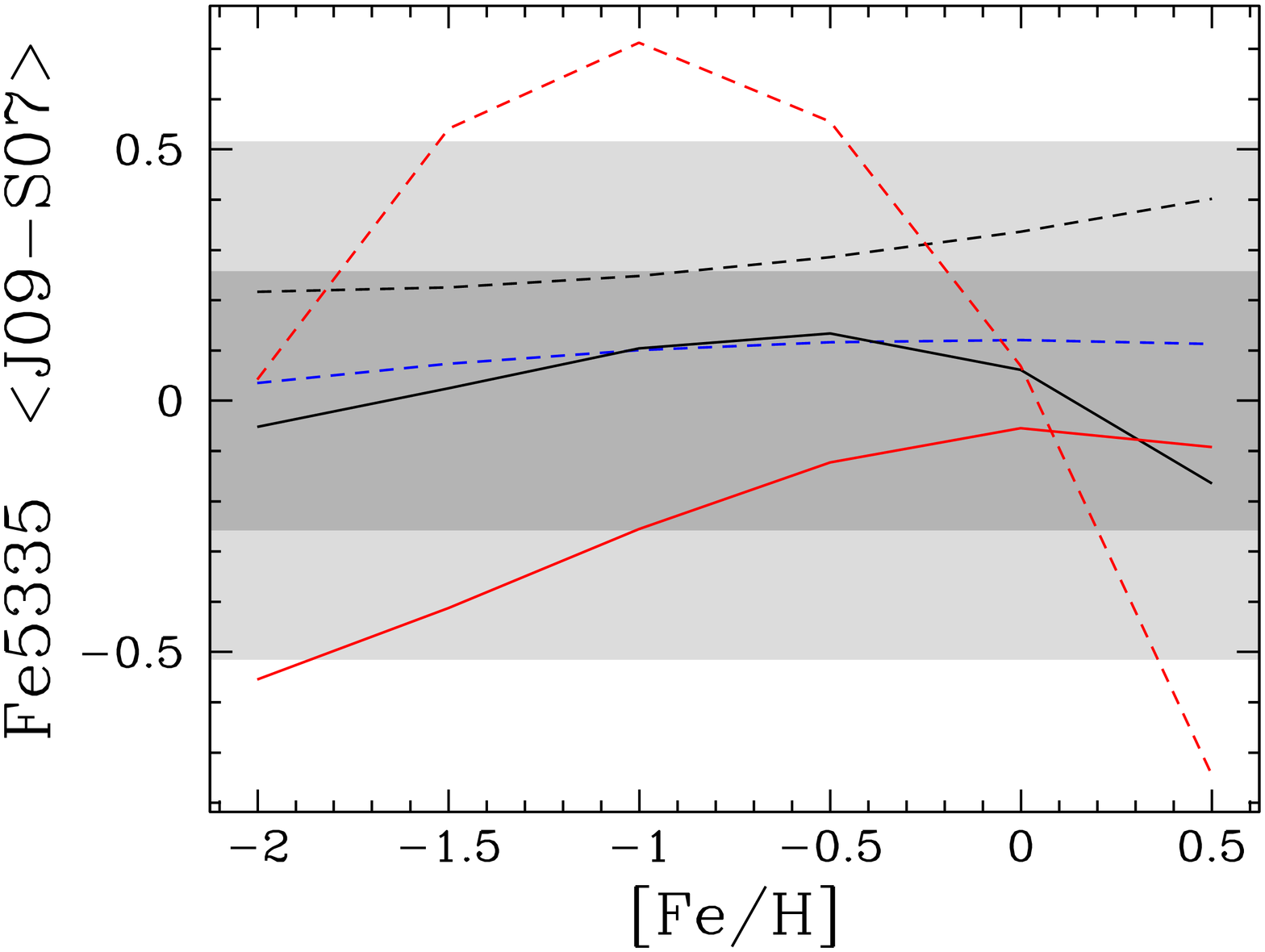}\\
\contcaption{}
\end{minipage}
\end{figure*}

\begin{flushleft}\textbf{Ca4455}\\\end{flushleft}
\emph{WFF comparison (Fig.~\ref{WFFcomp}):} \emph{Cold} \emph{Giants} show stronger indices for this work, 
extending well beyond the 2rms level in the metal-rich regime. \emph{Cold} \emph{Dwarfs} show weaker indices for this work,
extending well beyond the 2rms level. The \emph{Warm} regime show stronger indices for this work, extending beyond the
1rms level in the metal-poor regime.\\

\begin{flushleft}\textbf{Fe4531}\\\end{flushleft}
\emph{WFF comparison (Fig.~\ref{WFFcomp}):} \emph{Cold} \emph{Dwarfs} show stronger indices for this work 
at the 1rms level for the metal-poor end and increasing well beyond the 2rms 
level at the metal-rich end. \emph{Intermediate} temperature \emph{Giants} show stronger indices for this work, out to
the 2rms level at the metal-rich end.\\

\begin{flushleft}\textbf{C$\mathbf{_2}4668$}\\\end{flushleft}
\emph{WFF comparison (Fig.~\ref{WFFcomp}):} The metal-rich end show 
weaker indices for this work, mainly down to the 1rms level. The metal-poor
end show weaker and stronger indices for this work extending beyond the 1rms level
for the \emph{Warm} bins and \emph{Cold} \emph{Dwarfs}, respectively.\\
\emph{SFF comparison (Fig.~\ref{SFFcomp}):} \emph{Cold} \emph{Giants} show weaker indices for this work beyond the
1rms level in the metal-rich regime. \emph{Intermediate} temperature \emph{Dwarfs} show stronger indices for this work
at the metal-rich end out to the 2rms level.\\ 

\begin{flushleft}\textbf{H$\mathbf{\beta}$}\\\end{flushleft}
\emph{WFF comparison (Fig.~\ref{WFFcomp}):} We find this work to show
weaker indices for \emph{Warm} \emph{Giants} well beyond the 2rms level. \emph{Cold} \emph{Dwarfs} show 
stronger indices for this work beyond the 2rms level. \emph{Cold} \emph{Giants} show 
weaker indices for this work beyond the 1rms level.\\
\emph{SFF comparison (Fig.~\ref{SFFcomp}):} \emph{Cold} \emph{Dwarfs} show the biggest differences right 
beyond the 1 rms level. Otherwise are differences within the 1rms level are found.\\ 

\begin{flushleft}\textbf{Fe5015}\\\end{flushleft}
\emph{WFF comparison (Fig.~\ref{WFFcomp}):} \emph{Cold} \emph{Giants} show weaker indices for this work, beyond 
the 2rms level in the metal-poor regime. \emph{Intermediate} temperature \emph{Giants} show stronger indices for this work beyond 
the 2rms level in the metal-rich regime.\\
\emph{SFF comparison (Fig.~\ref{SFFcomp}):} \emph{Cold} \emph{Giants} show significantly weaker indices for this work, well
beyond the 2rms level. This work shows weaker indices for \emph{Cold} \emph{Dwarfs} beyond the 1rms level in the metal-poor regime.
Otherwise are differences within the 1rms mainly found.\\ 

\begin{flushleft}\textbf{Mg$\mathbf{_1}$}\\\end{flushleft}
\emph{WFF comparison (Fig.~\ref{WFFcomp}):} The \emph{Cold} \emph{Dwarfs} show stronger indices
for this work, out to the 2rms level in the metal-poor regime. The \emph{Cold} \emph{Giants} show stronger indices for this
work at intermediate metallicities. No differences found beyond the 2rms level.\\

\begin{flushleft}\textbf{Mg$\mathbf{_2}$}\\\end{flushleft}
\emph{WFF comparison (Fig.~\ref{WFFcomp}):} \emph{Cold} \emph{Dwarfs} show stronger indices for this work beyond the
1rms level in the metal-poor regime. 
\emph{Cold} \emph{Giants} show stronger indices for this work beyond the 1rms level for intermediate metallicities.
Otherwise are mainly differences within the 1rms level found. \\ 
\emph{SFF comparison (Fig.~\ref{SFFcomp}):} The \emph{Cold} end show weaker indices for this work in the metal-poor regime, beyond the
2rms level. \emph{Cold} \emph{Dwarfs} instead show stronger indices for this work beyond the 2rms level at the metal-rich end. \\

\begin{flushleft}\textbf{Mgb}\\\end{flushleft}
\emph{WFF comparison (Fig.~\ref{WFFcomp}):} \emph{Cold} and \emph{Intermediate} temperature \emph{Giants} show 
weaker indices for this work, extending beyond the 2rms level. 
\emph{Intermediate} temperature and \emph{Cold} \emph{Dwarfs} show weaker and stronger indices, 
respectively, for this work in the Metal-poor regime, beyond the 1rms level.\\
\emph{SFF comparison (Fig.~\ref{SFFcomp}):} Due to problems implementing
the SFFs we can not make a reliable comparison.\\

\begin{flushleft}\textbf{Fe5270}\\\end{flushleft}
\emph{WFF comparison (Fig.~\ref{WFFcomp}):} The \emph{Warm} end show stronger indices beyond the 2rms level 
in the metal-poor regime. \emph{Cold} \emph{Giants} show weaker indices for this work beyond the 2rms level.
The metal-rich end show stronger indices for this work beyond the 1rms level for \emph{intermediate} temperature
\emph{cold} \emph{Dwarfs}.\\
\emph{SFF comparison (Fig.~\ref{SFFcomp}):} We find weaker indices
for this work beyond the 1rms level for \emph{Cold} \emph{Giants}. Stronger indices for this work beyond the 2rms level are found for 
\emph{Cold} \emph{Dwarfs} in the intermediate metallicity regime. Otherwise are differences well within the 1rms level found.\\ 

\begin{flushleft}\textbf{Fe5335}\\\end{flushleft}
\emph{WFF comparison (Fig.~\ref{WFFcomp}):} \emph{Cold} \emph{Giants} show weaker indices for this work well 
beyond the 2rms level. The \emph{Warm} end show stronger indices out to the 1rms level at intermediate
metallicities. \emph{Cold} \emph{Dwarfs} show weaker and stronger indices out to the 1rms level a the metal-poor
and metal-rich ends, respectively.\\
\emph{SFF comparison (Fig.~\ref{SFFcomp}):} \emph{Cold} \emph{Dwarfs} show stronger indices beyond the 2rms level at 
intermediate metallicities and weaker indices beyond the 2rms level at the metal-rich end. \emph{Cold} \emph{Giants}
show weaker indices for this work beyond the 1rms level. \emph{Intermediate} temperature \emph{Dwarfs} instead 
show stronger indices for this work, beyond the 1rms level in the metal-rich regime.\\ 

\begin{flushleft}\textbf{Fe5406}\\\end{flushleft}
\emph{WFF comparison (Fig.~\ref{WFFcomp}):} The \emph{Cold} end show stronger indices for this work beyond the
1rms level in the metal-rich regime. \emph{Intermediate} temperature \emph{Giants} show weaker indices for this work beyond the
1rms level. No differences found beyond the 2rms level.\\

\begin{flushleft}\textbf{Fe5709}\\\end{flushleft}
\emph{WFF comparison (Fig.~\ref{WFFcomp}):} \emph{Cold} \emph{Dwarfs} show stronger indices for this work, 
extending beyond the 2rms level at the metal-rich end. Otherwise, no significant differences beyond the 1rms level.\\

\begin{flushleft}\textbf{Fe5782}\\\end{flushleft}
\emph{WFF comparison (Fig.~\ref{WFFcomp}):} \emph{Warm} \emph{Giants} show stronger indices for this work in the
metal-poor regime, extending beyond the 2rms level at the metal-poor end. 
\emph{Cold} \emph{Giants} show stronger indices for this work regime, beyond the 1rms level. \emph{Warm} \emph{Dwarfs}
show stronger indices for this work beyond the 1rms level at the metal-poor end.\\

\begin{flushleft}\textbf{NaD}\\\end{flushleft}
\emph{WFF comparison (Fig.~\ref{WFFcomp}):} The \emph{Cold} end show stronger indices for this
work extending well beyond the 2rms level, especially in the metal-poor regime.
\emph{Intermediate} temperature \emph{Giants} show stronger indices for this
work, extending beyond the 2rms level in the metal-rich regime.\\

\begin{flushleft}\textbf{TiO$_1$}\\\end{flushleft}
\emph{WFF comparison (Fig.~\ref{WFFcomp}):} \emph{Cold} \emph{Dwarfs} show weaker indices for this work, 
extending well beyond the 1rms level at the metal-rich end. Otherwise, no differences 
found beyond the 1rms level.\\

\begin{flushleft}\textbf{TiO$_2$}\\\end{flushleft}
\emph{WFF comparison (Fig.~\ref{WFFcomp}):} We find significantly weaker indices for this
work for \emph{Cold} \emph{Dwarfs}, extending very far beyond the 2rms level. Otherwise, no significant differences 
found beyond the 1rms level.\\

\section{summary}
\label{concs}

We have derived new empirical fitting functions for the relationship between 
Lick absorption indices and stellar atmospheric parameters (T$_{\rm eff}$, [Fe/H] and $\log g$) 
described by the \emph{MILES} library of stellar spectra, both for the resolution of the 
\emph{MILES} library and for the resolution of the Lick/IDS library. The \emph{MILES} library 
consists of 985 stars selected to produce a sample with extensive stellar parameter coverage. 
The \emph{MILES} library was also chosen because
it has been carefully flux-calibrated, making standard star derived offsets unnecessary. This
becomes important when comparing stellar population models to high redshift data where
no resolved individual stars are available. 

We find the index measurements of the \emph{MILES} spectra to have very high quality in terms
of observational index errors. These errors are also found to be significantly smaller
than for the Lick/IDS library. This was expected since the \emph{MILES} library was observed nearly thirty
years after the Lick/IDS library. Given the high quality of the index measurements, index errors
 should not be the major error sources for the final fitting functions. We instead find indications 
that the stellar parameter estimates are significant error sources.  

Lick Index offsets between the \emph{MILES} library and the classic
Lick/IDS library are derived in order to be able to compare stellar population models based
on this work with models in the literature. We find these offsets to be dependent on index strength
and have therefore derived least-square fits for the residual between the two libraries.
Offset to the Lick/IDS library are also derived for the flux-calibrated \emph{ELODIE} and \emph{STELIB} libraries.
We find clear offset deviations between the libraries. The largest deviations are found for the 
\emph{STELIB} library compared to the other two libraries, which is also the library having least 
stars in common with the Lick/IDS library. The deviations in offsets found between the three libraries
undermine the derivation of universal offsets between the Lick/IDS and these flux-calibrated systems.

We compare the fitting functions of this work to fitting functions 
in the literature, namely the fitting functions of \citet{worthey94}, \citet{worthey97} and 
\citet{schiavon07}. Generally we find good agreement within the rms of the residuals between the data and  
the fitting functions of this work. The differences found in the comparisons vary significantly from 
index to index and especially from one stellar parameter region to another for individual indices.
However, the major differences are found in the outskirts of stellar parameter
space, i.e. at the temperature and metallicity ends. This is probably due to a low number of data points in 
these regimes for the stellar libraries, inducing uncertainties which result in the major differences found.

In a forthcoming paper (Thomas et al. in prep.) the fitting functions of this work will be 
implemented in stellar population models following the techniques of \citet{maraston05} and \citet{TMB03}.

A user friendly Fortran 90 code is available online at www.icg.port.ac.uk/$\sim$johanssj
to easy the implementation of our fitting functions in population synthesis codes.

\section*{ACKNOWLEDGMENTS}
We thank Patricia S{\'a}nchez-Bl{\'a}zquez for help with the \emph{MILES} library. 
We thank Ricardo Schiavon and Rita Tojeiro for useful discussions regarding the empirical fitting method. 
We also thank an anonymous referee for a prompt and constructive report.


{}

\clearpage

\appendix

\begin{figure*}
\section[]{Fitting functions for Lick/IDS resolution}
\label{ffapp}
Short version, full appendix can be found at www.icg.port.ac.uk/$\sim$johanssj
\begin{minipage}{17cm}
\centering
\includegraphics[scale=0.4]{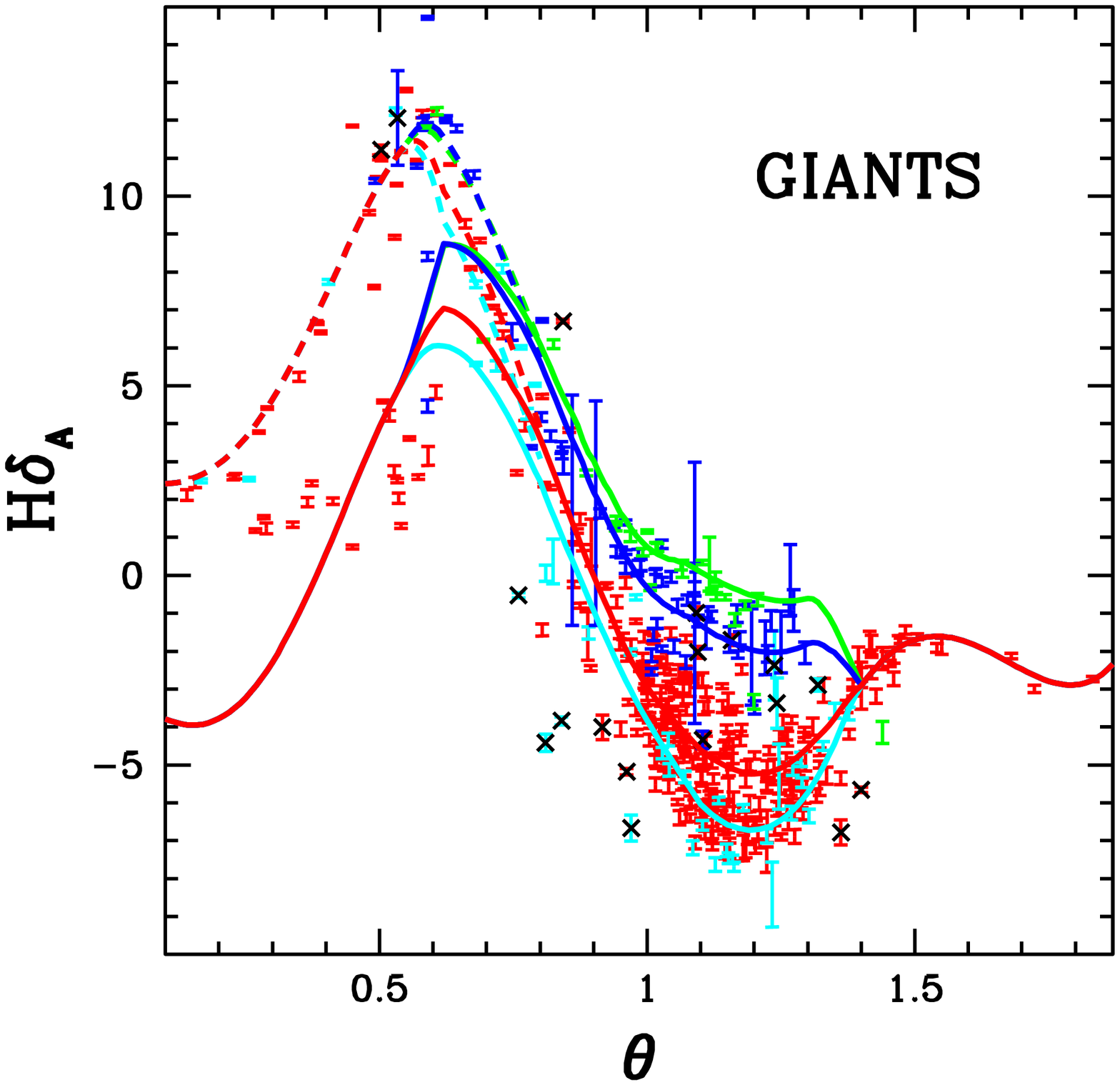}\includegraphics[scale=0.4]{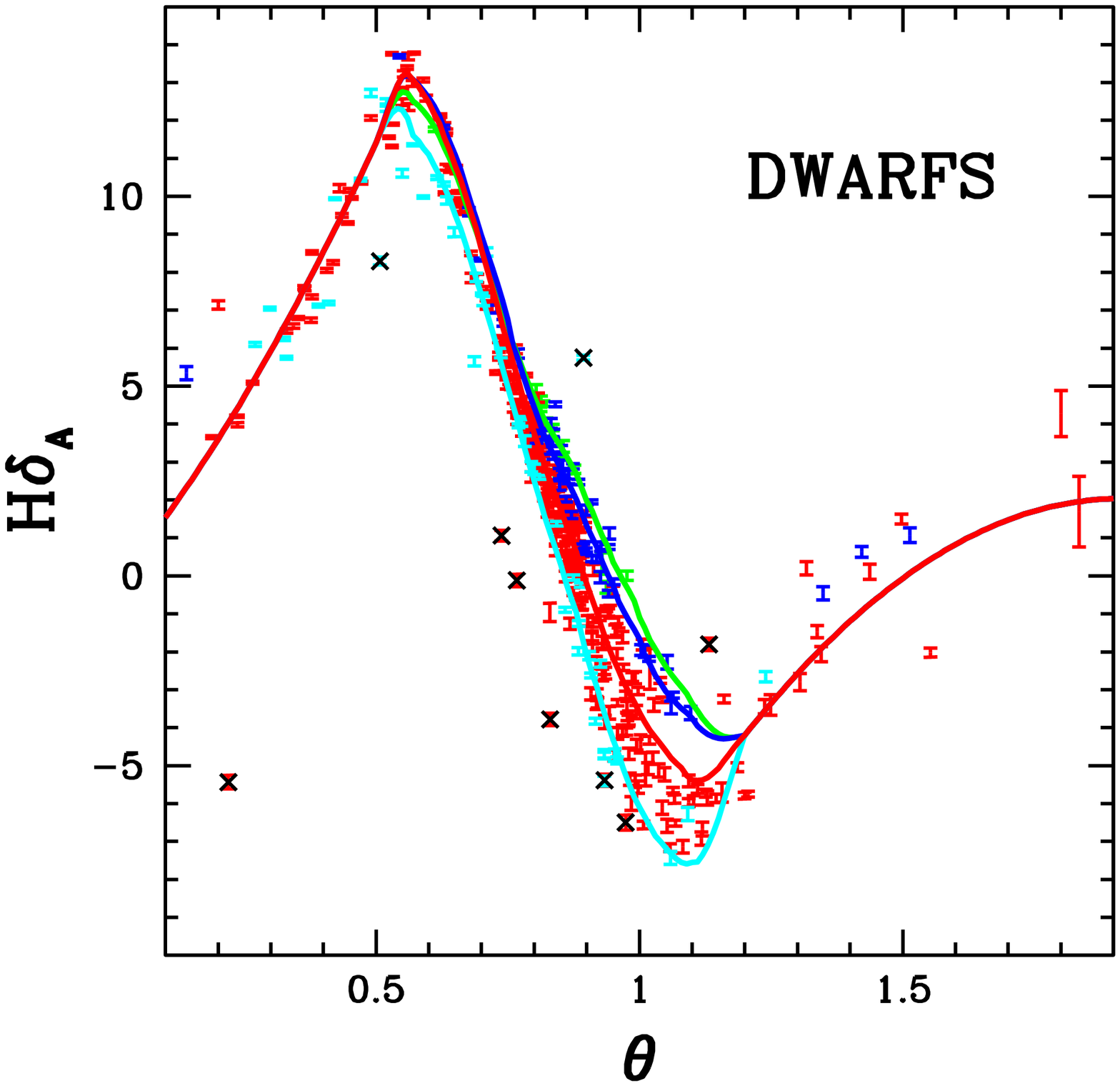}\\
\includegraphics[scale=0.4]{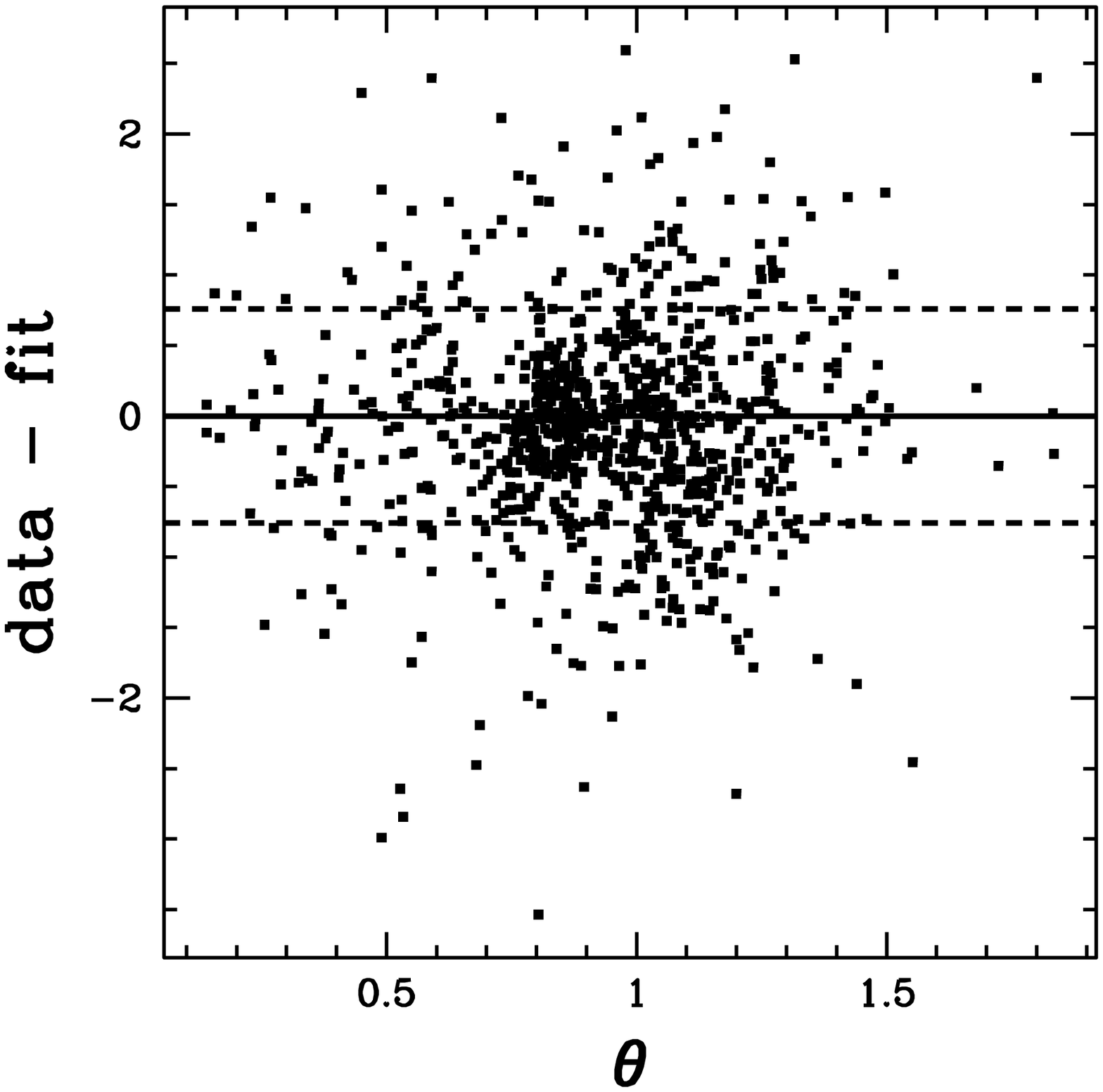}\includegraphics[scale=0.4]{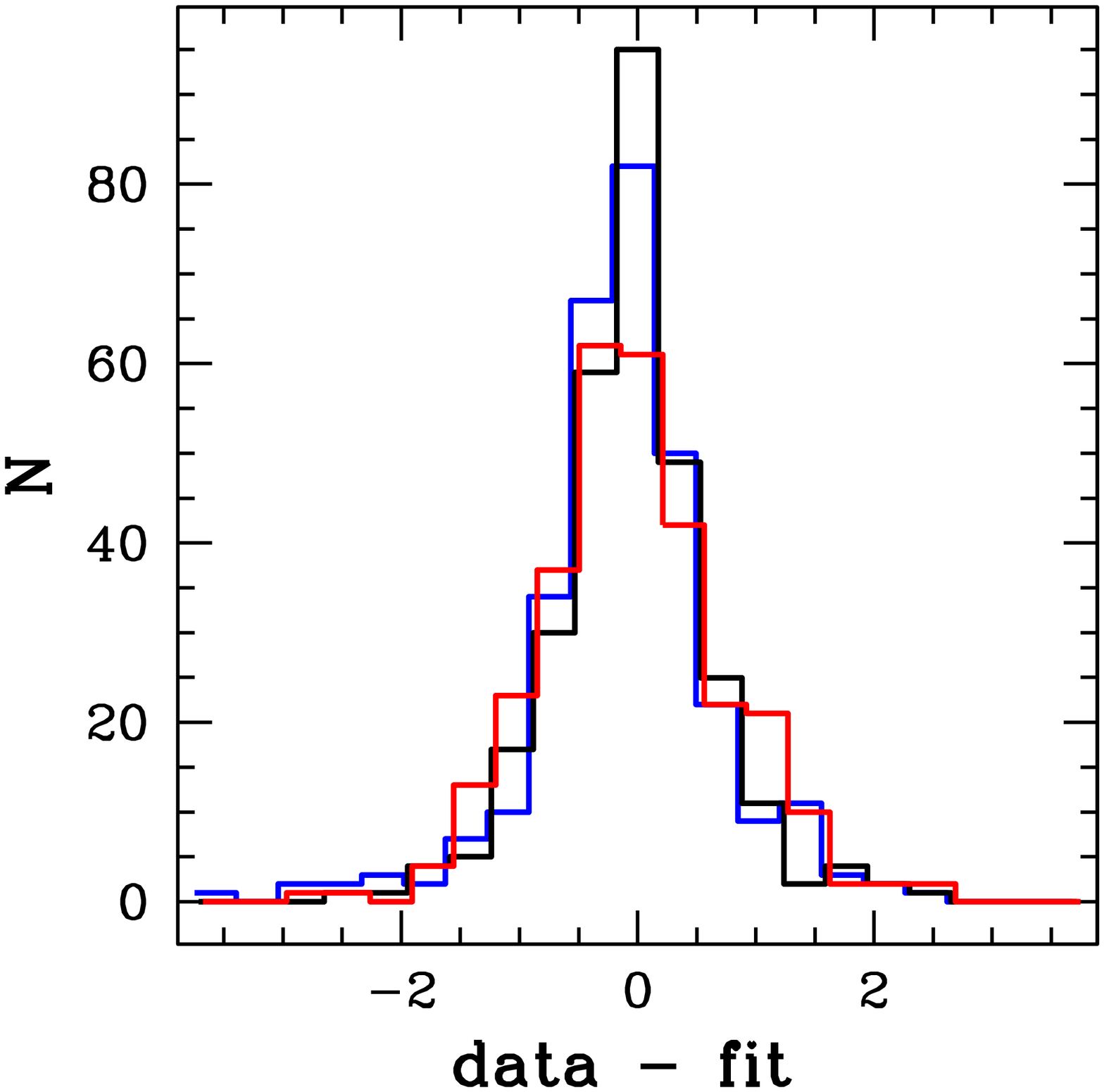}\\
\caption{H$_{\delta}$A, same as Fig. 3. Dashed-lines for additional $\log g$-value ($\log g=3.3$) cover the data points 
in strong $\log g$-dependent regions.}
\label{indfigs_hda}
\end{minipage}
\end{figure*}

\clearpage
\begin{table*}
\section[]{Fitting functions for \emph{MILES} resolution}
\label{ffapp_mr}
Short version, full appendix can be found at www.icg.port.ac.uk/$\sim$johanssj
\caption{\small{\textit{H$\delta_A$ fitting function coefficients for \emph{MILES} resolution}}}
\label{hdatable_mr}
\begin{tabular}{|c|c|c|c|c|c|c|c|c|}
\hline
\multicolumn{9}{|c|}{\bf \small overall rms=0.8009}\\
  & \multicolumn{4}{|c|}{\small $\log g\le 4.0$ and $\theta$ limits:} & \multicolumn{4}{|c|}{\small $\log g\ge 3.6$ and $\theta$ limits:} \\
\small Term & \small $\le 0.62$ & \small $0.53-1.1$ & \small $0.90-1.4$ & \small $\ge 1.3$ & \small $\le 0.57$ & \small $0.50-0.90$ & \small $0.80-1.2$ & \small $\ge 1.08$ \\
\hline     
\scriptsize Const.               &\scriptsize -8.002  &\scriptsize -136.8 &\scriptsize 36.39  &\scriptsize -127.4 &\scriptsize -3.369 &\scriptsize -134.7  &\scriptsize -49.75  &\scriptsize -41.96 \\
\scriptsize $\theta$             &\scriptsize -12.19  &\scriptsize 487.8  &\scriptsize x      &\scriptsize 23.06  &\scriptsize 16.05  &\scriptsize 596.3   &\scriptsize 297.8   &\scriptsize 42.42 \\
\scriptsize [Fe/H]               &\scriptsize x       &\scriptsize -2.367 &\scriptsize 41.22  &\scriptsize x      &\scriptsize x      &\scriptsize -13.31  &\scriptsize 31.61   &\scriptsize x          \\
\scriptsize $\log g$             &\scriptsize -4.330  &\scriptsize 21.43  &\scriptsize 2.662  &\scriptsize x      &\scriptsize x      &\scriptsize x       &\scriptsize -0.5889 &\scriptsize x                \\
\scriptsize $\theta^{2}$         &\scriptsize x       &\scriptsize -550.1 &\scriptsize -88.27 &\scriptsize 103.1  &\scriptsize 13.26  &\scriptsize -922.4  &\scriptsize -423.0  &\scriptsize -9.752  \\
\scriptsize [Fe/H]$^{2}$         &\scriptsize x       &\scriptsize x      &\scriptsize 0.5080 &\scriptsize x      &\scriptsize x      &\scriptsize -3.267  &\scriptsize x       &\scriptsize x  \\
\scriptsize $\log g^{2}$         &\scriptsize 5.072   &\scriptsize x      &\scriptsize x      &\scriptsize x      &\scriptsize 0.1957 &\scriptsize x       &\scriptsize x       &\scriptsize x  \\
\scriptsize $\theta$[Fe/H]       &\scriptsize x       &\scriptsize x      &\scriptsize -78.94 &\scriptsize x      &\scriptsize x      &\scriptsize 31.04   &\scriptsize -68.15  &\scriptsize x        \\
\scriptsize $\theta \log g$      &\scriptsize x       &\scriptsize -44.24 &\scriptsize -6.149 &\scriptsize x      &\scriptsize 0.2840 &\scriptsize 26.90   &\scriptsize x       &\scriptsize x                \\
\scriptsize [Fe/H]$\log g$       &\scriptsize x       &\scriptsize 0.2318 &\scriptsize x      &\scriptsize x      &\scriptsize x      &\scriptsize x       &\scriptsize x       &\scriptsize x                \\
\scriptsize $\theta^{3}$         &\scriptsize 273.0   &\scriptsize 195.6  &\scriptsize 48.49  &\scriptsize 49.23  &\scriptsize x      &\scriptsize 448.91  &\scriptsize 175.3   &\scriptsize x \\
\scriptsize [Fe/H]$^{3}$         &\scriptsize x       &\scriptsize 0.5286 &\scriptsize 0.3213 &\scriptsize x      &\scriptsize x      &\scriptsize -0.1733 &\scriptsize -0.1461 &\scriptsize x \\
\scriptsize $\log g^{3}$         &\scriptsize -0.8813 &\scriptsize x      &\scriptsize x      &\scriptsize x      &\scriptsize x      &\scriptsize x       &\scriptsize x       &\scriptsize x  \\
\scriptsize $\theta^{2}$[Fe/H]   &\scriptsize x       &\scriptsize x      &\scriptsize 34.99  &\scriptsize x      &\scriptsize x      &\scriptsize -20.90  &\scriptsize 33.01   &\scriptsize x                \\
\scriptsize $\theta^{2}\log g$   &\scriptsize -16.26  &\scriptsize 22.55  &\scriptsize 3.167  &\scriptsize x      &\scriptsize x      &\scriptsize -17.20  &\scriptsize -1.042  &\scriptsize x                \\
\scriptsize $\theta$[Fe/H]$^{2}$ &\scriptsize x       &\scriptsize 1.297  &\scriptsize x      &\scriptsize x      &\scriptsize x      &\scriptsize 2.919   &\scriptsize -0.7101 &\scriptsize x                \\
\scriptsize [Fe/H]$^{2}\log g$   &\scriptsize x       &\scriptsize x      &\scriptsize x      &\scriptsize x      &\scriptsize x      &\scriptsize 0.02043 &\scriptsize x       &\scriptsize x                \\
\scriptsize $\theta \log g^{2}$  &\scriptsize 2.150   &\scriptsize x      &\scriptsize x      &\scriptsize x      &\scriptsize x      &\scriptsize -1.628  &\scriptsize x       &\scriptsize x                \\
\scriptsize $\theta^{4}$         &\scriptsize -275.2  &\scriptsize x      &\scriptsize x      &\scriptsize -108.1 &\scriptsize x      &\scriptsize x       &\scriptsize x       &\scriptsize x     \\
\scriptsize $\theta^{5}$         &\scriptsize x       &\scriptsize x      &\scriptsize x      &\scriptsize 31.52  &\scriptsize x      &\scriptsize x       &\scriptsize x       &\scriptsize x     \\
\hline
\small rms                       &\small 1.033        &\small 1.076       &\small 0.8149      &\small 0.6510      &\small 0.8884      &\small 0.5558       &\small 0.5954       &\small 1.477   \\     
\small N                         &\small 96           &\small 346         &\small 359         &\small 41          &\small 49          &\small 278          &\small 276          &\small 33      \\     
\hline
\end{tabular} 
\label{lastpage}
\end{table*}

\end{document}